\documentclass[aps,prd,preprint,groupedaddress,nofootinbib,byrevtex]{revtex4-1}
\usepackage{amsmath,amssymb}
\usepackage{mathrsfs}
\usepackage[dvips]{graphicx}

\def\Slash#1{\ooalign{\hfil/\hfil\crcr$#1$}}
\begin{document}



\title{ Matter-enhanced  transition probabilities  in quantum field theory }

\author{Kenzo Ishikawa 
and Yutaka Tobita}

\affiliation{Department of Physics, Faculty of Science, Hokkaido
University, Sapporo 060-0810, Japan}

\begin{abstract}
The relativistic quantum field theory is the unique theory that combines
 the relativity and quantum theory and is  invariant under   the
 Poincar\'e transformation. The ground state, vacuum, is singlet and one
 particle states  are transformed as elements of irreducible
 representation of the group.  The  covariant one
 particles are momentum eigenstates expressed by plane waves and
 extended in space. Although the S-matrix defined with  initial and
 final states of these states  hold the symmetries and are applied
 to isolated states, out-going
 states  for the 
amplitude of the event
 that they are detected at a finite-time interval T in   
experiments are expressed by   microscopic states  that they interact
with, and are surrounded by matters in detectors and are not  plane
 waves.  These matter-induced  effects  modify the probabilities  observed in  
realistic  situations.    The transition 
amplitudes and  probabilities   of the events  are studied with the
 S-matrix,  $S[\text T]$, that 
satisfies  the boundary condition at T.  Using $S[\text T]$,  
the finite-size corrections of the form of ${1/\text T}$  are found.  
The corrections to  the Fermi's golden rule become  larger than the 
original values in some situations for light particles. They break 
 Lorentz invariance even in high energy region of short  de Broglie
wave lengths. 
\end{abstract}



%\end{keyword}
\preprint{EPHOU-12-004}
\maketitle
% \linenumbers

\section{Introduction}
\label{}
 In high energy scattering experiments,  
initial states  formed in accelerator  are approximately plane waves of 
finite spatial sizes, and final states  identified through their
reactions with 
atoms or nucleus in detector have the microscopic sizes of these
objects.  Hence the S-matrix  for  in- or out-going states  of wave functions
of finite
sizes, wave packets,  are necessary for  the realistic experiments \cite{LSZ,Low}. The 
ordinary S-matrix is defined at the infinite-time interval, which is  
denoted by $S[\infty]$, and the total probability from  $S[\infty]$  
defined with the wave packets agrees
with that defined with plane waves. As far as they form  complete sets, 
the probability is unique and independent from the base
functions.  Computation is easiest with
the planes waves. Accordingly, the transition probability at $\text T =\infty$
has been  computed  with plane waves, and compared with the experiment. 
 Measurements  are made with   large-time intervals of
macroscopic lengths, which were  approximated with $\infty$. The
approximation was  considered very good,  because that is much larger than 
both of de Broglie wave length and  Compton wave length, ${\hbar \over p}$
and ${\hbar \over mc}$,
 for a particle of the mass and momentum
$m$ and $p$. In a previous paper, it was studied if this approximation
is always verified  using an S-matrix of satisfying  boundary
conditions at  finite-time interval T, denoted as $S[\text T ]$, \cite{Ishikawa-Tobita-ptep}.
It was found that the probabilities of the events that the decay products are 
detected at T  are different from those at $\text T \rightarrow \infty$ 
in various
situations, and that the deviation is determined by a new length  
$({\hbar \over mc})\times({E \over mc^2}) $, which  becomes large for light
particles or at high energy.     
Transition  probability $P$ of these  events at T then has  the  form,
\begin{eqnarray}
P=\text T \Gamma+P^{(d)},
\label{finite-T-probability}
\end{eqnarray}
where  $\Gamma$ is computed with
Fermi's golden rule and fulfills  the
conservation law of kinetic energy and  Poincar\'e invariance.  $\Gamma$ from $S[\text T]$ agrees with that from 
$S[\infty]$, and  
for  $\Gamma \neq 0$ and $\text T \rightarrow
\infty$,  the first term is dominant and the second
term is negligible. 
Now in a  situation  where the second term is not
negligible,  $P$ behaves differently.  
Especially  $\Gamma \approx 0$ or  $\text T
\leq \tau$, where $\tau={\hbar \over \Gamma}$ is the average life-time of 
parent,  are such cases. Because the  state is
described by a superposition of the parent and daughters,
 they have   a finite interaction energy, if they
overlap.
Then  kinetic energy becomes different from  that of the initial state. The
rate 
and other 
physical quantities in this region 
which have been unclear  \cite{goldberger-watson-paper}, are affected by
 $P^{(d)}$.
Thus the probability at T deviates from 
the value   at  $\text T = \infty$, 
which  we call   a  finite-size correction, by an amount that is proportional 
to 1/T in  $\text T < \tau$. The corrections depend on  the particles
that are detected, and  are large for  light particles but small for
heavy particles. 
They  were found extremely large in
\begin{eqnarray}
& &\pi \rightarrow l+\nu, \label{neutrino-detection}\\
& &\mu \rightarrow e+\gamma\label{dipole-transition},
\end{eqnarray}
and violate  Poincar\'e invariance \cite{Ishikawa-Tobita-ptep}.  The
present 
paper clarifies the reasons and presents
quantitative analysis of the process
Eq. $(\ref{neutrino-detection})$. The detailed study of process
Eq. $(\ref{dipole-transition})$ and others  will be given elsewhere.

At $\text T \leq \tau$,   the wave functions of the 
parent and daughters  retain the wave nature of the probability wave,
which can not be studied with  $S[\infty]$
\cite{goldberger-watson-paper}, 
but with  the $S[\text T]$.    $S[\text T]$ satisfies
\begin{eqnarray}
[S[\text T],H_0] \neq 0,
\label{non-comuting}
\end{eqnarray}
and couples with the states of non-conserving the kinetic energy. The
  kinetic-energy conserving states give $\Gamma$ and non-conserving
  states give $P^{(d)}$.  
  A  state  that starts from an eigenstate of $H_0$ of an
eigenvalue $E_0$  evolves  with  $H_0+H_{int}$ to become  a superposition 
of waves of kinetic energies $E_{\beta}$ that includes  $\omega_\beta = E_{\beta} - E_0 \neq 0$ at 
a finite $t$, which is similar to  diffraction of classical wave. The amplitude and probability show a diffraction
  pattern.
Diffraction in classical physics,  appears  in its intensity in systems 
of disorder of  violating  a translational 
symmetry. Now 
the  diffraction  in the probability amplitude  is caused by  
non-constant kinetic energy  at a finite 
time, and  appears  in the system without  disorder even in vacuum, hence 
depends on the physical constants of Lagrangian.  From
  Eq. $(\ref{non-comuting})$, $P^{(d)}$ necessarily violates 
   Poincar\'e invariance.

A complete set of wave functions are necessary to compute
the probability correctly. A complete set is constructed with those 
functions that are  translated in space and having   position coordinates \cite{Ishikawa-Shimomura}, and the $S[\text T]$ is 
formulated as an extension of LSZ \cite{LSZ}  formula. LSZ  and 
textbooks on scattering \cite{Goldberger,newton,taylor,Sasakawa}, 
quantum field theory including axiomatic field theory
 \cite{qft-texts1,qft-texts2} have given   $S[\infty]$ with the 
large wave packets in 
 Poincar\'{e} invariant  manner. These works have solved of obtaining 
 scattering amplitudes  in general manners, and proved Poincar\'e
 invariance.   Finite T corrections are  not Poincar\'{e} invariant, 
although the asymptotic  values are Poincar\'e invariant. 
The  infinitely large 
wave packets combined with Poincar\'{e}  invariance are not suitable   
to find if the observed quantities in experiments are subject to the 
finite-size corrections. 
So in this paper    we do not assume the Poincar\'{e} invariance in ad hoc
 manner but compute the corrections  from 
Schr\"odinger equation. 

Weak decays mainly have been studied with massive particles.  For the 
probabilities of the events that the  charged leptons  are
detected \cite{Sakai-1949,Jack,Ruderman,LSZ,Low}, theoretical values of
  decay rates, average life times, and various  distributions  have
 agreed   with experiments \cite{Anderson}. 
Thus they do not have 
 finite-size corrections. 
  Neutrinos are extremely
 light and show large $P^{(d)}$   at near detectors of much shorter
 distance than the  flavor oscillation length.
Flavor oscillations are observed at $\text T \gg \tau$ and are computed 
with $S[\infty]$. Neutrino's mass squared differences 
were determined  from experiments~\cite{SK-Atom,SK-Solar,SNO-NC,KamLAND-Reactor,Borexino,K2K}
 of using neutrinos from the sun, accelerators, reactors, and atmosphere as
\cite{particle-data},
\begin{align}%PDG reference
\Delta m^2_{21} &= m_2^2 - m_1^2 = (7.58^{+0.22}_{-0.26})\times 10^{-5}~[\text{eV}^2/c^4],\label{mass-squared-difference1}\\
|\Delta m^2_{32}| &= |m_3^2 - m_2^2|= (2.35 ^{+0.12}_{-0.09})\times 10^{-3}~[\text{eV}^2 /c^4]\label{mass-squared-difference2},
\end{align}
where $m_i~(i=1-3)$ are mass values.
  Oscillation
experiments are useless for determining absolute masses, and tritium beta decays \cite{Tritium}~have
been used  but an existing upper bound for an effective electron 
neutrino mass-squared  is of the order  of $  2 ~[\text{eV}^2/c^4]$. 
The mass 
is $ 0.3 - 1.3 ~ [\text{eV}/c^2]$ from cosmological
observations~\cite{WMAP-neutrino}.  
The neutrino spectrum at $\text T \leq \tau$  is irrelevant to flavor 
oscillations due to such small $\Delta m^2$ of
 Eqs. $(\ref{mass-squared-difference1})$ and $(\ref{mass-squared-difference2})$. It
  will be shown that $P^{(d)}$  is directly connected with the absolute
  neutrino masses.

 $P^{(d)}$ is connected with scattering-into-cones theorem \cite{dollard}. The amplitude and probability 
of the event that the particle of a certain momentum is detected at a certain 
position, ${\vec X}$,  are computed in field theory   
\cite{Ishikawa-Tobita-ptep,Ishikawa-Shimomura}, using the
small wave packets.
 It is found that ${\vec X}$ dependence of the probability per unit time, 
$C({ \vec X},\vec{p}\,)$, is written   in the form,
\begin{eqnarray}
C^{(0)}+C^{(1)}{ 2 l_0 \over|{\vec X}-{\vec X}^{(i)}| },  ~\text {for large}  ~|{\vec X}-{\vec X}^{(i)}|, \label{finite-T-probability-in}
\end{eqnarray}
where ${\vec X}^{(i)}$ is the position of the initial state, $C^{(0)}$
and $C^{(1)}$ correspond to $\Gamma$ and $P^{(d)}$ and  
\begin{eqnarray}
l_0=\left({2|\vec{p}\,| \hbar c \over m_o^2}\right).
\end{eqnarray}
   $l_0$ determines  a typical length that the finite-size correction 
remains  and is   called a coherence length of the process. 
For a  pion
and an electron of energy $1$ [GeV], they are 
\begin{eqnarray}
& &l_0^{pion}={2 \hbar c \over 0.13^2}\,[\text{GeV}^{-1}]
= 2\times 10^{-14}\,[\text{m}],\label{pion-coherence }\\
& &l_0^{electron}={2 \hbar c \over 0.5^2}\,[\text{GeV}^{-1}]
= 1.6 \times 10^{-9}\, [\text{m}].\label{electron-coherence }
\end{eqnarray} 
Other hadrons are heavier and have  shorter lengths than that of the  pion.
Thus $l_0$ for hadrons and  charged leptons are  microscopic
lengths. The coherence length for a neutrino of mass $1$ [\text{eV}/$c^2$] and energy $1$
[\text{GeV}] is
\begin{eqnarray}
l_0^{neutrino}=\frac{2 \hbar c}{1^2}\times 10^{18}\,[\text{GeV}^{-1}]=  10^{2}-10^{3}
 \,[\text{m}],\label{neutrino-coherence }
\end{eqnarray} 
and  is macroscopic length of  a few
hundred meters.  The details of the derivation of Eq. $( \ref{finite-T-probability-in})$ will be presented in Section 5.

In a spatial region where  $C({X},\vec{p}\,)$ is independent of ${\vec
 X}$, particle-zone,  
 the probability shows particle-like behavior, and its flux follows   that of classical 
particles. The  particle is  treated  as a classical particle of a  flux determined
  by the  distribution functions of  the decay process.
From Eq. $(\ref{electron-coherence })$,  ordinary scattering experiments 
of the charged leptons and hadrons  belong the particle-zone, 
and  are treated with  the ordinary S-matrix, $S[\infty]$.
 From the delta function of an energy and 
momentum conservation,  a total  probability becomes  
proportional to T, and  the decay rate becomes constant.

Now in the region  ${|{\vec X}-{\vec X}^{(i)}| \leq l_0}$,  where
$C({ X},\vec{p}\,)$ depends on ${\vec X}$ and  disagrees with 
the asymptotic value,
the final state  behaves  like a correlated wave \cite{Gell-mann}.
\begin{figure}[t]
%\centering{
\includegraphics[scale=.8,angle=0]{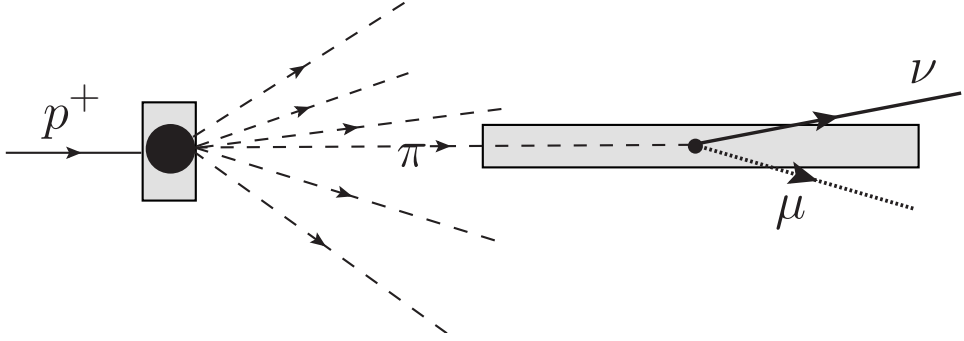}
\caption{The whole process in a high-energy neutrino experiment is
illustrated. By a collision of a proton with a target, a pion is
 produced. The pion
 propagates a macroscopic distance and decays in a decay tunnel. A
 neutrino is  produced and is detected.}
%}
\label{fig:whole-geo1}
\end{figure}

The whole process expressed in Fig.\,\ref{fig:whole-geo1} is 
studied. We study first the  probabilities of the events that the pion
is detected,  and second that of the event that the 
decay products  of the pion are detected,  in  regions, $\ |{\vec X}-{\vec X}^{(i)}| <l_0$.  
 A pion has short $l_0^{pion}$, and  
  retains wave nature only in microscopic  region and the 
ordinary experiments  are 
treated with $S[\infty]$.  The neutrino has  long $l_0^{neutrino}$, and retains 
wave nature in a macroscopic region and  is treated with $S[\text T]$.
 The finite-size correction depends on the neutrino 
energy and  size of wave packet, and is not invariant under Lorentz 
transformation.    
The large finite-size correction in a macroscopic area  can be 
used to  test  quantum mechanics and must be included for comparisons 
of the theory with experiments in this region.

The spectrum of  neutrino reveals an unusual
 macroscopic behavior of  an  interference  
pattern determined by   the wave function of entire decay process.
Quantum 
mechanics has been verified  
from many tests with the electron, photon, and other elements and  most of them 
are restricted  to 
 microscopic areas. 
In  electron bi-prism 
experiments by
Tonomura et al \cite{tonomura},  single quantum  interference  becomes
visible as  a total number of events  becomes significant. Even though  initial 
electrons are  created randomly,  a clear single quantum interference is
 seen  
when   a  signal  exceeds a statistical fluctuation,
 and 
 can be
 used as a new test  in macroscopic area.

This paper is organized in the following manner. 
In section 2,  S-matrix of a finite-time interval, $S[\text{T}]$,
is introduced. In section 3, pions in hadron reactions  are studied.  
In section 4,   we
study  an amplitude of the event that  a neutrino is detected  in a  
pion decay  
and  compute a position-dependent probability 
in section 5.
For a rigorous calculation of 
the position-dependent probability, a correlation  function is
introduced.  Using an expression with the correlation function and its
singular structure at a light-cone region,  the finite-size correction
is computed  in section 6. Implication to neutrino experiments, features
of the finite-size corrections, and 
summary and prospects 
are given in section 7, 8, and 9. 
 Various properties of
wave packets including the size, shape, and completeness are studied in
Appendix A.
\section{ S-matrix in the overlapping region: $S[\text T]$ }
\label{}
Poincar\'{e} invariant system described by the action integral 
\begin{eqnarray}
\mathcal{S}=\int d^4x \mathscr{L}[\varphi(x),\psi(x)]
\end{eqnarray}
with the  Lagrangian density 
\begin{eqnarray}
\mathscr{L}=\mathscr{L}_0+\mathscr{L}_{int}, \label{Lagrangian}
\end{eqnarray}
where $\mathscr{L}_0$ is a free part and $\mathscr{L}_{int}$ is an 
interaction part has
conserved tensors,
\begin{eqnarray}
& &\mathscr{T}_{\mu \nu},\ M_{\mu \nu \lambda}, \\
& &\partial^{\mu}\mathscr{T}_{\mu \nu}=0,\ \partial^{\mu} M_{\mu \nu \lambda}=0,
\end{eqnarray}  
and conserved charges
\begin{eqnarray}
& &\mathcal{P}_{\mu}=\int d{\vec x} \mathscr{T}_{0 \mu}, \\
& &\mathcal{L}_{\mu \nu}=\int d{\vec x}M_{0\mu \nu} . 
\end{eqnarray}
Thus the physical system is invariant under Poincar\'{e} transformation, 
the symmetry generated by
$\mathcal{P}_{\mu}$ and   $\mathcal{L}_{\mu \nu}$.

\subsection{Wave function at a finite time }
\subsubsection{Wave functions at a finite time }

A time evolution of  the state vector  $|\Psi(t)\rangle$ is described 
with   $H$ composed of a free and interaction parts,  $H_0$ and  $H_{int}$, 
derived from $\mathscr{L}_0$ and $\mathscr{L}_{int}$ of Eq. ($\ref{Lagrangian}$) 
\begin{eqnarray}
H=H_0+H_{int},
\end{eqnarray}
as \cite{Tomonaga,Schwinger},
\begin{eqnarray}
i \frac{\partial}{\partial t}|\Psi(t)\rangle=(H_0+H_{int})
 |\Psi(t)\rangle,
\end{eqnarray} 
in a unit of $\hbar=1$.  
Unitary operators 
\begin{eqnarray}
U(t)=e^{-iHt},\ U_0(t)=e^{-iH_0t}, \label{ translation-operators}
\end{eqnarray}
give time evolutions of the state vectors. A state vector in the 
interaction representation defined by
\begin{eqnarray}
| \tilde \Psi(t)\rangle =U_0 | \Psi(t)\rangle,\ \tilde{H}_{int}(t)=U_0(t) H_{int} U_0^{\dagger}(t),
\end{eqnarray}
satisfies 
\begin{eqnarray}
 i \frac{\partial}{\partial t}| \tilde \Psi(t)\rangle= \tilde{H}_{int}(t)
 |\tilde \Psi(t)\rangle,
\end{eqnarray}
and a solution is given by a time-ordered  product, 
\begin{eqnarray}
|\tilde \Psi(t)\rangle=\mathbf{T} \int_0^t dt' e^{ \tilde
 H_{int}(t')t'/{i}}|\Psi(0)\rangle
=|\Psi(0)\rangle +\int_0^t dt' \tilde A (t') |\Psi(0)\rangle,
\end{eqnarray}
where
\begin{eqnarray}
 \tilde A(t')=  \tilde{H}_{int}(t')/{i}+\int_0^{t'} d{t''} (\tilde
{H}_{int}(t')/{i})( \tilde{H}_{int}(t'')/{i} +\cdots).  
\end{eqnarray}

Divergences due to ultraviolet components in  higher order corrections are 
controlled with the methods of Refs.\,\cite{Tomonaga,Schwinger}. In the first 
order in $H_{int}$ and in tree levels, there is no ambiguity in the computations. 
The solution in the first order  is 
\begin{align}
|\tilde \Psi(t)\rangle&=\left\{1+\int_0^t dt'  {\tilde
 H_{int}(t')}/{i} \right\}|\Psi(0)\rangle \nonumber\\
&=|\Psi(0)\rangle+\int d{\beta} \frac{e^{i\omega t}- 1}{\omega}  |\beta \rangle
 \langle \beta |\tilde H_{int}(0)  |\Psi(0)\rangle,\label{function-finite}
\end{align}
where
\begin{eqnarray}
& &H_0 |\Psi(0)\rangle =E_0 |\Psi(0),~H_0 |\beta \rangle =E_{\beta} |\beta \rangle, \\
& &\omega= E_{\beta}-E_{0}. \nonumber
\end{eqnarray}

At   $t \rightarrow \infty$, the formula
\begin{eqnarray}
\frac{e^{i\omega t}-1}{\omega}=2ie^{i\frac{\omega t}{2}}\left(\frac{\sin(\omega t/2)}{
 \omega}\right) \approx 2\pi i\delta (\omega),
\label{large-t-delta}
\end{eqnarray} 
is substituted into Eq. ($\ref{function-finite}$), and the  state becomes  
\begin{align}
& |\tilde \Psi(t)_{\infty}\rangle=|\Psi(0)\rangle+  2\pi i\int d{\beta}|\beta \rangle
 \langle \beta |\tilde{H}_{int}(0)
  |\Psi(0)\rangle|\delta ({E_{\beta}-E_0}), \label{asymptotic-state}\\
&H |\tilde \Psi(t)\rangle=E_0 |\Psi(t)\rangle,\nonumber \\
&H_0 |\tilde \Psi(t)_{\infty}\rangle=E_0 |\Psi(t)_{\infty}\rangle. \nonumber 
\end{align}
At a finite $t$, the function has a finite peak at $\omega =0$ and a 
tail  $\omega \neq 0$.

Thus  $E_{\beta}=E_0$ at $t=\infty$
in Eq. ($\ref{asymptotic-state}$)  and the
kinetic energy is  constant, and  the  physical quantities at the asymptotic regions  such as the  cross section
and decay rate  are computed within this space. 
Now,   at a finite $t$, $E_{\beta} \neq E_0$  in Eq. $(\ref{function-finite})$, and    
the kinetic  energy  is not constant,  and the  state, Eq. $(\ref{function-finite})$, is 
 a superposition of waves of different kinetic energies.

A  transition rate at a finite T is computed from
Eq. $(\ref{function-finite})$ \cite{Schiff and Landau}. A probability of
the event that   $\beta$ is detected is given in the form,
\begin{eqnarray}
& &|F_{0,\beta}|^2
\frac{4\sin^2\left[(E_{\beta}-E_{0})\text{T}/2\right]}{(E_{\beta}-E_{0})^2},\label{finite-T-expression}\\
& &F_{0,\beta}=\langle \beta |\tilde{H}_{int}(0)|\Psi(0) \rangle. \nonumber
\end{eqnarray}
For  continuous $E_{\beta}$, a formula  
\begin{eqnarray}
\lim_{\text{T} \rightarrow \infty}
\frac{4\sin^2\left[(E_{\beta}-E_{0})\text{T}/2\right]}{(E_{\beta}-E_{0})^2}=\lim_{\text{T}
 \rightarrow \infty}2\pi
{\text T} \delta(E_{\beta}-E_{0})
\label{asymptotic-expression}
\end{eqnarray}
is used normally \cite{Dirac,Schiff-golden}.
At a finite  T, however,  the tail at $|E_{\beta}-E_{0}| \neq 0$ 
 gives a correction proportional to   ${1/\text T}$ \cite{peierls}.
The correction is computed   from  Eqs. $(\ref{large-t-delta})$ 
and $(\ref{asymptotic-expression})$.
Taylor expansion of $g(\omega)=|F_{0,\beta}|^2$ leads
\begin{align}
\int d \omega g(\omega)  
& \left({2\sin [\omega \text{T}/2] \over \omega }\right)^2\nonumber\\
&=2\pi\text T g(0)\left\{1+ {1 \over \text T}
 {2g'(0) \over \pi g(0)}\int dx x \left({2\sin x \over x }\right)^2+O(1/{\text T^2})\right\}, 
\label{diverging-integral}
\end{align}
and  the  integral 
over $x={\omega \text T \over 2}$ of the second term of the right-hand side in 
Eq. $(\ref{diverging-integral})$ diverges. So the tail gives the $1/
\text{T}$ correction of the diverging  coefficient.  The divergence
suggests a proper method is necessary, which we find  
in the following.

\subsection{Scattering operator of a finite-time interval}

The  state, Eq. $(\ref{function-finite})$, is 
 a superposition of waves of different kinetic energies, 
so  is non-uniform in space. 
$H_{int}$ which initially makes 
a transition of a particle to many  particles  gives an interaction energy
at a finite $t$ of Eq. $(\ref{function-finite})$. 
 Accordingly, the wave Eq. $(\ref{function-finite})$   shows
diffraction that is characteristic of a sum of waves of different wave 
lengths and reveals a position dependent probability.
The diffractive pattern  depends upon  the spectrum and states at all $E_{\beta}$, 
even though this is the phenomenon of  tree level.

\subsubsection{Boundary conditions}

The  scattering process of the finite-time interval T 
is computed with an S-matrix $S[\text T]$ that satisfies the boundary
conditions at T. 
For a scattering of a scalar field from an initial state 
 $|\alpha \rangle $ to a final state
$|\beta \rangle$,     the coefficients $\varphi^f(t)$   \cite{LSZ}
 given in the form, 
\begin{eqnarray}
\varphi^{f}(t)=i \int d^3 x
f^{*}({\vec x},t)\overleftrightarrow{\partial_0} \varphi({\vec x},t),\label{expansion-wave-packets} 
\end{eqnarray}
and  $\varphi_{in}^{f}(x)$ and $\varphi_{out}^{f}(x)$
 defined in the equivalent manner, where $f({\vec x},t)$ is a set of  normalized solutions of the free wave equation, are used. 
Operators, $\varphi(x)$, $\varphi_{in}(x)$, 
and $\varphi_{out}(x)$ stand interacting and free fields. The  
boundary conditions   
\begin{align}
 &\lim_{t \rightarrow -\text{T}/2}\langle \alpha |\varphi^{f}(t)|\beta \rangle
= \langle \alpha|\varphi^{f}_{in}|\beta \rangle,\label{bounday-LSZT1}\\
 &\lim_{t \rightarrow +\text{T}/2}\langle \alpha |\varphi^{f}(t)|\beta
 \rangle= \langle \alpha |\varphi^{f}_{out}|\beta \rangle.\label{bounday-LSZT2}
\end{align}   
The states $|\alpha  \rangle$
 and  $|\beta  \rangle$ are defined with $\varphi_{in}(x)$ 
and $\varphi_{out}(x)$. Since the
wave packets have finite spatial sizes and decrease fast   at 
large $|{\vec x}-{\vec x}_0|$ around a center ${\vec x}_0$, they  ensure  
the boundary  conditions at a finite $\text {T}$. The complete set 
formed as
\begin{eqnarray}
f({\vec x}-{\vec X},t)=|{\vec p},{\vec X},\beta \rangle;~ \text {all}~ {\vec X},
\label{non-covariant-wave-packet }
\end{eqnarray}
of the center coordinates of position and momentum, although this is not
covariant under Poincar\'e transformation, is used. 

 A  covariant wave  packet   defined as
\begin{eqnarray}
|\vec{p},{\vec X},\text{T} ;\text {cov.} \rangle
=U_L(\Lambda) U_T({\vec X},\text{T})|{\vec p}={\vec 0}, 
{\vec X}={\vec 0},\text{T}=0 ;\text {cov.} \rangle, \label{covariant-wave packet}
\end{eqnarray}
with unitary operators  $U_L(\Lambda)$ and $ U_T({\vec X},\text{T})$
 defined from generators  $\mathcal{L}_{\mu\nu}$ 
and $\mathcal{P}_{\mu}$ and  c-number values  $ \Lambda, {\vec X},\text{T}$,
is not convenient for practical calculation of experimentally observed 
quantities.
This   $ |{\vec 0},{\vec X},\text{T} ;\text {cov.} \rangle$  is 
located at ${\vec X}$, and 
  the momentum and position of $ |{\vec p},{\vec X},\text{T} ;\text
 {cov.} \rangle$  are  defined by 
the following Lorentz transformation, and  are  located 
at the four-dimensional position 
\begin{eqnarray}
{X}_i'=\Lambda_{i0} \text{T}+  \Lambda_{ij}X_j, \\
\text{T}'=\Lambda_{00} \text{T}+  \Lambda_{0j}X_j,
\end{eqnarray}
where $\Lambda_{\mu \nu }$ are defined by
\begin{eqnarray}
{P}_i'=\Lambda_{i0} M+  \Lambda_{ij}P_j, 
\end{eqnarray}
and depend on ${\vec P}$. Thus  the  positions are changed depending 
on the momentum. 

In experiments, the positions of the events are not measured.  
The probabilities of the events that the  particles are detected  
within detectors are given normally. These positions are independent from
their momenta. 
Thus the  states defined 
with Eq. $(\ref{non-covariant-wave-packet })$ are appropriate but 
those of  Eq. $(\ref{covariant-wave packet})$ are not for those states  of 
the experiments.  Equation $(\ref{non-covariant-wave-packet })$ are  used here.

For the large wave packet $\sigma = \infty$,    coordinates are
un-necessary and $S[\infty]$ is constructed with
\begin{eqnarray}
|{\vec p};\text {cov.} \rangle=U_L(\Lambda) |{\vec p}={\vec 0} ;\text {cov.} \rangle. \label{covariant-wave packet2}
\end{eqnarray}
The positions do not appear, and the position-independent analysis is
made with Eq. $( \ref{covariant-wave packet2})$. 

\subsubsection{Properties of S[\text T]}
 $S[\text T]$ satisfies various 
unique properties and is defined by
  M{\o}ller operators,  $\Omega_{\pm}(\text{T})$ \cite{Moller}.
  $\Omega_{\pm}(\text{T}) $ are
defined from $U(t)$ and $U_0(t)$ of Eq. $(\ref{ translation-operators})$ in 
the form
\begin{eqnarray}
\Omega_{\pm}(\text{T})=\lim_{t \rightarrow \mp \text{T/2}}U^{\dagger}(t)U_0(t),
\end{eqnarray}
and satisfy
\begin{eqnarray}
e^{iH \epsilon_t} \Omega_{\mp}(\text{T})=\Omega_{\mp}(\text{T} \pm\epsilon_t)
 e^{iH_0 \epsilon_t}.
\label{algebra-moeller-op }
\end{eqnarray}

Scattering operator of a finite-time interval $\text{T}$ is the product
\begin{eqnarray}
S(\text{T})=\Omega_{-}^{\dagger}(\text{T})\Omega_{+}(\text{T}),
\label{finite-time-S}
\end{eqnarray} 
and satisfies, from Eq. ($\ref{algebra-moeller-op }$), 
\begin{eqnarray}
[S(\text{T}),H_0]=i\left(\frac{\partial}{
 \partial \text{T}}\Omega_{-}(\text{T})\right)^{\dagger} \Omega_{+}(\text{T})-i\Omega_{-}^{\dagger}(\text{T}){\partial \over \partial\text{T}
 }\Omega_{+}(\text{T}).
\label{commutation-relation-S(T)  }
\end{eqnarray}
Hence, $S(\text{T})$ does not commute with $H_0$  and has two components, 
\begin{eqnarray}
S[\text T]=S^{(0)}[\infty]+S^{(1)}[\text T],
\label{decomposition-S[T]}
\end{eqnarray}
where 
\begin{eqnarray}
[S^{(0)}[\infty],H_0]=0,\ [S^{(1)}[\text T],H_0] \neq 0.
\label{commutation-S[T]}
\end{eqnarray} 
Matrix elements of $S^{(0)}[\infty]$ for the states defined by the boundary 
conditions  Eqs. $(\ref{expansion-wave-packets}),$
 $(\ref{bounday-LSZT1}),$ and $(\ref{bounday-LSZT2})$ are equivalent to those of  
 momentum states,
\begin{eqnarray}
& &\langle  \beta |S^{(0)}[\infty]| \alpha \rangle =\langle \beta | p_f\rangle \langle  p_f|S^{(0)}[\infty]| p_i \rangle \langle p_i| \alpha \rangle, \\
& &\langle  p_f|S^{(0)}[\infty]| p_i \rangle=\delta_{p_f,p_i}+(2\pi)^4\delta^{(4)} (p_{f}-p_{i}) f_{p_f,p_i},
\end{eqnarray}
where $| p_f\rangle $ and $| p_i\rangle $  are initial and final 
states of plane waves and $f_{p_f,p_i}$ is the matrix element.  The matrix element of $S^{(1)}[\text T]$ is not equivalent to  the standard  one and  written as, 
\begin{eqnarray}
 \langle \beta |S^{(1)}[\text T]| \alpha \rangle= \delta f(\text{T}).  
\label{two-compoent S-matrix}
\end{eqnarray}
Since the kinetic energy $E_{\beta}$ of $S^{(1)}[\text T]$ 
 is different from that of $S^{(0)}[\infty]$, the total transition 
probability  becomes a sum of  T-independent and dependent
 probabilities.
A magnitude of $\delta f$ depends on  a dynamics of
the system and satisfies  for the states of energies   $E_{\alpha}$ and 
$E_{\beta}$,  $|\alpha \rangle$ and $|\beta \rangle$, 
\begin{align}
&(E_{\alpha}-E_{\beta})  \langle \beta |S^{(1)}(\text{T})| \alpha \rangle=  \langle
 \beta |O(\text{T})| \alpha \rangle,\\
&O(\text{T})=i\left(\frac{\partial}{
 \partial \text{T}}\Omega_{-}(\text{T})\right)^{\dagger}
 \Omega_{+}(\text{T})-i\Omega_{-}^{\dagger}\frac{\partial}{\partial\text{T}
 }\Omega_{+}(\text{T}).
\end{align}
Hence 
\begin{eqnarray}
\delta f(\text{T})=\frac{1}{E_{\alpha}-E_{\beta} }\langle
 \beta |O(\text{T})| \alpha \rangle.
\label{positive-correction}
\end{eqnarray}
We have the probability,
\begin{align}
&P=P^{(0)}+P^{(1)}, \\
&P^{(0)}= V\text{T} (2\pi)^4 \int dp_f\delta^4(p_f-p_i) |f|^2,\\
&P^{(1)}= \int d{\beta} |\delta f(\text{T})|^2,
\end{align}
 and $P^{(1)}$  is written as,
\begin{eqnarray}
\int d{\beta}|\delta f|^2=\int d{\beta}\left({1 \over E_{\alpha}-E_{\beta} }\right)^2|\langle
 \beta |O(\text{T})| \alpha \rangle|^2 \geq 0,
\end{eqnarray}
where the equality is satisfied at $\text{T} \rightarrow \infty$. Thus  
$S^{(1)}[\text T]$ and  states of $E_{\beta} \neq E_0$ give the
finite-size corrections of $1/\text T$.

The   whole set of wave packets forms a complete set \cite{Ishikawa-Shimomura},
\begin{eqnarray}
\sum_{{\vec p},{\vec X}}|{\vec p},{\vec X},\beta \rangle
 \langle {\vec p},{\vec X}, \beta |=1,  
\label{complete-set}
\end{eqnarray}
hence an expectation value of the physical quantity $\mathcal{O}$,
\begin{eqnarray}
\sum |\langle \alpha |\mathcal{O}||{\vec p},{\vec X},\beta \rangle|^2
 \label{rep-independence }
\end{eqnarray}
is independent of a choice of the set, if the operator is defined
uniquely independent of the set.  Normal physical quantities  of microscopic
 physics  obtained from 
 $S[\infty]$  are independent from the used  basis  and the probabilities 
agree.    Now $S[\text T]$ is  defined according to the
boundary conditions Eqs. $(\ref{expansion-wave-packets}),\ (\ref{bounday-LSZT1}),\text{ and}~ (\ref{bounday-LSZT2})$ which depend on the wave packets, and is not 
independent of the wave packets.  Hence the matrix elements of 
non-invariant component, $S^{(1)}[\text T]$,  accordingly  the probability  
 depend on the choice of the basis, and 
the finite-size correction depends on the wave packet size $\sigma$.  
      
\subsection{Symmetry of $S[\text{T}\,]$}

\subsubsection{ Poincar\'{e} invariance}

% \subsection{ Poincar\'{e} invariance}
    The wave packets  reflect the boundary conditions of 
experiments and are defined in laboratory frame.
The wave packet for out-state   shows a wave function of minimum physical 
system  which  an
outgoing particle makes reactions  in a detector, and that for in-state 
shows a wave function of beam. The former part gives data and 
both are not symmetric. 
The wave packets 
, hence, are not necessary  Poincar\'{e} invariant.  Accordingly $S[\text T]$ 
defined with non-invariant wave packets is not Poincar\'{e} invariant 
 even in the invariant system.

 $S[\infty]$ has no explicit space-time parameter and is manifestly 
covariant and can be  used  
for computing the asymptotic values.   The probability and matrix elements 
are connected  
\begin{align}
\sum_{f,{i}} |\langle {\vec p}_f;cov|S[\infty]|{\vec p}_i;cov \rangle|^2 = &\sum_{f,{i}} |\langle {\vec p}_f,{\vec X}_f;\sigma_f|S[\infty]|{\vec p}_i,{\vec X}_i;\sigma_i \rangle|^2,\label{total-probability} \\
\langle {\vec p}_f;cov|S[\infty]|{\vec p}_i;cov \rangle=&\langle {\vec p}_f;cov
|{\vec p}_f,{\vec X}_f;\sigma_f\rangle \langle {\vec p}_f,{\vec X}_f;\sigma_f| S[\infty]|{\vec p}_i,{\vec X}_i;\sigma_i \rangle \nonumber\\
&\times\langle {\vec p}_i,{\vec X}_i;\sigma_i |{\vec p}_i;cov \rangle,\\
\langle {\vec p}_f,{\vec X}_f;\sigma_f| S[\infty]|{\vec p}_i,{\vec X}_i;\sigma_i \rangle= &\langle {\vec p}_f,{\vec X}_f;\sigma_f| {\vec p}_f;cov \rangle \langle {\vec p}_f;cov|S[\infty]|{\vec p}_i;cov \rangle\nonumber\\
& \times\langle {\vec p}_i; cov |{\vec p}_i,{\vec X}_i;\sigma_i\rangle,
\end{align}
where the final states are summed over and the same average are taken for 
the  initial states in Eq. $(\ref{total-probability})$, from 
the completeness of the states.

\subsubsection{Space-time symmetry}
The generators of  Poincar\'{e} transformations
\begin{eqnarray}
\mathcal{P}_{\mu}(\mathcal{P}_0=H),\ \mathcal{L}_{\mu \nu}
\end{eqnarray} 
are conserved charges and $S^{(0)}[\infty]$ satisfy commutation relations 
\begin{eqnarray}
& & [S^{(0)}[\infty], H_0] = 0,\ [S^{(0)}[\infty], {\vec{\mathcal{P}}}\,] = 0,
\label{momentum-non-conservation2} \\
& & [S^{(0)}[\infty], \mathcal{L}_{\mu \nu}] = 0.
\end{eqnarray} 
 $S^{(1)}[\text T]$, from its
 definition, Eqs. $(\ref{decomposition-S[T]})$ and $(\ref{commutation-S[T]})$,
 satisfy 
\begin{eqnarray}
& & [S^{(1)}[\text T], H_0] \neq 0,\ [S^{(1)}[\text T],\ {\vec{\mathcal{P}}}\,] \neq 0,
\label{momentum-non-conservation1}
\\
& & [S^{(1)}[\text T], \mathcal{L}_{\mu \nu}] \neq 0.
\end{eqnarray} 

$S[\text T]$  does not conserve the kinetic energy,  momentum, and  angular 
momentum, so shows  different properties from those of $S[\infty]$.
The finite-size corrections 
to  transition rates  are  computable   
with $S[\text T]$,  but not with  $S[\infty]$. They  are  necessary 
to find if the experimental values  are subject to  the finite-size 
corrections.

{\bf Inversion}

 Space inversion 
\begin{eqnarray}
I_{space}:{\vec x} \rightarrow -{\vec x},
\end{eqnarray}
and   time inversion,
\begin{eqnarray}
I_{time}:t \rightarrow -t
\end{eqnarray}
are defined at a finite T. Hence $S[\text T]$ of invariant system 
such as electromagnetic and strong 
interactions defined in symmetric region of  space and time satisfies
\begin{eqnarray}
[S[\text T],I_{space}]=0,\ [S[\text T],I_{time}]=0.
\end{eqnarray}
$I_{space}$ is a linear operator and $I_{time}$ is an anti-linear operator. 

\subsubsection{Internal symmetry}

{\bf Exact symmetry}

A  charge $Q$ of internal symmetry satisfies 
\begin{eqnarray}
& &[Q,H]=0, \\
& &[Q,H_0]=0,
\end{eqnarray}
and
\begin{eqnarray}
[Q,S(\text{T})] =0.
\end{eqnarray}
Hence $Q$  is conserved in
$S(\text{T})$, and a state $|\psi \rangle $ and  $S[\text{T}]|\psi \rangle $  
have a same charge
\begin{eqnarray}
& &Q |\psi \rangle =q |\psi \rangle, \\
& &Q S[\text{T}]|\psi \rangle =q S[\text{T}]|\psi \rangle. 
\end{eqnarray}

If $Q$ is  a charge of non-compact group, its eigenvalue 
\begin{eqnarray}
Q|q \rangle =q|q \rangle,
\end{eqnarray}
is continuous and the eigenstates are  normalized with Dirac delta function,
\begin{eqnarray}
\langle q_1| q_2 \rangle=2\pi \delta(q_1-q_2).
\end{eqnarray}
 Then   the matrix element is written in the diagonal form in $q$, 
\begin{eqnarray}
\langle q_2|S[\text T]| q_1 \rangle=2\pi \delta(q_1-q_2) \tilde S[\text
 T](q_1),
\end{eqnarray}
and the matrix element between any states is written with the reduced
matrix $\tilde S[\text T]$
\begin{eqnarray}
\int d q_2 d q_1 F(q_2) \langle q_2|S[\text T] | q_1 \rangle
 G(q_1)=2\pi\int d q_1 F(q_1)
 \tilde S[\text T](q_1) G(q_1).
\end{eqnarray}

{\bf Approximate  symmetry}

For an  approximate  symmetry of satisfying 
\begin{eqnarray}
[S[\text T],Q] \neq 0, \label{symmetry-breaking}
\end{eqnarray}
the finite-size correction  is non-invariant. Because the correction depends on  the mass, mass difference causes a large  symmetry breaking.
  The masses are very different in neutrinos and charged leptons, hence 
they have different finite-size corrections.   
%%%%%%%%%%%%%%%%%%%%%%%%%%%%%%%%%%%%%%%%%%%%%%%%%%%%%%%%%%%%%%%%%
\subsection{Unitarity }
%%%%%%%%%%%%%%%%%%%%%%%%%%%%%%%%%%%%%%%%%%%%%%%%%%%%%%%%%%%%%%%%
The $S[\text{T}]$ is defined with M{\o}ller operators,
Eq. $({\ref{finite-time-S}})$, and 
satisfies a unitarity 
relation,
\begin{eqnarray}
 S^{\dagger}[\text{T}]S[\text{T}]=S[\text{T}]S^{\dagger}[\text{T}]=1,
\label{unitarity}
\end{eqnarray}
and  an optical theorem
\begin{align}
&i(\mathcal{T}[\text T]-\mathcal{T}[\text T]^{\dagger})=
\mathcal{T}[\text T]\mathcal{T}^{\dagger}[\text T], \\
& S[\text{T}]=1+i\mathcal{T}[\text T]. 
\end{align}
The probability is preserved in $S[\text{T}]$ and the imaginary part of
the amplitude at T is determined by the total probability measured at T.  

$S[\text{T}]$ is decomposed into the energy conserving term $\mathcal{T}_1[\text
T]$ and non-conserving term $\mathcal{T}_2[\text T]$
\begin{eqnarray}
S[\text{T}]=1+i(\mathcal{T}_1[\text{T}]+\mathcal{T}_2[\text T]),
\end{eqnarray}
then the unitarity Eq. $(\ref{unitarity})$ is written in the form,
\begin{eqnarray}
 (1+i(\mathcal{T}_1[\text{T}]+\mathcal{T}_2[\text T]))(1-i(\mathcal{T}_1^{\dagger}[\text{T}]
+\mathcal{T}_2^{\dagger}[\text{T}]))=1.
\end{eqnarray}
Hence we have 
\begin{align}
-i(\mathcal{T}_1[\text{T}]-\mathcal{T}_1^{\dagger}[\text{T}])-i(\mathcal{T}_2[\text{T}]
-\mathcal{T}_2^{\dagger}[\text{T}])
=&\mathcal{T}_1[\text{T}]\mathcal{T}_1^{\dagger}[\text{T}]
+\mathcal{T}_2[\text{T}]\mathcal{T}_2^{\dagger}[\text{T}]\nonumber\\
&+\mathcal{T}_1[\text{T}]\mathcal{T}_2^{\dagger}[\text{T}]+\mathcal{T}_2[\text{T}]
\mathcal{T}_1^{\dagger}[\text{T}],
\end{align}
and 
\begin{eqnarray}
& &-i(\mathcal{T}_1[\text{T}]-\mathcal{T}_1^{\dagger}[\text{T}])
=\mathcal{T}_1[\text{T}]\mathcal{T}_1^{\dagger}[\text{T}],\\
& &-i(\mathcal{T}_2[\text{T}]-\mathcal{T}_2^\dagger[\text{T}])
=\mathcal{T}_2[\text{T}]\mathcal{T}_2^{\dagger}[\text{T}]
+\mathcal{T}_1[\text{T}]\mathcal{T}_2^\dagger[\text{T}]+\mathcal{T}_2[\text{T}]
\mathcal{T}_1^\dagger[\text{T}].
\end{eqnarray}
The total transition probability from a state $\alpha$ is
\begin{eqnarray}
P=\sum_{E_\beta \approx E_{\alpha}}  | \langle  \beta|\mathcal{T}_1|\alpha
 \rangle |^2 
+\sum_{E_{\beta} \neq E_{\alpha}} | \langle  \beta|\mathcal{T}_2|\alpha \rangle|^2,
\end{eqnarray}  
where the energy conserving term is proportional to T.

It is noted that the unitarity connects  physical quantities measured 
at T.  Optical theorem proves that the imaginary part of 
 forward amplitude at T is written by the total probability at T. Hence the
 life time at T, depends on T if the finite-size correction to the total
 probability is finite. The unitarity does not connect the probability
 at T with those at a different T.

 %%%%%%%%%%%%%%%%%%%%%%%%%%%%%%%%%%%%%%%%%%%%%%%%%%%%%%
\section{Pion in $NN$ collisions  }
%%%%%%%%%%%%%%%%%%%%%%%%%%%%%%%%new-part%%%%%

Applying $S[\text T]$, we study   pions in nucleon reactions in this section. It is found that the finite-size correction is negligibly small because the pion's mass is not small.   Iso-triplet  
pions and doublet nucleons    are 
expressed with 
 fields ${\vec \varphi}(x) $  and
$\psi_N(x)$, and this  system is described in term of the 
renormalizable  Lagrangian,
\begin{eqnarray}
\mathscr{L}=\bar \psi_N(\Slash{p} - m_N)N+g\bar \psi_N \gamma_5 {\vec \tau}\cdot {\vec
 \varphi}(x) \psi_N+\frac{1}{2}\left({\partial
 _{\mu}}{\vec \varphi}\right)^2- \frac{1}{2} m_{\pi}^2\vec{\varphi}^{\,2}(x),
\label{pi-N-lagrangian}
\end{eqnarray}  
where $m_N$ and $m_{\pi}$ are  masses of the nucleons and pions. A
mass difference between a proton and a neutron and that of neutral and
charged pions are ignored and $SU(2)$
symmetry is assumed in most places. Second term in the right-hand side
shows an 
interaction between a nucleon and a pion.  Due to this interaction, a nucleon 
emits and absorbs a pion in intermediate states. These physical
processes are 
treated by a renormalization prescription in  a nucleon self-energy  
and others.

\subsection{Relativistic wave packets}
Wave packets are normalizable solutions of free wave
equations, and  those of Dirac equation are similar to
that of non-relativistic Schr\"odinger equation.     
\subsubsection{ Nucleon}
Plane waves of the Dirac equation,
\begin{eqnarray}
(\Slash{p} - m_N)\psi_N(x)=0,
\end{eqnarray} 
are 
\begin{eqnarray}
& &u(p,s) e^{ip\cdot x};\ (\bar u(p,s_1),u(p,s_2))=\delta_{s_1s_2}, \\  
& &v(p,s) e^{ip\cdot x};\ (\bar v(p,s_1),v(p,s_2))=-\delta_{s_1s_2},\\
& &\sum_su_{\alpha}({p},s) \bar u_{\beta}({p},s)=
\left(\frac{\Slash{p}+ m_N}{2m_N}\right)_{\alpha \beta},\\
& &\sum_sv_{\alpha}({p},s) \bar v_{\beta}({p},s)=
\left(\frac{\Slash{p} - m_N}{2m_N}\right)_{\alpha \beta}.
\end{eqnarray}
The nucleon field operator is expanded with annihilation and creation operators as
\begin{eqnarray}
& &\psi_N(x)=\sum_i\int \frac{d\vec{p}}{(2\pi)^{\frac{3}{2}}}\left(\frac{m_N}{E({\vec p}\,)}\right)
^{\frac{1}{2}}\left\{u({p},s) b({\vec p},s)e^{-ip\cdot x}+v({p},s) d^{\dagger}({\vec p},s)
e^{ip\cdot x}\right\},\nonumber \\
& & \\
& &\left\{ b({\vec p}_1,s_1), b^{\dagger}({\vec p}_2,s_2)\right\}
=\delta({\vec p}_1-{\vec p}_2)\delta_{s_1,s_2},
\end{eqnarray}
and one particle states are constructed from creation operators 
\begin{eqnarray}
& &b^{\dagger}({\vec p},i) |0 \rangle, \\
& &d^{\dagger}({\vec p},i) |0 \rangle .
\end{eqnarray}
They satisfy
\begin{eqnarray}
& &\mathcal{P}_{\mu}b^{\dagger}({\vec p},i) |0 \rangle 
=p_{\mu}b^{\dagger}({\vec p},i) |0 \rangle,\\
& &\mathcal{P}_{\mu}d^{\dagger}({\vec p},i) |0 \rangle
=p_{\mu}d^{\dagger}({\vec p},i) |0 \rangle,
\end{eqnarray}
and are  expressed as 
\begin{eqnarray}
& &b^{\dagger}({\vec p},i) |0 \rangle =U(\Lambda) b^{\dagger}({\vec
 0},i) |0 \rangle \label{unitary-transformed-expression1},\\
& &d^{\dagger}({\vec p},i) |0 \rangle =U(\Lambda)d^{\dagger}({\vec 0},i) |0\rangle,
\label{unitary-transformed-expression2}
\end{eqnarray}
with  a unitary operator of transforming the state ${\vec p}=0$ to that of ${\vec p}$.  
 
One particle states of wave packets are constructed with c-number functions as  
\begin{eqnarray}
& &|{\vec p},  {\vec X},\text{T} \rangle= \int d{\vec k} 
e^{i({\vec k}\cdot{\vec X}-E({\vec k})\text{T})}f({\vec k}-{\vec p};i)
 b^{\dagger}({\vec k},i) |0 \rangle \label{wave-packet-Dirac1},\\
 & &|{\vec p}\,',  {\vec X},\text{T} \rangle= \int d{\vec k} 
e^{i({\vec k}\cdot{\vec X}-E({\vec k})\text{T})}g({\vec k}-{\vec p}\,';j)
 d^{\dagger}({\vec k},j) |0 \rangle.\label{wave-packet-Dirac2}
\end{eqnarray}
The functions Eqs. ($\ref{wave-packet-Dirac1}$) and ($\ref{wave-packet-Dirac2}$)
 satisfy the normalization 
conditions,
\begin{eqnarray}
& &\int d{\vec k} f^{*}({\vec k},i) f({\vec k},j) =\delta_{i,j}, \\  
& &\int d{\vec k}g^{*}({\vec k},i) g({\vec k},j)=\delta_{i,j},
\end{eqnarray}
and the states  form a complete set \cite{Ishikawa-Shimomura},
\begin{align}
&\int \frac{d{\vec p} d{\vec X}}{(2\pi)^3} |{\vec p},  {\vec X},\text{T} \rangle
 \langle {\vec p},  {\vec X},\text{T}| \nonumber\\
&= \int d{\vec p} \sum_{i,j} \int d{\vec k} f^{*}({\vec
 k}-{\vec p};i)f({\vec
 k}-{\vec p};j)  b^{\dagger}({\vec p},j)|0 \rangle
 \langle 0|b({\vec p},i)\nonumber \\
&=\sum_i \int d{\vec p}\, b^{\dagger}({\vec p},i)|0 \rangle
 \langle 0|b({\vec p},i)=1,
\end{align}
within one particle states. Many particle states constructed as  direct
products of one particle states of wave packets form a
complete set.  

Invariant wave packets under space-time inversions are expressed by those that satisfy
\begin{eqnarray}
& &I_{space}:  f({\vec k}-{\vec p},i)=f(-{\vec k}+{\vec p},i), \\
& &I_{time}: f({\vec k}-{\vec p},i)=f^{*}(-{\vec k}+{\vec p},-i).
\end{eqnarray}
A spin independent Gaussian wave packet,
\begin{eqnarray}
f({\vec k}-{\vec p};i)=Ne^{-\frac{\sigma}{2}({\vec k}-{\vec p})^2},
\end{eqnarray}
satisfies these conditions
and used here. That gives
\begin{eqnarray}
\langle 0|\psi_N(x) |{\vec p},{\vec X},\text{T} \rangle 
&=& \sum \int \frac{d\vec{p}\,'}{(2\pi)^{\frac{3}{2}}}
 \sqrt{\frac{m}{E({\vec{p}\,'})}} e^{ip'\cdot x}u({{p}\,'})
 d{\vec k} f({\vec k}-{\vec p}\,)e^{-ik\cdot X} \delta({\vec k}-{\vec{p}\,'}) \nonumber \\
 &=&\sum \int \frac{d\vec{p}\,'}{(2\pi)^{\frac{3}{2}}}
 \sqrt{\frac{m}{E({\vec{p}\,'})}} e^{ip'\cdot(x-X)}
u({{p}\,'})  f({\vec{p}\,'}-{\vec p}\,) \nonumber \\
 &\approx&  {\tilde N} \frac{1}{(2\pi)^{\frac{3}{2}}} 
\sqrt{\frac{m}{E({\vec{p}}\,)}} 
e^{-\frac{1}{2\sigma}({\vec x}-{\vec v}(t-\text{T}) -{\vec X})^2+ip\cdot(x-X)}u({p}),
\end{eqnarray}
where $\tilde N= N(2\pi/\sigma)^{\frac{3}{2}} $.
Thus the state is the approximate eigenstate  of  $\mathcal{P}_{\mu}$ of average
  value ${\vec p}$ and its  center   moves with a constant velocity
  ${\vec v}$ as  
\begin{eqnarray} 
& &{\vec x}={\vec v}(t-\text{T})+{\vec X}, \\
& &{\vec v}={{\vec p} \over E({\vec p}\,)}.
\end{eqnarray}  
\subsubsection{Pion}
 Klein-Gordon equation,
\begin{eqnarray}
(p^2 -m_\pi^2) \varphi(x)=0
\end{eqnarray}
has solutions
\begin{eqnarray}
e^{i(E({\vec p})t-{\vec p}\cdot{\vec x})},~E({\vec p}\,)=\sqrt{{\vec p}\,^2+m_\pi^2},
\end{eqnarray}
and the field is expanded as 
\begin{eqnarray}
& &\varphi(x)=\int d{\vec p} \left(\frac{1}{2E({\vec p}\,)(2\pi)^{3}}\right)^{\frac{1}{2}}
(a({\vec p}\,) e^{ip\cdot x}+ a^{\dagger}({\vec p}\,) e^{-ip\cdot x}),\\
& &\left[a({\vec p}_1),a^{\dagger}({\vec p}_2)\right]=\delta({\vec p}_1-{\vec p}_2).
\end{eqnarray}
%\subsubsection{Neutrino}

\subsection{Pion emitted  from a nucleon}

\subsubsection{Fluctuations }
Fluctuations of a  relativistic field is expressed by 
 the  Green's function $\Delta(x_1-x_2)$,    
\begin{eqnarray}
\Delta(x_1-x_2)= \frac{1}{(2\pi)^4} \int d^4p\, e^{ip\cdot(x_1-x_2)}\frac{1}{p^2-m^2},
\end{eqnarray}
where $m$ is a particle's mass. From  the pole of positive frequency, we have
a component of on-mass shell waves of positive frequency   
\begin{eqnarray}
\Delta_0(x_1-x_2)= \frac{1}{(2\pi)^3} \int \frac{d\vec{p}}{2E(\vec{p}\,)}
 e^{ip\cdot(x_1-x_2)},\ E(\vec{p}\,)=\sqrt{\vec{p}^{\,2} +m^2}.
\label{positive-frequency-part}
\end{eqnarray}

 $\Delta_0(x_1-x_2)$ is composed of a singular part  and 
 Bessel functions \cite{Wilson-OPE}, 
\begin{align}
\Delta_0(x_1-x_2)=&i\left[{1
 \over 4\pi}\delta(\lambda)\epsilon(\delta t) +f_{short}\right],\label{singular-function-f-p} \\
f_{short}=&-{i { m} \over
 8\pi \sqrt{-\lambda}} \theta(-{\lambda})\left\{N_1\left( m \sqrt{
 -\lambda}\right)-i\epsilon(\delta t) J_1\left( m \sqrt{ -\lambda}\right)\right\} \nonumber \\
 &-\theta(\lambda){i
  m \over
 4\pi^2\sqrt{\lambda}}K_1\left(
 m\sqrt{\lambda}\right),~\lambda=(x_1-x_2)^2,\delta t=\delta x^0,
\end{align}
where $N_1$, $J_1$, and $K_1$ are Bessel functions.
The latters damp or oscillate rapidly and are short range functions  of order  
Compton wave length 
$\text{L}_{cw}={\hbar/(mc)}$. $\text{L}_{cw}={\hbar/(mc)}$  becomes 
\begin{eqnarray}
& &\text{L}_{cw}\geq\begin{cases}
2\times 10^{-15}~[\text{m}]~~~pion,\\
2 \times 10^{-16}~[\text{m}]~~~proton,
	       \end{cases}\label{compton-wl}
\end{eqnarray}
and  de Broglie wave length for $1$\,[GeV/c] is
\begin{eqnarray}
\lambda_{dB}=2\times 10^{-16}.
\end{eqnarray}
Lengths become,  $ 2\times 10^{-7}~[\text{m}]$,
$4\times 10^{-13}~[\text{m}]$, and $1\times 10^{-15}~[\text{m}]$
for  neutrino$\ (m_{\nu}\leq 1\text{eV}/c^2)$,  electron, and muon, respectively.
 The first term in the right-hand side of Eq. $(\ref{singular-function-f-p})$  
is  called the light-cone singularity and is 
long-range in $|t_1-t_2|$ or
 $|{\vec x}_1-{\vec x}_2|$.  This singularity reflects  relativistic 
invariance, i.e.,  an   energy of 
a mass $m$ and a  momentum ${\vec p}$ is $E(\vec{p}\,)=
\sqrt{\vec{p}^{\,2}+m^2}$ and approaches 
$E(\vec{p}\,) \rightarrow |\vec{p}\,|$ at $|\vec{p}\,| \rightarrow \infty $. Hence  
the   phase in Eq. $(\ref{positive-frequency-part})$,  $p\cdot(x_1-x_2)$, 
 cancels at a light-cone, $|t_1-t_2|=|{\vec x}_1-{\vec x}_2|$ in the 
direction ${\vec p}$ then. Consequently the wave  
becomes singular and  real  along the light cone. 

 If $m$ is pure imaginary in Eq. $(\ref{singular-function-f-p})$, the behaviors at $\lambda >0$ and $\lambda<0$ are 
interchanged, but the light-cone  singularity is the same. It is shown that 
 correlation functions of many particle states also have the light-cone 
singularity.
  
\subsection{Position-dependent amplitude from $S[\text T]$}
We study  the amplitude of a charged pion  produced in a hadron reaction with $S[\text T]$.
 The position-dependent amplitude of a  pion  is expressed  with  a 
wave packet.

A nucleon of a momentum ${\vec p}_{N_i}$  is prepared at time $t=\text{T}_{N_i}$, and  makes a
transition to a pion of  average values of the  
momentum ${\vec p}_{\pi}$  at  a four dimensional position
$({\text{T}_{\pi},{\vec X}_{\pi}})$ and other particles \cite{Ishikawa-Shimomura}. 
The  amplitude from this nucleon to   a nucleon 
 of   ${\vec
p}_{N_f}$ and a pion is,  
\begin{eqnarray}
\mathcal{M}=\int d^4x \, \langle {N}_f,{pion}   |H_{int}(x)| N_i \rangle,\ 
H_{int}=g \bar \psi_N \gamma_5 {\vec \tau}\cdot{\vec \varphi}(x)\psi_N, 
\label{amplitude0}
\end{eqnarray}
where the time and space are  integrated over the region  $\text{T}_{N_i} \leq x^0 \leq
 \text{T}_{\pi},X_{N_i}^j \leq x^j \leq X_{\pi}^j $. 
 The initial and final states are either   plane waves or the wave packet, 
\begin{eqnarray}
|N_i \rangle=   | {\vec p}_{N_i},\text{T}_{N_i}  \rangle,\ 
|N_f ,pion  \rangle=   |N_f,{\vec p}_{N_f};pion ,{\vec p}_{pion},{\vec X}_{pion},\text{T}_{pion}          \rangle.
\end{eqnarray}
The matrix elements of these states are defined in the forms, 
\begin{eqnarray}
& &\langle{\vec p}_{\pi},{\vec X}_{\pi},\text{T}_{\pi}  |\varphi(x)|0   \rangle
= N_{\pi}\rho_{\pi}\int d{\vec k_\pi} \, e^{-{\sigma_{\pi} \over 2}({\vec
 k}_\pi-{\vec p}_{\pi})^2}e^{i\left(E({\vec
 k}_\pi)(t-\text{T}_{\pi}) - i{\vec
 k}_\pi\cdot({\vec x}-{\vec X}_{\pi})\right)}  \nonumber \\
& &\approx 
N_{\pi}\rho_{\pi}\left({2\pi \over \sigma_{\pi}}\right)^{\frac{3}{2}}e^{-{1 \over 2 \sigma_{\pi} }\left({\vec x}-{\vec X_{\pi}}-{\vec v}_{\pi}(t-\text{T}_{\pi})\right)^2}
e^{i\left(E({\vec p}_{\pi})(t-\text{T}_{\pi}) - {\vec
 p}_{\pi}\cdot({\vec x}-{\vec X}_{\pi})\right)} \label{amplitude11}
%\label{pion-wf1}
,\\
%\label{pion-wf}\\
& &\langle N_f,{\vec p}_{N_f}|\bar
 u(x) \gamma_5 u(x) |N_i,{\vec p}_{N_i},\text{T}_{N_i} \rangle = \frac{1}{(2\pi)^3}
 \left({m_{N}
 \over E({\vec p}_{N_f})}\right)^{\frac{1}{2}}\left({m_{N} \over E({\vec
 p}_{N_i})}\right)^{\frac{1}{2}}{1 \over \sqrt V_{i}}
\nonumber\\ 
& &\times  \bar u({\vec p}_{N_f}) \gamma_5 u
 ({\vec p}_{N_i})
 e^{i\left((E({\vec p}_{N_f})-E({\vec p}_{N_i}))t-({\vec
 p}_{N_f}-{\vec
 p}_{N_i })\cdot{\vec x}+E(\vec{p}_{N_i})\text{T}_{N_i} \right)}
 \label{pi-wf1},
\end{eqnarray}
where
\begin{eqnarray}
& &N_{\pi} = \left(\frac{\sigma_{\pi}}{\pi}\right)^{\frac{3}{4}},\ 
\rho_{\pi}=
 \left(\frac{1}{2E_{\pi}(2\pi)^3}\right)^{\frac{1}{2}}, \nonumber
\end{eqnarray}
and  $V_i$ is normalization volume for 
 initial state. In this paper, the spinor's normalization is
\begin{eqnarray}
\sum_{s} u(p,s)\bar u(p,s)=\frac{\Slash{p} + m}{2m}.
\end{eqnarray}
In the above equation  the pion  life time is ignored.

Substituting Eqs. $(\ref{amplitude11})$ and $(\ref{pi-wf1})$, we have the amplitude for the pion detected at the space-time position $({\vec X}_{\pi},T_{\pi})$, which 
satisfies the boundary condition at T. It is important that the 
${\vec k}_{\pi}$ was integrated in Eq. $(\ref{amplitude11})$.  If the
integration 
over ${\vec x}$ is made prior to  the integration over 
${\vec k}_{\pi}$, the amplitude does not satisfy the boundary condition. That 
becomes,
\begin{align}
N&\int d t d{\vec x}d {\vec k}_{\pi}e^{-i({\vec p}_{N_i}-{\vec
 p}_{N_f}-{\vec k}_{\pi} )\cdot{\vec x}}
 e^{i\left(E({\vec p}_{N_f})-E({\vec p}_{N_i})+E({\vec k}_{\pi})\right)t+iE({\vec p}_{N_i})\text{T}_{N_i}} \nonumber \\
&\times \bar u({\vec p}_{N_f}) \gamma_5 u({\vec p}_{N_i}) e^{-{\sigma_{\pi} \over 2}({\vec
 k}_\pi-{\vec p}_{\pi})^2}e^{-i(E({\vec k}_{\pi})\text{T}_{\pi}-{\vec
 k}_{\pi}\cdot{\vec X}_{\pi})} \nonumber\\
= &N\int d t d {\vec k}_{\pi}(2\pi)^3\delta_{\text{L}}({\vec p}_{N_i}-{\vec p}_{N_f}-{\vec k}_{\pi} )
 e^{i\left(E({\vec p}_{N_f})-E({\vec p}_{N_i})+E({\vec k})\right)t+iE({\vec p}_{N_i})\text{T}_{N_i}}\nonumber\\
&\times \bar u({\vec p}_{N_f}) \gamma_5 u({\vec p}_{N_i}) e^{-{\sigma_{\pi} \over 2}({\vec
 k}_\pi-{\vec p}_{\pi})^2}e^{-i(E({\vec k}_{\pi})\text{T}_{\pi}-{\vec
 k}_{\pi}\cdot{\vec X}_{\pi})}, \label{amplitude-not-satisfy}\\ 
&N=N_{\pi} \rho_{\pi} \left( {m_N \over (2\pi)^3 E({\vec p}_{N_f})}
			 {m_N \over (2\pi)^3 E({\vec p}_{N_i})}{1 \over
			 \sqrt V_i}\right)^{1 \over 2},
%& &\approx N\int dx^0(2\pi)^3 e^{i\left(E({\vec p}_{N_f})
% -E({\vec p}_{N_i})+E({\vec k})\right)t}\bar u({\vec
% p}_{N_f}) \gamma_5 u({\vec p}_{N_i}) \nonumber \\
% & &e^{-{\sigma_{\pi} \over 2}({\vec
% k}_\pi-{\vec p}_{\pi})^2}e^{-i(E({\vec k}_{\pi})\text{T}_{\pi}-{\vec
% k}_{\pi}{\vec X}_{\pi})}|_{{\vec k}_{\pi}={\vec p}_{N_f}-{\vec
% p}_{N_i}}, \nonumber
\end{align}
where $\delta_\text{L}(x)$ is an approximate  delta function of a finite size
L, where $\text{L}=|{\vec X}_{\pi}-{\vec X}_{N_i}|$. 
In the above expression,  ${\vec x}$ 
was  integrated over whole region at any $t$, hence Eq. $(\ref{amplitude-not-satisfy})$ 
does not satisfy the 
boundary condition at T. This   is not suitable and is not used
here. However this amplitude shows some features of the phenomenon, and
its feature is analyzed hereafter.

 The integrand  in Eq. $(\ref{amplitude-not-satisfy})$ has two stationary momenta, one is that of the real part and 
the other is that of the imaginary part of the exponent. The stationary momentum from 
the real part, 
\begin{equation}
{\vec k}_{\pi}={\vec p}_{\pi}
\end{equation}
gives
\begin{equation}
 {\vec p}_{N_f}\approx  {\vec p}_{N_i} -{\vec p}_{\pi},
\end{equation}
which  gives  a finite contribution to Eq. $(\ref{amplitude-not-satisfy})$.  The   momentum
is  conserved approximately within uncertainty ${1}/{\sqrt {\sigma_{\pi}}}$, so 
the probability  from this kinematical 
region is regarded as an energy-conserving  term, which  agrees with that  
computed in the standard method of plane waves and asymptotic boundary
conditions. 
Another stationary momentum obtained  from the  imaginary
part of the exponent satisfies  
\begin{align}
\frac{\partial}{\partial {\vec k}_{\pi}} \xi =0, \ 
\xi= ((E({\vec p}_{N_f})-E({\vec p}_{N_i})+E({\vec k}))t-E({\vec
 k})\text{T}_{\pi}+{\vec k}\cdot{\vec X}_{\pi})|_{{\vec k}={\vec p}_{N_i}-{\vec
 p}_{N_f}},
\end{align}
and is important  at large 
$\text{T}_{\pi}-\text{T}_{N_i}$ and $|{\vec X}_{\pi}-{\vec X}_{N_i}|$. The solution is,
\begin{eqnarray}
&{\vec v}_{N_f} t-{\vec v}_{\pi} t +{\vec v}_{\pi}\text{T}_\pi-{\vec X}_{\pi}=0, \label{second-root}\\
&{\vec v}_{N_f}={{\vec p}_{N_f} \over E({\vec p}_{N_f})},{\vec
 v}_{\pi}={{\vec p}_{N_i}-{\vec p}_{N_f} \over E({\vec p}_{N_i}-{\vec
 p}_{N_f} )},
\end{eqnarray}
and is determined with the space-time position. The latter   gives a new contribution
of  violating    the kinetic-energy conservation.
Thus two stationary momenta have different properties.    The
 first one corresponds to the normal term of 
Eq. $(\ref{two-compoent S-matrix})$ and the second one corresponds
 to  the finite-size correction $\delta f$ of  Eq. $(\ref{two-compoent
 S-matrix})$. 
The finite-size correction is computed rigorously next.

\subsection{Expressing probability with correlation function }
The total probability is expressed  in the form 
\begin{align}
&P=\int {d{\vec X}_{\pi} \over (2\pi)^3}d{\vec p}_{\pi} d{\vec
 p}_{N_f} |\mathcal M|^2=\int {d{\vec X}_{\pi} \over (2\pi)^3} {1 \over V_i}{1 \over E_{N_i}
 (2\pi)^3}\frac{d{\vec p}_{\pi}}{E_{\pi}(2\pi)^3} \tilde P,
\label{total-probability-pion}\\
%& &\tilde P=\int {d{\vec p}_{N_f}
% \over (2\pi)^3} |T|^2 \nonumber 
%\end{eqnarray} 
%\begin{eqnarray}
&\tilde P=\int dx_1 dx_2 \Delta(x_1,x_2)e^{-\sum_i{1 \over 2\sigma_{\pi}}({\vec x}_i-{\vec
 X}_{\pi}-{\vec v}(t_i-\text{T}_{\pi}))^2}  e^{iE_{\pi}({\vec p}_{\pi})(t_1-t_2)-i{\vec
 p}_{\pi}\cdot({\vec x}_1-{\vec x}_2)}, 
\label{pion-probability}
\end{align}
where
\begin{eqnarray}
& &\Delta(x_1,x_2)= \int N d{\vec p}_{N_f} d({\vec p}_{N_f},{\vec p}_{N_i})
 e^{-i(p_{N_i}-p_{N_f})\cdot(x_1-x_2)} ,\\
& &d({\vec p}_{N_f},{\vec p}_{N_i})=\frac{1}{2}\sum_{spin}(\bar u({\vec p}_{N_f})\gamma_5
 u({\vec p}_{N_i}))^{*}\bar u({\vec p}_{N_f})\gamma_5 u({\vec
 p}_{N_i}),\\
& &N={m_N^2 \over (2\pi)^3  E_{N_f}}.
\end{eqnarray}
Taking a sum over  the final spins and an average over the initial
spin, we have 
\begin{align}
&d({\vec p}_{N_i},{\vec p}_{N_f})=  \frac{p_{N_i}\cdot p_{N_f}+m_N^2}{2 m_N^2},\\
&\Delta(x_1,x_2)={1 \over (2\pi)^3}\int {d{\vec p}_{N_f} \over E_{N_f}}
 (m_N^2+p_{N_i}\cdot p_{N_f})e^{-i(p_{N_i}-p_{N_f})\cdot(x_1-x_2)}\nonumber   \\
 &=\int d^4 p  \theta(p^0)\text{Im}\left[{1 \over p^2-m_N^2+i\epsilon}\right]
 (m_N^2+p_{N_i}\cdot p)e^{-i(p_{N_i}-p)\cdot(x_1-x_2)}\nonumber \\
&=e^{-ip_{N_i}\cdot(x_1-x_2)}\int d^4 p  \theta(p^0)\text{Im}\left[{1 \over p^2-m_N^2+i\epsilon}\right]
 (m_N^2+p_{N_i}\cdot p)e^{ip\cdot(x_1-x_2)}.\label{pion-correlation}
\end{align}
The integral over ${\vec X}_{\pi}$ in
Eq. $(\ref{total-probability-pion})$ gives
\begin{eqnarray}
\int d{\vec X}_{\pi}\frac{1}{V_i}=1.
\end{eqnarray}
%%%%%%%%%%%%%%%%%%%%%%%%%%%%%%%%%%%%%%%%%%%%%%%%%%%%%%%%%%%%%%
\subsubsection{Symmetry of probability   }
%%%%%%%%%%%%%%%%%%%%%%%%%%%%%%%%%%%%%%%%%%%%%%%%%%%%%%%%%%%%
$\tilde P$ is written in the form,
\begin{eqnarray}
\tilde P&=&\int dx_1 dx_2 \Delta(x_1,x_2)
e^{-{1 \over 4\sigma_{\pi}}({\vec x}_1-{\vec x}_2-{\vec
v}(t_1-t_2))^2}  e^{iE_{\pi}({\vec p}_{\pi})(t_1-t_2)-i{\vec
 p}_{\pi}\cdot({\vec x}_1-{\vec x}_2)}\nonumber\\
& &~~ \times e^{-{1 \over \sigma_{\pi}}({{\vec x}_1+{\vec x}_2 \over
 2}-{\vec X}_{\pi}-{\vec
v}_{\pi}({t_1+t_2 \over 2} -T_{\pi}))^2}. 
\label{P-probability}
\end{eqnarray}     
The short-range part $\Delta_s(x_1,x_2) $ in $\Delta(x_1,x_2) $
becomes non-vanishing at
\begin{eqnarray}
(x_1^0-x_2^0,{\vec x}_1-{\vec x}_2) \approx (0,{\vec 0}),
\end{eqnarray}
and $\tilde P_s$ becomes the product of factorized integrals
\begin{align}
&\tilde P_s=N_s \int d(x_1-x_2)  \Delta_s(x_1-x_2)
e^{-{1 \over 4\sigma_{\pi}}({\vec x}_1-{\vec x}_2-{\vec
v}(t_1-t_2))^2}  e^{iE_{\pi}({\vec p}_{\pi})(t_1-t_2)-i{\vec
 p}_{\pi}\cdot({\vec x}_1-{\vec x}_2)},\\
&N_s=\int d({x_1+x_2 \over 2})e^{-{1 \over \sigma_{\pi}}({{\vec x}_1+{\vec x}_2 \over
 2}-{\vec X}_{\pi}-{\vec
v}_{\pi}({t_1+t_2 \over 2} -T_{\pi}))^2}, 
\end{align}  
where $N_s$ is a constant. $P_s$ is equivalent to that of
$\sigma_{\pi}=\infty$ and Lorentz invariant. 

The long-range part $\Delta_l(x_1,x_2) $ in $\Delta(x_1,x_2) $
becomes non-vanishing at large
\begin{eqnarray}
x_1-x_2,
\end{eqnarray}
and $\tilde P_l$ becomes 
\begin{eqnarray}
\tilde P_l&=& \int d(x_1-x_2)  \Delta_l(x_1-x_2)
e^{-{1 \over 4\sigma_{\pi}}({\vec x}_1-{\vec x}_2-{\vec
v}(t_1-t_2))^2}  e^{iE_{\pi}({\vec p}_{\pi})(t_1-t_2)-i{\vec
 p}_{\pi}\cdot({\vec x}_1-{\vec x}_2)}\nonumber \\
& & \times\int d({x_1+x_2 \over 2})e^{-{1 \over \sigma_{\pi}}({{\vec x}_1+{\vec x}_2 \over
 2}-{\vec X}_{\pi}-{\vec
v}_{\pi}({t_1+t_2 \over 2} -T_{\pi}))^2}. 
\label{P_l-probability}
\end{eqnarray}
  In $\tilde P_l$, two integrals are not factorized.  Accordingly
  $\tilde P_l$ is
  not Lorentz
  invariant and depends on $\sigma_{\pi}$. 
%%%%%%%%%%%%%%%%%%%%%%%%%%%%%%%%%%%%%%%%%%%%%%%%%%%%%%%%%%%%%%
\subsubsection{light-cone singularity 1}
%%%%%%%%%%%%%%%%%%%%%%%%%%%%%%%%%%%%%%%%%%%%%%%%%%%%%%%%%%%%
Here the formula for a relativistic field     
Eq. $(\ref{singular-function-f-p})$
is substituted into 
Eq. $(\ref{pion-correlation})$,
and we have  
\begin{eqnarray}
& &\Delta(x_1,x_2)=e^{-ip_{N_i}\cdot(x_1-x_2)}i\left(\frac{1}{4\pi}\delta
 (\lambda)\epsilon(t_1-t_2)+Bessel~ functions \right)
\label{pion-non-singular} \nonumber \\
& &=e^{-i\bar \phi_{N_i}(\delta t )} {i \over 4\pi}\delta
 (\lambda)\epsilon(\delta t)+Bessel~functions,\\
& &\bar \phi_{N_i}=(E_{N_i}-|{\vec p}_{N_{i}}|) \delta t,~ \delta t=t_1-t_2. \nonumber
\end{eqnarray}
%%%%%%%%%%%%%%%%%%%%%%%%%%%%% figure of Light Cone %%%%%%%%%%%%%%%%%%%%%%%%%%%%%%%%%%%%%%%%%%%%%%%%
%\begin{figure}[t]
%\centering{
% \includegraphics[scale=.5]{fig4_0.eps}
%}
%\caption{The space time region of the correlation function is shown. The 
%region I around the origin corresponds to short-range correlation where $t
% \sim x \sim 0$, while the region II corresponds to
% long-range correlation where $t^2 - x^2 \sim 0$ and both $t$
% and $x$ can be macroscopic.}
%\label{fig:light-cone}
%\end{figure}
%%%%%%%%%%%%%%%%%%%%%%%%%%%%%%%%%%%%%%%%%%%%%%%%%%%%%%%%%%%%%%%%%%%%%%%%%%%%%%%%%%%%%%%%%%%%%%%%%%
The first term in the right hand side of
Eq. $(\ref{singular-function-f-p})$ is the most singular term and the
second and third terms have  singularity  of the form $1/\lambda$ around
$\lambda =0$ and decrease as $e^{-\tilde m\sqrt{|\lambda}|}$ or
oscillates as  $e^{i\tilde m \sqrt{|\lambda}|}$. The singular functions
and regular functions
behave differently 
%and are expressed in Fig.\,\ref{fig:light-cone} for one space
%dimension.  
The singular functions are  non-vanishing   in  narrow
regions around  the light
cone and the regular functions have finite values in  a small area around
the origin. Since the light cone  singularity is extended  and decreases
slowly as ${1 \over |t_1-t_2|}$, the correlation function has  
long-range component. The correlation function  from other terms  
decreases exponentially or oscillates rapidly. 
In the above equation, ``Bessel functions'' are oscillating or
decreasing fast with $\lambda$ and  this property is sufficient to know  a 
large T behavior of the probability.
  In
latter sections, concrete forms of  these functions   are obtained  
and their properties are studied.
 
Equation $(\ref{pion-non-singular})$ is substituted into
Eq. $(\ref{pion-probability})$ and after a tedious calculation   the probability is written as the sum,
\begin{eqnarray}
& &\tilde P=\int_{\text{T}_{N_i}}^{\text{T}_\pi} dt_1 dt_2(\sigma_\pi)^{\frac{3}{2}}{\sigma_{ \pi} \over 2}{1 \over |\delta
 t|}\epsilon(\delta t) e^{i(\bar \phi_{\pi}(\delta t)-\bar
 \phi_{N_i}(\delta t))}+\tilde P^{(0)},
\label{pion-probability1}
\\
& &\bar  \phi_{\pi}=\omega_{\pi} \delta t,~\omega_{\pi}=(E_{\pi}-|\vec{p}_{\pi}|), 
\end{eqnarray}
where we use the notation $t=x^0$, and the first and second terms in 
right-hand side are  derived  from the long and short range parts.
The first term  is proportional to  the following
function of  $\omega \text T$, 
\begin{eqnarray}
& &\text{T} g(\text{T},\omega)=i  \int_0^{\text{T}} dt_1 dt_2  
\frac{e^{i {\omega }\delta t }}{|\delta t|}\epsilon(\delta t)=-2\left(\text{Sin}~ x-\frac{1 -\cos x}{x}\right),\label{probability1-1}\\ 
& &x=\omega \text{T}, \ \text{Sin}~x=\int_0^x dy \frac{\sin y}{y}, \nonumber
\end{eqnarray}
which  satisfies  
\begin{eqnarray}
& &{\partial \over \partial \text{T}}g(\text{T},\omega )|_{\text{T}=0}
= -\omega ,\\
& &g(\infty,\omega)= {-\pi}.
\end{eqnarray} 
Subtracting  the asymptotic value, we   define $\tilde g(\text{T},\omega_\nu) $ 
\begin{eqnarray}
 g(\text{T},\omega)= \tilde g(\text{T},\omega)+g(\infty,\omega),
\end{eqnarray}
which satisfies 
\begin{eqnarray}
{d \over dx}\tilde g(x)=-4\left(\frac{\sin (x/2)}{x}\right)^2,
\end{eqnarray}
and  oscillates rapidly at large $x$ with an average   
\begin{eqnarray}
 \tilde g(x) ={2 \over x}.
\end{eqnarray}
Thus  $\tilde g(\text{T},\omega )$ behaves as 
\begin{eqnarray}
\tilde g(\text{T},\omega ) \approx \frac{\text{T}_0}{\text{T}},\ \text{T}_0={2 \over \omega},
\end{eqnarray}  
and is given in Fig.\,$\ref{fig:gtilde}$ as a function of $x$. For particles 
of $1$ [GeV] and $cT=100$ [m]. $\omega $ is large  and 
$\tilde g(\text T,\omega)$ is extremely small for the electron and muon, but 
the value becomes around 1 for the neutrino.  
%\centering
{\begin{figure}[t]
\begin{center}
\includegraphics[scale=.55,angle=-90]{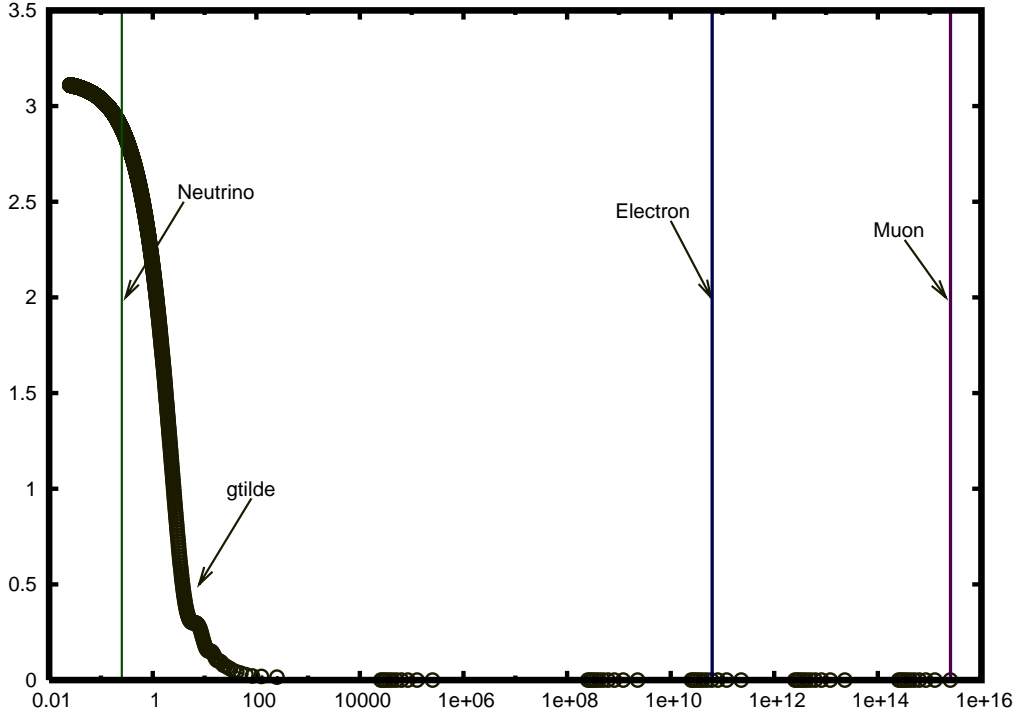}
\end{center}
\caption{The function $\tilde g(\text{T},\omega)$ is  given as a function of $\omega \text T$. The value of $\omega \text T$ for various
  particles of the energy $1$ [GeV] and $\text cT=100 $[M] are expressed in solid lines. The 
vertical line shows the magnitude of  $\tilde g(\text{T},\omega)$  and the horizontal line shows $\omega \text T$.
  The mass $m_{\nu}=1$ [eV] is used for the neutrino in this
  Figure. Since the electron
  and muon are massive, $\tilde g(\text{T},\omega)$ are negligibly small.  The values for the pion and other hadrons are
  smaller than that of muon.}
\label{fig:gtilde}
\end{figure}}

 The probability derived from the second term of
 Eq. $(\ref{pion-probability1})$ is a constant. Thus the probability is
 expressed in the form 
\begin{eqnarray}
& &\tilde P=C_0(\sigma_{\pi})\tilde g(\text{T},\omega)+\tilde P^{(0)},\label{pion-probability11}\\
& &\omega=(E_{\pi}-|\vec{p}_{\pi}|)-(E_{N_i}-|\vec{p}_{N_i}|),\ \text{T}=\text{T}_\pi-\text{T}_{N_i},\\
& &C_0(\sigma_{\pi})=\text T (\sigma_{\pi})^{5/2} /2 , \tilde P^{(0)}=\tilde P^{(0)}+\text T (\sigma_{\pi})^{5/2} /2  g(0),
\end{eqnarray}
$\tilde g(\omega, \text{T})$ is a function of $\omega \text{T}$ and 
is inversely proportional to  $\omega\text{T} $ at a 
large $\omega\text{T} $. Here    
$\omega$ is determined by the pion's mass and energy and the nucleon's
mass and energy and  is not very small. Hence, $\tilde g(\omega \text
T)$  
vanishes at a macroscopic
$\text{T}$. 
 The integrand of the probability in Eq. $(\ref{pion-probability1})$ 
oscillates rapidly in $\delta t$ 
with  a non-small angular 
velocity
$\omega $, and  the probability becomes
constant fast. 
%%%%%%%%%%%%%%%%%%%%%%%%%%%%%%%%%%%%%%%%%%%%%%%%%%%%%%%%%%%%%%%
\subsubsection{light-cone singularity 2}
%%%%%%%%%%%%%%%%%%%%%%%%%%%%%%%%%%%%%%%%%%%%%%%%%%%%%%%%%%%%%%%%
The integrand  of  Eq. $(\ref{pion-correlation})$  around a momentum 
 of  $(p_{N_f}-p_{N_i})^2=0$ does not oscillate 
 at $\lambda=0$ and  becomes real.  A sum of  these  waves becomes
 real and singular at $\lambda=0$ due to constructive interference and  
 forms  the light-cone  singularity.  Especially this function   does not 
accompany  the oscillating  function 
$e^{-i\bar \phi_{N_i}(\delta t)}$, hence gives different probability. This singularity  is extracted with a
 suitable expression of the integral. 
  Changing  the variable     from $p$ to
 $q=p_{N_f}-p$,  we write 
\begin{align}
\Delta(x_1,x_2)=&
\int d^4 q  \theta(p_{N_i}^0-q^0)\text{Im}\left[{1 \over (q-p_{N_i})^2-m_N^2+i\epsilon}\right] \nonumber\\
&\times
 (2m_N^2-p_{N_i}\cdot q)e^{-iq\cdot(x_1-x_2)}. \label{correlation-new-variable} 
\end{align}
Next we have the  expression  of the denominator in the form,
\begin{eqnarray}
{1 \over (q-p_{N_i})^2-m_{N_f}^2+i\epsilon}
 =\sum_l\left(2q\cdot p_{N_i}{\partial
 \over \partial \delta m_N^2}\right)^l {1 \over q^2+\delta m_N^2+i\epsilon},
\end{eqnarray}
where a small difference between $m_{N_i}$ and $m_{N_f}$ which has been
ignored so far is included here and $\delta m_N^2=m_{N_i}^2-m_{N_f}^2$ is the mass-squared difference 
between a proton and
a neutron. Then $\Delta(x_1,x_2)$ is written as 
\begin{eqnarray}
\Delta(x_1,x_2)
=i\frac{\epsilon({\delta t})}{4\pi}\delta (\lambda)
  +
 Bessel~functions+regular~function\label{pion-singular-term},
\end{eqnarray}
where the light-cone singular term  is derived from $l=0$, and the 
others are from $l \neq 0$. The  second and third terms of Eq. $( \ref{pion-singular-term})$ are not studied  in detail here, but the first term, which  gives  the large
$|\delta t|$ behavior of $\Delta(x_1,x_2)$ is studied hereafter. The regular  terms are 
from $0< q^0$. This method of applying the Taylor expansion is valid when 
the infinite series converges.
The  integrations of the series  in the expansion of the denominator converge 
in the kinematical region,
\begin{eqnarray}
2p_{N_i}\cdot p_{\pi} \leq {\delta m_N}^2.
\label{pion-convergence-condition}
\end{eqnarray}
Hence the present method is valid and the  light-cone singularity exists
in this  narrow kinematical  region. The convergence of the series will
be studied later in detail.

Equation $(\ref{pion-singular-term})$ is substituted into
Eq. $(\ref{pion-probability})$ and after  tedious calculations,
we have the probability  in the
form,
\begin{eqnarray}
\tilde P&=&\int_{\text{T}_{N_i}}^{\text{T}_\pi} dt_1dt_2(\sigma_\pi)^\frac{3}{2}
{\sigma_{ \pi} \over 2}\frac{\epsilon(\delta t)}{|\delta t|}
 e^{i\bar \phi_{\pi}(\delta t)}+\tilde P^{(0)},
%\bar \phi_{\pi}=(E_{\pi}-p_{\pi}) \delta t
\label{pion-probability2} \nonumber\\
&=&C_0(\sigma_{\pi})\tilde g(\text{T},\omega_{\pi})+\tilde P^{(0)}. 
\end{eqnarray} 
In Eq. $(\ref{pion-probability2})$, $\tilde P^{(0)}$ is the asymptotic value,
 and the first  term  vanishes at $\text T \rightarrow \infty$ and 
is the finite-size correction. The correction is the product of the 
universal function
$\tilde g(\text{T},\omega_{\pi},\text)$ that
 is independent of the wave packet and $C_0(\sigma_{\pi})$.

Equation (\ref{pion-probability2}) has almost the same form as
Eq. ($\ref{pion-probability11}$) but  a different  angular velocity
$\omega_{\pi}$ is used. $\omega_{\pi}$ is smaller than $\omega$ of 
Eq. $(\ref{pion-probability11})$. Hence this   is more 
convenient to study the large T behavior than the former
one since it has the slowest oscillation. At 
large $|\delta t|$ region,
the frequency in time is given from
\begin{eqnarray}
\bar \phi(\delta t) =(E_{\pi}-|\vec{p}_{\pi}|) \delta t={m_{\pi}^2 \over
 2E_{\pi}} \delta t.
\end{eqnarray}
In the last calculation, the large momentum expansion $E(\vec{p}\,)=|\vec{p}\,|+\frac{m^2}{
2|\vec{p}\,|}$ was made. The pion's mass $m_{\pi}$,  $139\ [\text{MeV}/c^2]$, makes   the 
angular velocity ${m_{\pi}^2}/{(2E_{\pi})}$  small only in an extreme high-energy region. 
Thus a coherence length of a pion emitted from  a  nucleon is $l_0={2cE_{\pi}}/{m_{\pi}^2}$, which is microscopic due to the large mass, and the pion in
the kinematical region, Eq. (\ref{pion-convergence-condition}) can be 
observed inside this length.
 If the
 pion's mass $m_{\pi}$ were  $1\,[\text{eV}/c^2]$, then $l_0$ would have been 
macroscopic.    
%%%%%%%%%%%%%%%%%%%%%%%%%%%%%%%%%%%%%%%%%%%%%%%%%%%%%%%%%%%%%%%%%%%%%
\subsection{Pion from $NN$ collisions }
%%%%%%%%%%%%%%%%%%%%%%%%%%%%%%%%%%%%%%%%%%%%%%%%%%%%%%%%%%%%%%%%%%% 
A probability of the event that  one pion produced  in $NN$ collision is
detected is studied in this section. 
\begin{figure}[t]
\centering{\includegraphics[scale=.6]{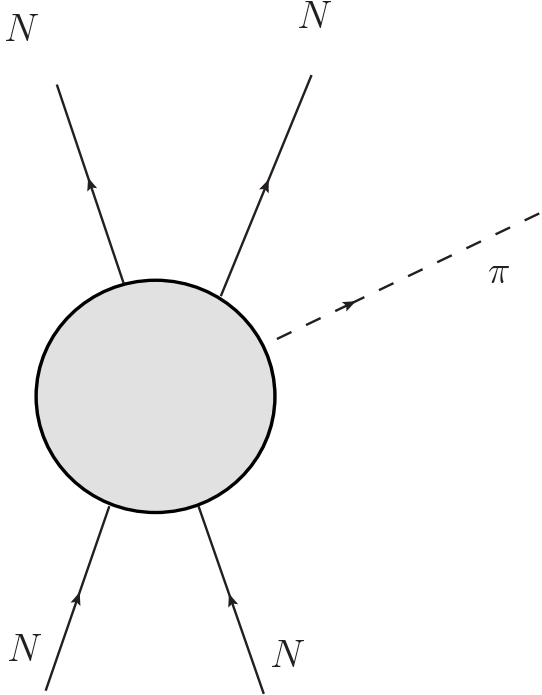}
\caption{A Feynman diagram of a pion production in $NN$ scattering. The 
amplitude for one pion  is  computed in the text.}}
\end{figure}
In the  Feynman diagram 
Fig.\,4, the dot line  shows the pion that   is detected at a
position $X^{\mu}$ and other particles are momentum eigen states.  
$\mathcal{M}(p_{N_f},\cdots, p_n;X_T)$ is the amplitude of the event
that  one pion of a
momentum ${\vec p}_\pi$  is detected,  
\begin{align}
\mathcal{M}=\int d^4 x 
 w(\vec{x}_{\pi}-\vec{v}_{\pi}(t-\text{T}_{\pi});X_{\pi})e^{i(p_{N_f}+p_{\pi}-p_{N_i})
\cdot x}\mathcal{M}(p_{N_i},\cdots, p_n;X_T),
\end{align}
where $p_{N_i}$ and 
$p_{1},p_{2},\cdots$ are  momenta of the initial and final state, and $X_T$
is the position of the target, 
Using $\mathcal{M}$, the probability of the event is expressed  in the form, 
\begin{align}
C(X_{\pi},\vec{p}_{\pi})=&\int d^4x_1 d^4x_2 
w(\vec{x}_1-\vec{v}_{\pi}(t_1-\text{T}_{\pi});X_{\pi})
w^{*}(\vec{x}_2-\vec{v}_{\pi}(t_2-\text{T}_{\pi});X_{\pi}) \nonumber\\
 &\times e^{i(p_{N_f}-p_{N_i})
\cdot (x_1-x_2)}e^{ip_{\pi}
\cdot (x_1-x_2)}
|\mathcal{M}(p_{N_i},\cdots, p_n;X_T)|^2d \vec{p}_{N_f} \label{Probability-general}.
\end{align}
A sum of the final states is decomposed to  a light-cone singularity and
regular term, 
\begin{eqnarray}
& &\int \sum d\vec{p}_{N_f} e^{i(p_{N_f}-p_{N_i})
\cdot (x_1-x_2)} |\mathcal{M}(p_{N_i},\cdots, p_n;X_T)|^2 \nonumber\\
& &=D \delta(\lambda)\epsilon(\delta t)+\text {regular term}\label{light-cone-general} ,
\end{eqnarray}
where  $D$ is the coefficient. Substituting
Eq. $(\ref{light-cone-general} )$ to Eq. $(\ref{Probability-general})$
 we have 
\begin{eqnarray}
& &C(X_{\pi},\vec{p}_{\pi})=  DC_0(\sigma_{\pi}) \tilde
 g(\text{T},\omega_{\pi})  ,
\end{eqnarray}
which is equivalent to that of
$\tilde P(X_{\pi},p_{\pi};p_{N_i})$ of Eq. $(\ref{pion-probability2})$ 
and  has a length $l_0$  of
Eq. ($\ref{pion-coherence }$). The length  $l_0={c \hbar E_{\pi}}/{m_{\pi}^2}$
 is much longer than the de Broglie wave length, $\lambda={h}/{|\vec{p}\,|}$,
 but is microscopic in ordinary high-energy experiments.

Thus, the pion's coherence length $l_0$ is of microscopic size and
the probability of the event that the pion is detected in the region $|{\vec X}_{\pi}-{\vec
X}_T| \gg l_0$ 
agrees with that of $S[\infty]$ of $\sigma$, hence that  of
$S[\infty]$ of $\sigma=\infty$ 
from Eq. $( \ref{rep-independence })$. The probability of the event
that the pion is detected     at a macroscopic T agrees with the
asymptotic value. The finite-size correction is negligible.  

%%%%%%%%%%%%%%%%%%%%%%%%%%%%%%%%%%%%%%%%%%% Sec1
%%%%%%%%%%%%%%%%%%%%%%%%%%%%%%%%%%%%%%%%%%% %%%%%%%%%%%%%%%%%%%%%%%%%%%%%%%%%%%%%%%%%%%%%%%%%%%%
%%%%%%%%%%%%%%%%%%%%%%%%%%%%%%%%%%%%%%%%%%% Sec1 %%%%%%%%%%%%%%%%%%%%%%%%%%%%%%%%%%%%%%%%%%%%%%%%%%%%
\section{Probability of the event that the neutrino is detected}
%%%%%%%%%%%%%%%%%%%%%%%%%%%%%%%%%%%%%%%%%%%%%%%%%%%%%%%%%%%%%%%%%%%%
The pion decays to a neutrino and lepton by weak interaction. Hereafter,
the event that  this  neutrino  is detected is studied. 
%%%%%%%%%%%%%%%%%%%%%%%%%%%%%%%%%%%%%%%%%%%%%%%%%%%%%%%%%%%%%%%%%%%

\subsection{Pion decays}

 The correction of probability of the event that  the neutrino is detected 
becomes large due to tiny neutrino mass. Particularly a large correction
is induced in   the electron and electron neutrino mode that is
suppressed in the asymptotic region due to kinetic energy and angular
momentum conservations.  
   Since a neutrino, charged lepton, and pion are described by a 
many-body wave function that follows  Schr\"{o}dinger  equation,  
the kinetic energy  of the final state at a finite time deviates from
that of the initial energy. That  is not a constant and takes a wide 
range of values.   If the  initial  pion is expressed by a wave function
of large size,  the 
neutrino wave
 overlaps with the pion in wide area \cite{goldberger-watson-paper}, and $S[\text T]$
    that 
satisfies boundary conditions  of experiments
is appropriate  \cite{LSZ,Low} and  are  used here 
\cite{Ishikawa-Shimomura,Ishikawa-Tobita-ptp,Ishikawa-Tobita}.
The entire processes is analyzed with $S[\text T]$ expressed by wave packets.

 %%%%%%%%%%%%%%%%%%%%%%%%% geometric picture %%%%%%%%%%%%%%%%%%%%%%%%%%%%%%%%%%%%%%%%
\begin{figure}[t]
 \includegraphics[scale=.65,angle=-90]{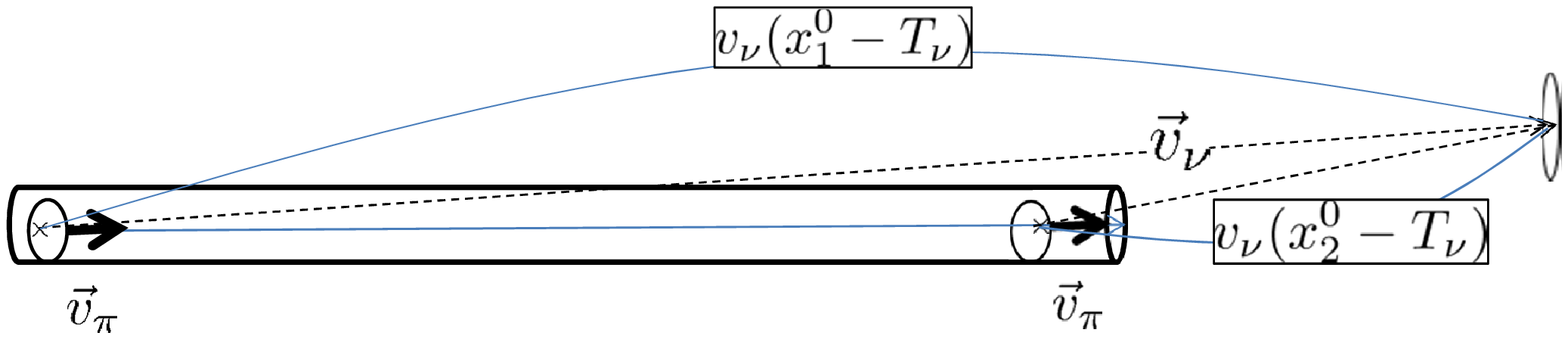}
\caption{The geometry of the event that the neutrino is detected. The neutrino is 
produced in the pion decay and is detected by the detector.
Since the decay occurs between the position of the initial pion and that of the 
detector,  the amplitude of the event  is the overlap between 
the superposed initial wave
 and a final state expressed by the wave packet of the small size.
 The probability measured by the detector at $T_{\nu}$  shows an
 interference pattern.  }
\label{fig:geo}
\end{figure}
%%%%%%%%%%%%%%%%%%%%%%%%% geometric picture %%%%%%%%%%%%%%%%%%%%%%%%%%%%%%%%%%%%%%%%

 A 
neutrino propagates with almost  constant velocity.  In an event that  a 
neutrino is detected at one position, the position 
where the neutrino is produced   varies,  and  a neutrino wave 
at the detector is a  superposition of those waves
that are produced at different space-time positions. When these space-time
positions is extended in the wide area, as in Fig.\,$\ref{fig:geo}$,
 the neutrino waves keep  their coherence, and the probability amplitude  and probability reveal  
interference patterns. A
  condition for the interference phenomenon to occur for    
 the pion expressed by a wave function of the size
 $\sqrt{\sigma_{\pi}}$ 
and  a 
 velocity   $\vec{v}_{\pi}$ is easily obtained for one dimensional motion. Let a
 neutrino be produced either at  
time $t_1$ or $t_2$  from the  pion prepared  at $\text{T}_{\pi}$ and travel
for some period and 
be finally  detected  at $\text{T}_{\nu}$,   
then the waves  overlap if 
\begin{eqnarray}
|(c(\text{T}_{\nu}-t_1)+v_{\pi}(t_1-\text{T}_{\pi}))-(c(\text{T}_{\nu}-t_2)+v_{\pi}(t_2-\text{T}_{\pi}))|
 \leq \sqrt{ \sigma_{\pi}},
\label{coherence-condition}
\end{eqnarray}
is fulfilled, where the speed of light
 is used for the speed of neutrino  $v_{\nu}=c$.
So  Eq. $(\ref{coherence-condition})$ is one of the necessary conditions for the 
neutrino interference to occur in the one-dimensional space. 
For a plane wave of pion $\sigma_{\pi}=\infty$ and the above condition
is satisfied. For a high-energy pion of a finite $\sigma_{\pi}$, its
 speed is close to the speed of light  and the 
left hand side of Eq. $(\ref{coherence-condition})$ becomes 
$c(m_{\pi}^2/2E_{\pi}^2)(t_1-t_2)$. Hence this
condition Eq. $(\ref{coherence-condition})$ is written in the form, 
 $c(t_1-t_2) \leq 
\sqrt{\sigma_{\pi}}(2E_{\pi}^2/m_{\pi}^2)$.
 When this length $c(t_1-t_2)$ is a macroscopic size, the
interference  phenomenon  occurs at a macroscopic length. 
We  estimate the 
lengths of these particles 
in Appendix A and confirm that this condition in three-dimensional space 
is fulfilled even in a macroscopic
distance.  It is noted that  $\sqrt{\sigma_{\pi}}$  is the size of pion wave function in laboratory frame and is not related with $l_0$ of the pion 
discussed in the previous section.

%\newpage
%%%%%%%%%%%%%%%%%%%%%%%%%%%%%%%%%%%%%%%%%%% Section 2 %%%%%%%%%%%%%%%%%%%%%%%%%%%%%%%%%%%%%%%%%%%%%%%%%%%%%%%%%%%
%%%%%%%%%%%%%%%%%%%%%%%%%%%%%%%%%%%%%%%%%%%%%%%%%%%%%%%%%%%%%%%%%%%%%%%%%%%%%%%%%%%%%%%%%%%%%%%%%%%%%%%%%%%%%%%%%

%\newpage
\subsection{Amplitude of the event that the neutrino is detected at T}
  Probabilities of the events that a neutrino and a charged
  lepton   
are detected at T are computed  with $S[\text T]$.  When the wave functions
  of pion and
  daughters overlap, they have a finite interaction energy. Consequently,
the kinetic energy of daughters  deviates  from that of the pion.
\subsubsection{Leptonic decay of the pion }
A leptonic decay of a pion is described with  the weak Hamiltonian  
\begin{align}
&H_{int}=g \int d{\vec x}\, {\partial_{\mu}
 }\varphi(x)J_{V-A}^{\mu}(x)=-igm_{\mu}\int d{\vec x} \,
 \varphi(x)J_5(x)+\delta L_{int},
\label{weak-hamiltonian}\\
&J_{V-A}^{\mu}(x)=\bar
 \mu(x)\gamma^{\mu}(1-\gamma_5)\nu(x),J_5(x)=\bar
 \mu(x)(1-\gamma_5)\nu(x),
\end{align}
where ${\varphi(x)}$, $\mu(x)$, and $\nu(x)$ are the pion field, muon
field, and neutrino field. In Eq. $(\ref{weak-hamiltonian})$,    $g$ is the coupling
strength, $J_{V-A}^{\mu}(x)$, and $J_5(x)$ are a
leptonic charged $V-A$ current, and a leptonic  pseudoscalar.  $G_F$ is Fermi 
constant  and $f_\pi$ is a pion decay 
constant, and
\begin{eqnarray} 
g=\frac{G_F}{\sqrt 2}f_{\pi}.
\end{eqnarray}
Here 
\begin{eqnarray}
\delta L_{int}={\partial \over \partial x_{\mu}}G^{\mu}, G^{\mu}=g 
\varphi(x)J_{V-A}^{\mu}(x),
\label{total-derivative}
\end{eqnarray}
 is a total derivative and does not give bulk
effects. The equation of motion and 
$\Gamma$ are kept intact. Nevertheless, this contributes to the non-bulk
probability $P^{(d)}$. Especially $P^{(d)}$ becomes important for  the 
electron mode, because $m_{electron} \ll m_{\mu}$. A computation will be
made later for the electron mode using the $(V-A)\times(V-A)$ form of interaction. 

A pion and decay products  are expressed  by   the    Schr\"{o}dinger equation
\begin{eqnarray}
i\hbar {\partial \over \partial t}|\Psi(t)\rangle= H
 |\Psi(t)\rangle,\ H=H_0+H_{int},
\end{eqnarray}   
 and the solution   of satisfying   
the initial condition 
\begin{eqnarray}
 |\Psi(t) \rangle|_{t=\text{T}_{\pi}}=|\text {pion} (t)\rangle|_{t=\text{T}_{\pi}}=e^{-i\frac{E_{\pi}}
{\hbar} \text{T}_{\pi}}|{\vec p}_{\pi},\text{T}_{\pi} \rangle,
\end{eqnarray}
 in the first order in $H_{int}$ is   
\begin{eqnarray}
& &|\Psi(t) \rangle= |\text {pion} (t)\rangle+|\text {muon,~neutrino}(t)
 \rangle\label{state-vector},
\end{eqnarray}
where   
\begin{eqnarray}
& &|\text{pion} (t) \rangle = a(t)  |\text {pion},{\vec p}_{\pi}
 (t)\rangle, a(t)=1+O(g^2) \\
& &|\text {muon,~neutrino}(t) \rangle =\int_{\text{T}_{\pi}}^{t} {d t' \over i\hbar}H_{int}(t')
 |\text{pion}
 (t')\rangle. \label{lowest-order mu-nu state}
\end{eqnarray}
 The state is written as     
\begin{align}
|\text {muon,~neutrino}(t) \rangle= &g  e^{-i\frac{E_{\pi}}{\hbar}t}\int
  d{\vec p}_{\mu} d{\vec p}_{\nu}\sqrt{m_{\mu} m_{\nu}
 \over E_{\mu}({\vec p}_{\mu})E_{\nu}({\vec p}_{\nu})}
\frac{e^{-i{\omega t}/ \hbar}-1}{\omega}\nonumber\\
&\times \delta^{(3)}({\vec p}_{\pi}-{\vec
 p}_{\mu}-{\vec p}_{\nu})
({p_{\pi}})_{\mu}\bar \mu({\vec
 p}_{\mu})\gamma^{\mu}(1-\gamma_5)\nu({\vec p}_{\nu})|{\vec
 p}_{\mu},{\vec p}_{\nu} \rangle  \label{wave-function-t},
\end{align}
where $\omega =E_{\mu}+E_{\nu}-E_{\pi}$ and   $|{\vec
 p}_{\mu},{\vec p}_{\nu} \rangle $ is a two-particle state composed of 
the muon
 and neutrino of momenta $ {\vec
 p}_{\mu}$ and  $\ {\vec p}_{\nu} $.

At $t=\infty$, 
\begin{align}
|\text {muon, neutrino} (t)\rangle= &-i g  e^{-i\frac{E_{\pi}}{\hbar}t}\int
  d{\vec p}_{\mu} d{\vec p}_{\nu} \sqrt{m_{\mu} m_{\nu} \over E_{\mu}({\vec p}_{\mu})E_{\nu}
({\vec p}_{\nu})}(2\pi)\delta^{(4)}(p_{\pi}-p_{\mu} -p_{\nu}) \nonumber\\
&\times {(p_{\pi})}_{\mu}\bar \mu({\vec
 p}_{\mu})\gamma^{\mu}(1-\gamma_5)\nu({\vec p}_{\nu})|{\vec
 p}_{\mu},{\vec p}_{\nu} \rangle,
\end{align} 
and the norm of this state  is given by, 
\begin{eqnarray}
&\langle   \text {muon,~neutrino} (\text{T})|\text
 {muon,~neutrino}(\text{T}) \rangle =\text{T}\Gamma ,
\end{eqnarray}
where $\Gamma$ is the average decay rate \cite{Dirac,Schiff-golden}.
A neutrino and muon are produced simultaneously, but  they propagate  
differently. 
 The neutrino propagates  long distance with the speed of light and  is 
detected afterword.  Hence the probability  for the neutrino  is affected   
by a retarded  effect  similar to the classical electric field
caused by a moving charge, and  is different from the probability of the
 events that the muon is detected.

The events of  neutrino are  identified in experiments  with
its reaction products  with nucleus in  detector, hence  the wave 
function of the final states
is the one of the nuclei. The simplest form of Gaussian function  
of the  size $\sigma$ of  nuclear wave function is used.  Using them,
we 
express  
$S[\text T]$ following  the formula of
Ref.\,\cite{LSZ}. In Ref. \cite{LSZ}, $S[\infty]$ was computed, and the wave 
functions were replaced with
plane waves. Here $S[\text T]$  is computed with the amplitude  of finite 
$\sigma$.

Previous works  studied flavor
oscillations with $S[\infty]$ \cite{Kayser,Giunti,Nussinov,Kiers,Stodolsky,Lipkin,Akhmedov,Asahara}
described with large wave packets in Poincar\'{e} invariant manner, where  the 
position dependence is ignorable and non-computable. 
Now 
$S[\text T]$  depends  on the position 
and T, and the finite-size corrections are computable.  
 For the small wave packets,   
wave packets of different positions \cite{Ishikawa-Shimomura} are necessary 
from the  completeness.

The element of $S[\text T]$ is defined  with  a  wave function of 
an initial pion located at a position ${\vec
X}_\pi$, a neutrino at a  position ${\vec
X}_{\nu}$ and a  muon as  
\begin{eqnarray}
\mathcal{M}=\int d^4x \, \langle {\mu},{\nu}   |H_{int}(x)| \pi \rangle,\label{amplitude1}
\end{eqnarray}
where the pion and neutrino  are expressed 
in the form   
\begin{eqnarray}
|\pi \rangle=   | {\vec p}_{\pi},{\vec X}_{\pi},\text{T}_{\pi}  \rangle,\ 
|\mu ,\nu \rangle=   |\mu,{\vec p}_{\mu};\nu,{\vec p}_{\nu},{\vec X}_{\nu},\text{T}_{\nu}          \rangle,
\end{eqnarray}
and with   the matrix elements, 
\begin{eqnarray}
& &\langle 0|\varphi(x)|{\vec p}_{\pi},{\vec X}_{\pi},\text{T}_{\pi} \rangle 
\approx 
N_{\pi}\rho_{\pi}\left({2\pi \over \sigma_{\pi}}\right)^{\frac{3}{2}}e^{-{1 \over 2 \sigma_{\pi} }\left({\vec x}-{\vec X_{\pi}}-{\vec v}_{\pi}(t-\text{T}_{\pi})\right)^2-i\left(E({\vec p}_{\pi})(t-\text{T}_{\pi}) - {\vec
 p}_{\pi}\cdot({\vec x}-{\vec X}_{\pi})\right)} ,
\label{pion-wf}\nonumber \\
& & \\
& &\langle \mu,{\vec p}_{\mu};\nu,{\vec p}_{\nu},{\vec X}_{\nu},\text{T}_{\nu}|\bar
 \mu(x) (1 - \gamma_5) \nu(x) |0 \rangle \nonumber\\
& &= \frac{N_{\nu}}{(2\pi)^{{3}}}({2\pi \over \sigma_{\nu}})^{3/2} e^{-{1
 \over 2 \sigma_{\nu}}({\vec x}-{\vec X}_{\nu}-{\vec
 v}_{\nu}(t-T_{\nu}))^2}
\left({m_{\mu}
 \over E({\vec p}_{\mu})}\right)^{\frac{1}{2}}\left({m_{\nu} \over E({\vec
 p}_{\nu})}\right)^{\frac{1}{2}}  \bar u({\vec p}_{\mu}) (1 -\gamma_5) \nu
 ({\vec p}_{\nu})\nonumber\\
 & &\times e^{i\left(E({\vec p}_{\mu})t-{\vec
 p}_{\mu}\cdot{\vec x}\right)+ i\left(E({\vec p}_{\nu})(t-\text{T}_{\nu})-{\vec
 p}_{\nu}\cdot({\vec x}-{\vec X}_{\nu})\right)},\label{mu-nu-wf}
\end{eqnarray}
where
\begin{align}
& N_{\pi} = \left(\frac{\sigma_{\pi}}{\pi}\right)^{\frac{3}{4}},~N_{\nu}
 = \left(\frac{\sigma_{\nu}}{\pi}\right)^{\frac{3}{4}},\rho_{\pi}=
\left(\frac{1}{2E_{\pi}(2\pi)^3}\right)^{\frac{1}{2}}.
\end{align}

In the above equation,  the pion's life time is ignored but that is easily 
introduced and its effect is included later.
 $\sigma_{\pi}$ and
$\sigma_{\nu}$,  in Eqs. $(\ref{pion-wf})$ and 
  $(\ref{mu-nu-wf})$   are sizes of the pion  and   
neutrino wave packets. Minimum wave packets are used in majorities  of the
present  paper but non-minimum wave packets are studied and it is shown
that  main  results are the same 
\footnote{For  non-minimal wave packets which  have larger uncertainties, 
Hermite  polynomials of ${\vec k}_{\nu}-{\vec p}_{\nu}$ are multiplied to
the right-hand side of
Eq. $(\ref{mu-nu-wf})$. A completeness of the wave packet states is
also satisfied for the
non-minimum case \cite{Ishikawa-Shimomura} and the  probabilities   
 are  the same as far as the wave packet is almost symmetric. This
 condition 
is guaranteed in the high energy neutrino which this paper studies, but 
may not be so in the low energy neutrino.  We will 
confirm in the text and 
appendix that the universal long-range correlation  is independent 
of  the wave packet shape as far as the wave packet is invariant under the time
inversions. Low energy neutrinos such as solar or reactor neutrinos will
be presented in the next paper.}.
 From the result of Appendix,
 the pion produced in proton nucleon collision can be regarded 
as a free particle.  The effect due to a mean free 
path   estimated in the Appendix    is
used as a size of wave packet. The size of pion wave packet is of the 
order of $0.5-1.0$\,[m] and  a momentum spreading is small. So ${\vec
k}_{\pi}$ is integrated easily, and is 
replaced with  its  central value ${\vec p}_\pi$ and the final 
expression  of Eq. $(\ref{pion-wf})$ is obtained. 
We use the  nuclear size for $\sigma_{\nu}$.

\subsection{Wave of observed neutrino: small angular velocity of a
  center motion }

We have 
 \begin{eqnarray}
& &\langle {\mu},{\nu}   |H_{int}(x)| \pi \rangle 
=igm_{\mu}\tilde N   e^{-{1 \over 2 \sigma_{\pi} }\left({\vec x}-{\vec X_{\pi}}-{\vec v}_{\pi}(t-\text{T}_{\pi})\right)^2
-iE({\vec p}_{\pi})(t-\text{T}_{\pi}) +i {\vec
 p}_{\pi}\cdot({\vec x}-{\vec X}_{\pi})+iE({\vec p}_{\mu})t-i{\vec
 p}_{\mu}\cdot{\vec x}}\nonumber\label{neutrino-position-amplitude}
\\
& &\times\bar u({\vec p}_{\mu}) (1 - \gamma_5) \nu ({\vec p}_{\nu})
e^{i\phi(x)}  e^{-{1 \over 2\sigma_{\nu} }({\vec x}-{\vec X}_{\nu} -{\vec
v}_{\nu}(t-\text{T}_{\nu}))^2}, \\
& &\tilde N=N_{\pi}
 N_{\nu}\left(\frac{2\pi}{\sigma_{\pi}}\right)^{\frac{3}{2}}\left(\frac{2\pi}{\sigma_{\nu}}\right)^{\frac{3}{2}}N_0,\ 
N_0=\rho_{\pi}{1
 \over (2\pi)^3}\left({m_{\mu} m_{\nu} \over E_{\mu}E_{\nu}}\right)^{1/2},\\
& &\phi(x)=E({\vec
 p}_{\nu})(t-\text{T}_{\nu})-{\vec  p}_{\nu}\!\cdot\!({\vec x}-{\vec X}_{\nu}).
\label{neutrino-phase} %\label{neutrino-position-amplitude}
\end{eqnarray}
where  ${\vec v}_{\nu}$ is a
neutrino velocity.
 It is important to notice that the integration over ${\vec k}_{\nu}$ is made prior to the integration over $(t,{\vec x})$  in order to satisfy the boundary 
condition of $S[\text T]$.

The neutrino wave function evolves with time in a specific manner. At
$t=\text{T}_{\nu}$, 
\begin{eqnarray}
\psi_{\nu}(\text{T}_{\nu},{\vec x})=e^{i\phi(x)-{1 \over 2\sigma_{\nu} }({\vec x}-{\vec X}_{\nu})^2},
\end{eqnarray}
which is localized around the position ${\vec X}_{\nu}$ and has the
phase $\phi=-{\vec p}_{\nu}\cdot({\vec x}-{\vec X}_{\nu})$. At a time $t <\text{T}_{\nu}$, 
\begin{eqnarray}
\psi_{\nu}(t,{\vec x})=e^{i\phi(x)-{1 \over 2\sigma_{\nu} }({\vec
 x}-{\vec X}_{\nu}-{\vec v}_{\nu}(t-\text{T}_{\nu}))^2},
\end{eqnarray}
which is localized around the position
\begin{eqnarray}
{\vec x}_G={\vec X}_{\nu}+{\vec v}_{\nu}(t-\text{T}_{\nu}),
\label{center-coordinate}
\end{eqnarray} 
and has the phase $\phi(x)$.  
  $\phi(x)$ is written at a  position ${\vec r}={\vec
x}-{\vec x}_G$ in the form, 
\begin{eqnarray}
\phi(x)=\bar \phi_G+\phi({\vec r}\,),
\end{eqnarray}
where 
\begin{eqnarray}
\bar \phi_G
\label{phase}
&=&E({\vec p}_{\nu})(t-\text{T}_{\nu})-{\vec p}_{\nu}\cdot{\vec
 v}_{\nu}(t-\text{T}_{\nu})\nonumber\\
&=&{E^{2}_{\nu}(\vec p_{\nu})-{\vec p}_{\nu}^{~2} \over E_{\nu}({\vec
 p}_{\nu})}(t-\text{T}_{\nu})= {m_{\nu}^2 \over E_{\nu}(\vec{p}_{\nu})}
 (t-\text{T}_{\nu}),\\
\phi({\vec r}\,)&=&-{\vec p}\cdot{\vec r}.
\end{eqnarray}
A  phase at the center,  $\bar \phi_G$, has a typical form of the relativistic
particle. Since the position is moving with the velocity ${\vec v}_{\nu}$, the time-dependent phase is almost 
cancelled with the space-dependent phase.

When the position is moving with the light velocity in the parallel
direction to the momentum ${\vec p}_{\nu}$, instead of Eq. ($\ref{center-coordinate}$), 
\begin{eqnarray}
{\vec x}={\vec X}_{\nu}+{\vec c}(t-\text{T}_{\nu}),{\vec c}={{\vec
 p}_{\nu} \over p_{\nu}},
\end{eqnarray}
the phase  is given by
\begin{eqnarray}
\bar \phi_c(t - \text{T}_\nu)=E({\vec p}_{\nu})(t-\text{T}_{\nu})-{\vec p}_{\nu}\!\cdot{\vec
 c}\,(t-\text{T}_{\nu})
=\frac{m_{\nu}^2}{2 E_{\nu}(\vec{p}_{\nu})} (t-\text{T}_{\nu}),\label{light-phase}
\end{eqnarray}
and becomes a half of $\bar{\phi_G}$. We will see that this phase plays the important role later.

The coordinate ${\vec r}$ is integrated in the amplitude
Eq. $(\ref{amplitude1})$, where   the rapidly changing phase 
$\phi({\vec r})$ is combined with those phases of the pion and muon fields,
and the slow phase $\bar{\phi_c}$  remains and gives a characteristic behavior 
to the transition amplitude.  The emergence
   of slow phase occurs   independently of   the detail of wave packet.

 The phase in the
   longitudinal direction is not affected by        
 a  spreading of the wave packet, and    does not change the behavior of the amplitude. 
 So the spreading effect   has been  ignored for 
simplicity in this section  and will be studied in the latter section and
   Appendix. It will be shown there that the spreading  in the transverse 
direction modifies the ${\vec k}_{\nu}$ integration but the final result 
turns actually into the same.

A  neutrino velocity  is slightly smaller than  the speed of 
light. A
 neutrino of energy  $1~[\text{GeV}]$ and  a mass 
$1~[\text{eV}/c^2]$    has a velocity
\begin{eqnarray}
v/c=1-2\epsilon, \ \epsilon=\left({m_{\nu}c^2 \over E_{\nu}}\right)^2=5\times 10^{-19},
\end{eqnarray} 
hence the neutrino propagates a  distance $l$, where  
\begin{eqnarray}
l=l_0(1-\epsilon)=l_0-\delta l, \ \delta l= \epsilon l_0,
\end{eqnarray}
while  the light propagates  a distance $l_0$. This difference of distances,
$\delta l$, becomes
\begin{eqnarray}& &\delta l=5\times 10^{-17}~[\text{m}]; ~l_0=100~[\text{m}], \\ 
\label{neutrino-ovelapp}
& &\delta l=5\times 10^{-16}~[\text{m}]; ~l_0=1000~[\text{m}] ,
\end{eqnarray}
which are much smaller than the sizes of the neutrino  wave packets.
 %and $(\ref{mfp-muon})$.
Since the  difference of velocities is small,  the neutrino amplitude 
at the nuclear  or atom
targets  show interference. The geometry of the neutrino interference 
is shown in
Fig.~\ref{fig:geo}. The neutrino wave produced at a time $t_1$ arrives
at one nucleus or atom in the detector and is superposed  to the wave
produced  at $t_2$ and arrives to the same    nucleus or atom same
time. A constructive interference of waves is shown in the text.

%\newpage
%%%%%%%%%%%%%%%%%%%%%%%%%%%%%%%%%%%%%%%% sub section %%%%%%%%%%%%%%%%%%%%%%%%%%%%%%%%%%%%%%%%%%%%%%%%%%%%%%%%%%
\section{Position-dependent  probability}
 
  The  probability of the event that the neutrino is 
detected  at a finite-distance is computed   and its deviation from the 
asymptotic value, the finite-size correction, is found. The correction
 has   a universal
property unique to  the relativistically invariant system and is determined by 
the absolute neutrino mass.

\subsection{Transition probability }
The  case of  $\sigma_{\pi}=\infty$  is studied  first, and 
that  of large $\sigma_{\pi}$ is later.

In Eq. $(\ref{neutrino-position-amplitude})$, the integrand is a Gaussian
function around the center ${\vec x}_0(t)={\vec
v}_{\nu}(t-\text{T}_{\nu})+{\vec X}_{\nu}$ and is invariant under 
\begin{eqnarray}
& &{\vec x} \rightarrow {\vec x}+{\vec v}_{\nu} \delta t, \\
& &t \rightarrow t+\delta t.
\end{eqnarray}
Thus  a shifted energy given by
\begin{eqnarray}
H_0-{\vec v}_{\nu}\cdot{\vec{\mathcal{P}}},
\end{eqnarray} 
 satisfies
\begin{eqnarray}
\left[S,H_0-{\vec v}_{\nu}\cdot{\vec{\mathcal{P}}}\right]=0,
\end{eqnarray}
and is conserved.

 Integrating over ${\vec x}$ in
 Eq. $(\ref{neutrino-position-amplitude})$, we have the amplitude,
\begin{align}
&\mathcal{M}=Ce^{i\phi_0}\bar u(p_{\mu})(1-\gamma_5)u(p_{\nu})  e^{-\frac{\sigma_{\nu}
 }{2}{\delta {\vec p}}^{\,2}} e^{-i\omega \text T/2}2 \frac{\sin (\omega
 \text T/2)}{\omega} 
\label{integrated-amplitude-honbun}
,\\
&\omega=\delta E-\vec{v}_\nu\cdot\delta {\vec p},\,
\delta{\vec p}={\vec p}_{\pi}-{\vec p}_{\mu}-{\vec p}_{\nu},\, \delta E=E({\vec p}_{\pi})-E({\vec p}_{\mu})-E({\vec p}_{\nu}), \nonumber
\end{align}
where $\phi_0$ is a constant. Because the modulus of wave function does
not vanish in the finite space-time region around the moving 
center ${\vec x}_0(t)$  of
       velocity ${\vec
v}_{\nu}$, the angular velocity in 
Eq. $(\ref{integrated-amplitude-honbun})$, $\omega=E-{\vec v}_{\nu} \cdot \delta{\vec p}$,
is different from the energy difference $\delta E$ of the rest system.
 This causes the unusual properties of       transition      amplitude.

In Eq. $(\ref{integrated-amplitude-honbun})$,  due to the Gaussian 
factor $ e^{-\frac{\sigma_{\nu}
 }{2}{\delta {\vec p}}^{\,2}}$,   $|\delta {\vec p}\,|$ has a finite
 uncertainty. Hence  $\omega $ generally deviates from $\delta E$. At 
$\delta {\vec
p}=0$,  $\omega=0$ is the same as                $\delta E=0$, 
 whereas at $\delta {\vec
p} \neq 0$, $\omega=0$  gives the relation  
$\delta E={\vec
 v}_{\nu} \cdot\delta {\vec p}\neq0$. Kinetic energy takes broad range 
and the amplitude  at a finite T reflects this.  
%As was  shown in \cite{ishikawa-tobita}, 
A configuration of the momentum satisfying $\omega=0$ is a large
  ellipse,   on which  
 the normal solution of $\delta {\vec p}=0$, and the
new  solution of large $|\delta {\vec p}\,|$ are. $\omega$ varies
 rapidly with the change of momentum   around the former solution  and  $2\frac{\sin {(\omega \text{T}/2)}}
 {\omega}=2\pi\delta(\omega)$ \cite{Schiff and Landau} can be applied. This
 gives the asymptotic  term which satisfies the
 energy-momentum conservation. On the other hand, $\omega$ varies
 extremely slowly  around the latter
 momentum, and the states of $\omega \approx 0$ lead the slow convergence
 at large T  and give the finite-size correction.  Since $|\delta {\vec
 p}\,|$ and $\delta E$ are not small, the
 spectrum at the ultraviolet region, which exists  in the wave function 
at a finite time, gives a contribution to  the 
finite-size correction. 

For computing  the probability in a consistent manner with Lorentz
invariance, 
it  is convenient to write  $|\mathcal{M}|^2$ with 
a correlation function.   
   A transition probability   
of a pion of  a momentum ${\vec
p}_{\pi}$ located at a space-time position $(\text{T}_{\pi},{\vec
X}_{\pi})$, decaying to the  neutrino of the  momentum ${\vec p}_{\nu}$ at  
a space-time position $(\text{T}_{\nu},{\vec X}_{\nu})$ and a muon of momentum
${\vec p}_{\mu}$,
is expressed in the form 
\begin{eqnarray}
|\mathcal{M}|^2 &=& g^2 m_{\mu}^2 
|\tilde N|^2 \int d^4x_1 d^4x_2
S_{5}(s_1,s_2)
\nonumber\\
 &\times& e^{i( \phi(x_1) -\phi(x_2))}  
e^{-{1 \over 2\sigma_{\nu} }\sum_i \left({\vec x}_i-{\vec X}_{\nu} -{\vec
			     v}_{\nu}(t_i - \text{T}_{\nu})\right)^2}
\nonumber \\
&\times& e^{-i\left(E({\vec p}_{\pi})(t_1 - t_2)-{\vec p}_{\pi}\cdot({\vec x}_1-{\vec
x}_2)\right)}
\times e^{i\left(E({\vec p}_{\mu})(t_1-t_2)-{\vec p}_{\mu}\cdot({\vec
	     x}_1-{\vec x}_2\right))}
\nonumber \\
&\times&e^{-{1 \over 2 \sigma_{\pi} }\sum_j \left({\vec x}_j-{\vec X_{\pi}}-{\vec v}_{\pi}(t_j-\text{T}_{\pi})\right)^2},
\label{probability}
\end{eqnarray}
where $S_{5}(s_1,s_2)$ stands for  products of Dirac
spinors and their  complex conjugates,   
\begin{eqnarray}
S_{5}(s_1,s_2)=\left(\bar u({\vec p}_{\mu})
 (1 - \gamma_5) \nu ({{\vec p}_{\nu}})\right)\left(\bar u({\vec p}_{\mu})
 (1 -  \gamma_5) \nu ({{\vec p}_{\nu}})\right)^{*},
\label{spinor-1}
\end{eqnarray}
and its spin summation is  given by
\begin{eqnarray}
S^{5}&=&\sum_{s_1,s_2}S^{5}(s_1,s_2)
=\frac{2p_{\mu}\!\cdot\! p_{\nu}}{m_{\nu}m_{\mu}}.\label{spinor-2}
\end{eqnarray}
Now the
probability is finite and an order of integrations are interchangeable. 
Integrating  momenta of the final state 
and taking average over the initial momentum, we have the total probability 
 in the form 
 \begin{eqnarray}
& &\int d{\vec p}_{\pi}\rho_{exp}({\vec p}_{\pi}) {d{\vec X}_{\nu} \over (2\pi)^3}d{\vec
 p}_{\mu}d{\vec p}_{\nu}  \sum_{s_1,s_2}|\mathcal{M}|^2 
\nonumber\\
& &= g^2 m_{\mu}^2 |N_{\pi\nu}|^2\frac{2}{(2\pi)^3}\int {d{\vec X}_{\nu}
  \over (2\pi)^3}d{\vec p}_{\nu} \rho_{\nu}^2 d^4x_1 d^4x_2
e^{-{1 \over 2\sigma_{\nu} }\sum_i\left({\vec x}_i-{\vec X}_{\nu} -{\vec
v}_\nu(t_i-\text{T}_{\nu})\right)^2}
\nonumber \\
&  &\times 
\Delta_{\pi,\mu}(\delta t,\delta {\vec x})
e^{i \phi(\delta x_{\mu})} 
e^{-{1 \over 2 \sigma_{\pi} }\sum_j \left({\vec x}_j-{\vec X_{\pi}}-\bar{\vec
 v}_{\pi}(t_j- \text{T}_{\pi})\right)^2} ,
 \label{probability-correlation1}\\
& &
N_{\pi\nu} =
\left(\frac{4\pi}{\sigma_{\pi}}\right)^{\frac{3}{4}}\left(\frac{4\pi}{\sigma_{\nu}}\right)^{\frac{3}{4}},~\rho_{\nu}=\left(\frac{1}{2E_{\nu} (2\pi)^3 }\right)^{\frac{1}{2}},~\delta  x= x_1-x_2,
\end{eqnarray}
with a  correlation function $\Delta_{\pi,\mu}(\delta t,\delta {\vec x})$. The correlation
function is defined with   a pion's momentum distribution $\rho_{exp}({\vec p}_{\pi})$, by
\begin{align}
\Delta_{\pi,\mu} (\delta t,\delta {\vec x})=
 {\frac{1}{(2\pi)^3}}\int
{d {\vec p}_{\pi} \over E({\vec p}_{\pi})}\rho_{exp}({\vec p}_{\pi})
{d {\vec p}_{\mu} \over E({\vec p}_{\mu})}  (p_{\mu}\!\cdot\! p_{\nu})
 e^{-i\left(\{E({\vec
 p}_{\pi})-E({\vec p}_{\mu})\}\delta t-({\vec p}_{\pi}-{\vec
 p}_{\mu})\cdot \delta {\vec x})\right)}.%  \nonumber \\
\label{pi-mucorrelation1}
%& &\delta t=t^1-t^2,\delta {\vec x}={\vec x}^1-{\vec x}^2.\nonumber
\end{align} 

 In the above equation, the final states are integrated over a
 complete set \cite{Ishikawa-Shimomura}.   The muon and neutrino momenta are integrated over  
entire positive energy  regions, and the neutrino position  is 
integrated over the region of the detector. The pion in the initial
 state is assumed to be the statistical ensemble of the 
distribution $\rho_{exp}({\vec p}_{\pi})$.
If the momentum distribution is narrow around the central value, the velocity ${\vec v}_{\pi} $ in the pion Gaussian factor was
replaced with its average $\bar {\vec v}_{\pi}$. This is verified from
the large spatial size of the pion wave packet discussed in the 
previous section. 
For    the probability of  a fixed pion momentum, the correlation function 
 \begin{align}
\tilde \Delta_{\pi,\mu} (\delta t,\delta {\vec x})=
 {\frac{1}{(2\pi)^3}}
{1 \over E({\vec p}_{\pi})}
\int {d {\vec p}_{\mu} \over E({\vec p}_{\mu})}  (p_{\mu}\!\cdot\! p_{\nu})
 e^{-i\left(\{E({\vec
 p}_{\pi})-E({\vec p}_{\mu})\}\delta t-({\vec p}_{\pi}-{\vec
 p}_{\mu})\cdot \delta {\vec x}\right)}
\label{pi-mucorrelation2}
\end{align}
is used instead of Eq. $( \ref{pi-mucorrelation1})$.    
\subsection{Light-cone singularity   }

The correlation function $\tilde \Delta_{\pi,\mu}(\delta t,\delta {\vec
x})$ is a standard form of Green's function and has the light-cone 
singularity that is real and decreases very slowly
along the light cone. The singularity  is generated by the states
at the ultraviolet energy region  near 
the light-cone region
\begin{eqnarray}
\lambda=\delta t^2-{\left|\delta\vec x\right|}^2 = 0,
\end{eqnarray}
and  is extended in a large  $|\delta {\vec x}|$  independently 
of ${\vec p}_{\pi}$. Thus  the probability Eq. $(\ref{probability-correlation1})$   gets a finite $\text T(=\text{T}_{\nu}-\text{T}_{\pi})$ 
correction  from the
integration over $t_1$ and $t_2$ at  $|t_1-t_2| \rightarrow
\text{T}  $. 
We find  that the
light-cone singularity of    $\tilde \Delta_{\pi,\mu}
(\delta t,\delta {\vec x})$ \cite{Wilson-OPE}  gives a large finite-size
correction to the probability in the following.

%%%%%%%%%%%%%%%%%%%%%%%%%%%%%%%%%%%%%%%%%%%% Sub section Probability %%%%%%%%%%%%%%%%%%%%%%%%%%%%%%%%%%%%%%%%%%%%%%
%%%%%%%%%%%%%%%%%%%%%%%%%%%%%%%%%%%%%%%%%%%%%%%%%%%%%%%%%%%%%%%%%%%%%%%%%%%%%%%%%%%%%%%%%%%%%%%%%%%%%%%%%%%%%%%%%%%
\subsubsection{Separation of singularity  }

For the particles  expressed by plane waves,  the integration over the
space time is made  over the infinite-time interval, and  the  
kinetic-energy is strictly conserved  and 4-dimensional  momenta satisfy  
\begin{eqnarray}
p_{\pi}=p_{\mu}+p_{\nu},\ 
(p_{\pi}-p_{\mu})^2=m_{\nu}^2 \approx 0.
\label{lightlike}
\end{eqnarray}
Conversely, $\tilde{\Delta}_{\pi,\mu} (\delta
t,\delta {\vec x})$ becomes, from an  integral over the momentum  in the 
region where   the 
momentum difference $p_{\pi}-p_{\mu}$ is almost
light-like,  to have  a singularity around  the light cone,
$\lambda=0$. 
%The light-cone singularity  gives a large
%contribution to  Eq. $(\ref{probability-correlation1})$. 
 In order to extract the singular term from  $\tilde \Delta_{\pi,\mu} (\delta
t,\delta {\vec x})$, we write  the integral in a four-dimensional form   
\begin{align}
&\tilde \Delta_{\pi,\mu} (\delta t,\delta {\vec x})=
 {\frac{1}{(2\pi)^3}} {1 \over E({\vec
 p}_{\pi})}I(p_{\pi},\delta x),\\
&I(p_{\pi},\delta x)={2 \over \pi} \int d^{4}p_{\mu} \, \theta(p_{\mu}^0)
(p_{\mu}\!\cdot\! p_{\nu}) \text {Im}\left[1 \over p_{\mu}^2-m_{\mu}^2-i\epsilon\right]
 e^{-i\left(\{E({\vec
 p}_{\pi})-E({\vec p}_{\mu})\}\delta t-({\vec p}_{\pi}-{\vec
 p}_{\mu})\cdot \delta {\vec x}\right)},
\end{align}
first, and change the integration variable   from $p_{\mu}$ to 
$q=p_{\mu}-p_{\pi}$ that is conjugate to $\delta x$. Next,  we separate the
integration region into two parts,  $0 \leq q^0$ and $-p_{\pi}^0 \leq
q^0 \leq 0$, 
 and have the expressions,     
\begin{align}
&I(p_{\pi},\delta x)=I_1(p_{\pi},\delta x)+I_2(p_{\pi},\delta x), 
\label{seperation-region}\\
&I_1(p_{\pi},\delta x)=\left\{p_{\pi}\! \cdot\! p_{\nu}+p_{\nu}\!\cdot\! \left(-i{\partial \over
 \partial \delta x}\right)\right\} \tilde I_1,\\
&\tilde I_1={2 \over \pi}\int d^4 q \,  \theta(q^0)\text {Im}\left[1 \over
 (q+p_{\pi})^2-m_{\mu}^2-i\epsilon\right] e^{iq \cdot \delta x },\\
& I_2(p_{\pi},\delta x)= {2 \over \pi} \int_{-p_{\pi}^0}^{0}d^4 q\, p_{\nu}\!\cdot\! (p_{\pi}+q)\text {Im} \left[1 \over
 (q+p_{\pi})^2-m_{\mu}^2-i\epsilon\right] e^{iq \cdot \delta x }.
\end{align}
$I_1(p_{\pi},\delta x)$ is the integral over the infinite region 
 and has the light-cone singularity and
$I_2(p_{\pi},\delta x)$ is the integral over  the finite region  and 
is regular. 

  $I_1(p_{\pi},\delta x)$ comes from the states of  non-conserving  kinetic
  energy and does not contribute to the total probability at an 
infinite-time interval.  $I_2(p_{\pi},\delta x)$, on the other hand, 
contributes to that at the infinite-time and finite-time intervals.   So
  the leading finite-size correction to a physical 
quantity  is computed using the 
most singular term of $I_1$.   

Next we compute  $\tilde I_1$.  Expanding the integrand with 
$p_{\pi}\!\cdot\! q$, we have  $\tilde I_1$  in the form 
\begin{align}
&\tilde I_1(p_{\pi},\delta x)\nonumber\\
=&{2 \over \pi}\int d^4 q \, \theta(q^0)~ \text {Im}\left[{1 \over
q^2+m_{\pi}^2-m_{\mu}^2+2q\!\cdot\! p_{\pi}-i\epsilon}\right] e^{iq
\cdot \delta x } \nonumber\\
=&{2 \over \pi}\int d^4 q  \,\theta(q^0)\left\{1+2p_{\pi}\!\cdot\! \left(i{\partial \over \partial \delta
 x}\right) {\partial  \over \partial {\tilde m}^2}+\cdots \right\}\,\text {Im}\left[ {1 \over
 q^2+{\tilde m}^2-i{\epsilon}} \right]e^{iq \cdot\delta x } \nonumber\\
=&2  \left\{1 +2p_{\pi} \!\cdot\!\left(i{\partial \over \partial \delta
		  x}\right) {\partial  \over \partial {\tilde m}^2}+\cdots \right\}
\int d^4 q \, \theta(q^0)\delta (q^2+{\tilde m}^2) e^{iq \cdot\delta x }, \label{singular-function} 
\end{align}
where 
\begin{eqnarray}
{\tilde m}^2=m_{\pi}^2-m_{\mu}^2.
\end{eqnarray}
The expansion in $2p_\pi\!\cdot\! q$ of 
Eq. $(\ref{singular-function})$  converges  in the region
\begin{eqnarray}
{2p_{\pi}\!\cdot\! q \over q^2+{\tilde m}^2} < 1.
\end{eqnarray}
Here $q$ is the integration variable and varies. So we  evaluate  
the series after the integration over $q$, and find a condition for its  
convergence. 
 We  find later that the series after the momentum integration 
converges in the region
${2p_{\pi}\cdot p_{\nu} \over {\tilde m}^2} \leq 1$.

     $\tilde I_1(p_{\pi},\delta
x)$  is written in the form 
\begin{align}
\tilde I_1(p_{\pi},\delta x)
=  &2(2\pi)^3i\left\{1 +2p_{\pi} \!\cdot\!\left(i{\partial \over \partial
 \delta x}\right) {\partial  \over
 \partial {\tilde m}^2}+\cdots \right\}
%\nonumber\\
%&\times
\left( {1 \over 4\pi}\delta(\lambda)\epsilon(\delta
 t)+f_{short}\right),
\end{align}
where $f_{short}$ is written by Bessel functions and a formula for a
relativistic field     Eq. $(\ref{singular-function-f-p})$  is used.

Next  $I_2$ is evaluated. For $I_2$, we use a momentum $\tilde
q=q+p_{\pi}$ and write in the form
\begin{align}
I_2(p_{\pi},\delta x)=
&\frac{2}{\pi} \int_{0< \tilde
 q^0<p_{\pi}^0} d^4 \tilde q \, (p_{\nu} \!\cdot\!\tilde q)  \text {Im}\left[\frac{1}
 {\tilde q^2-m_{\mu}^2-i\epsilon}\right] e^{i(\tilde q-p_{\pi}) \cdot\delta x } 
\nonumber \\
=&e^{-ip_{\pi}\cdot\delta x }\left\{p_{\nu} \!\cdot\!\left(-i{\partial \over \partial
 \delta x}\right)\right\}  {2 \over \pi}\int_{0< \tilde q^0
 <p_{\pi}^0 } d^4 \tilde{q}\,  \pi \delta( q^2-m_{\mu}^2)  e^{i\tilde q \cdot\delta x } 
\nonumber \\
=& e^{-i p_{\pi} \cdot\delta x}\left\{p_{\nu} \!\cdot\!\left(-i{\partial \over \partial
 \delta x}\right)\right\} \int\frac{ d\vec{q}}{
  \sqrt{\vec{q}^{\,2}+m_{\mu}^2}} \theta\left(p_{\pi}^0-\sqrt{\vec{q}^{\,2}+m_{\mu}^2} \right)
 e^{iq \cdot\delta x }.\label{normal-term} 
\end{align}
The regular part $I_2$ has no singularity because the integration domain is
finite and becomes short-range.  

Thus the first term in $\tilde I_1$ gives the most singular 
term  and the rests, the second
term in $I_1$ and $I_2$, give regular terms.  
The correlation function, $\tilde \Delta_{\pi,\mu}(\delta t,\delta {\vec x})$ 
 is written in the form 
\begin{align}
&\tilde \Delta_{\pi,\mu}(\delta t,\delta {\vec x})={1 \over (2\pi)^3} {1 \over E(p_{\pi})}\Biggl[\left\{p_{\pi}\! \cdot\! p_{\nu}-p_{\nu}\!\cdot\!\left(i\frac{\partial}{
 \partial \delta x}\right)\right\}2(2\pi)^3i\nonumber\\ 
&\times\left\{1 +2p_{\pi} \!\cdot\! \left(i{\partial \over \partial
 \delta x}\right) {\partial  \over
 \partial {\tilde m}^2}+\cdots \right\} \left( \frac{1}{4\pi}\delta(\lambda)  
\epsilon(\delta t)+f_{short}\right) + I_2 \Biggr],\label{muon-correlation-total}
\end{align}
where the dots stand for the higher order terms.

\subsection{Integration over spatial coordinates   }
Next, we integrate over the coordinates ${\vec x}_1$ and ${\vec x}_2$  in
\begin{eqnarray}
\int d{\vec x}_1 d{\vec x}_2e^{i\phi(\delta
 x)}e^{-\frac{1}{2\sigma_{\nu} } \sum_i
\left({\vec x}_i-{\vec X}_{\nu} -{\vec
v}_\nu(t_i-\text{T}_{\nu})\right)^2}
\tilde \Delta_{\pi,\mu}(\delta t,\delta{\vec x}).\label{lightcone-integration1}
\end{eqnarray}

\subsubsection{Singular terms: long-range correlation}
The most singular term of $\tilde \Delta_{\pi,\mu}(\delta t,\delta{\vec x})$ 
is substituted, then   Eq. $(\ref{lightcone-integration1})$ becomes    
\begin{eqnarray}
\label{singular-correlation}
J_{\delta(\lambda)}&=&\int d{\vec x}_1 d{\vec x}_2e^{i\phi(\delta
x)}e^{-{1 \over 2\sigma_{\nu} }\sum_i \left({\vec x}_i-{\vec X}_{\nu} -{\vec
v}_\nu(t_i-\text{T}_{\nu})\right)^2} \frac{\epsilon(\delta t)}{4 \pi}\delta(\lambda) ,
\end{eqnarray}
and is computed easily  using a center coordinate $R^\mu=\frac{
x_1^\mu+x_2^\mu}{2}$ and a relative coordinate
$\vec{r}=\vec{x}_1-\vec{x}_2$.
After the center coordinate ${\vec R}$ is integrated,
$J_{\delta(\lambda)}$ 
becomes the integral of the   transverse and longitudinal component $({\vec
r}_T,r_l)$ of the relative coordinates,   
\begin{eqnarray}
\epsilon(\delta t) (\sigma_{\nu}\pi)^{\frac{3}{2}} \int d{\vec r}_Td r_l \, e^{i\phi(\delta t,{\vec
 r})-\frac{1}{4\sigma_{\nu} }({{\vec r}_T}^{\,2}  +(r_l-{
 v}_{\nu}\delta t)^2)}\frac{1}{4\pi}\delta (\delta t^2-{{\vec r}_T}^{\,2} -{{r}_l}^{2}).
\label{lightcone-integration-s}
\end{eqnarray}
The transverse coordinate ${\vec r}_T$ is integrated using the Dirac
delta function and $r_l$ is integrated next.
Finally we have  
\begin{align}
J_{\delta(\lambda)}&={(\sigma_{\nu}\pi)}^{\frac{3}{2}}
 \frac{\sigma_{\nu}}{2}\frac{\epsilon(\delta t)}{|\delta t|
 }e^{i\bar \phi_c(\delta t)-\frac{m_{\nu}^4}{
 16\sigma_{\nu} E_{\nu}^4} {\delta t}^2}\nonumber\\
 &\approx {(\sigma_{\nu}\pi)}^{\frac{3}{2}} \frac{\sigma_{\nu}}{2}
  \frac{\epsilon(\delta t)}{|\delta t|
  }e^{i\bar \phi_c(\delta t)}\label{lightcone-integration2-2}. 
\end{align}

The next term  of $\tilde \Delta_{\pi,\mu}(\delta t,\delta{\vec x})$, of
the form ${1}/{\lambda}$, in Eq. $(\ref{lightcone-integration1})$
leads       
\begin{align}
J_{1/\lambda}=&\int d{\vec x}_1 d{\vec x}_2e^{i\phi(\delta x)}e^{-{1
 \over 2\sigma_{\nu} }\sum_i \left({\vec x}_1-{\vec X}_{\nu} -{\vec
v}_\nu(t_1-\text{T}_{\nu})\right)^2}
{i  \over 4\pi^2 \lambda},
\label{lightcone-integration4}
\end{align}
which becomes 
\begin{align} 
J_{1/\lambda}\approx& {(\sigma_{\nu}\pi)}^{\frac{3}{2}} \frac{\sigma_{\nu}}{2} \left(\frac{1}{
 \pi \sigma_{\nu} |\vec{p}_{\nu}|^2}\right)^{\frac{1}{2}} e^{-\sigma_{\nu}|\vec{p}_{\nu}|^2}
\frac{ 1}{|\delta t| }e^{i\bar
\phi_c(\delta t)}.
\label{lightcone-integration4-2} 
\end{align} 
This term also has the universal $|\delta t|$ dependence but its magnitude is much
smaller than that of $J_{\delta(\lambda)}$ and is negligible in the present decay mode.  

From Eqs. $(\ref{lightcone-integration2-2})$
and $(\ref{lightcone-integration4-2})$, the singular terms
$J_{\delta(\lambda)}$ and $J_{1/\lambda}$ have the slow  phase
$\bar \phi_c(\delta t)$ and the magnitudes that are inversely proportional
to $\delta t$. Thus these terms are long-range with the small
angular velocity and are
insensitive to the ${\tilde m}^2$.  These properties of the
time-dependent correlation functions $J_{\delta(\lambda)}$ 
 hold
for  the general wave packets,
and the following theorem is proved.

{\bf Theorem}

The singular part $J_{\delta(\lambda)}$ of the correlation function has
the slow  phase that is determined with  the absolute value of the neutrino
mass
%$\bar \phi(\delta t)$ and the small correction 
and the 
magnitude inversely proportional
to $\delta t$,  of the form 
Eq. $(\ref{lightcone-integration2-2})$, at the large distance. The phase is given in the form
of a sum of $\bar \phi_c(\delta t)$ and small corrections, which are 
inversely proportional to the neutrino energy in general systems 
and become $1/E^2$ if the neutrino wave
 packet is invariant under the time inversion.  

{\bf (Proof: General cases including spreading of wave packet)}

We prove the theorem for general wave packets. 
$J_{\delta(\lambda)}$  is written in the form,
\begin{align}
\label{singular-correlation-centerG}
J_{\delta(\lambda)}=\int  d{\vec r}\, e^{i\phi(\delta x)}
\tilde w \left({\vec r} -{\vec
v}_\nu\delta t\right)  \frac{\epsilon(\delta t)}
{4 \pi}\delta(\lambda),  
\end{align}
where $\tilde w({\vec x}-{\vec v}t)$ is expressed with  a  wave packet in the
coordinate representation $w({\vec x}-{\vec v}t)$ and its complex
conjugate as,
\begin{align}
\tilde w(r_l-v_{\nu}\delta t,{\vec r}_T)&=\int d{\vec R} w\left({\vec R}+\frac{\vec r}{2}\right)w^{*}\left({\vec R}-\frac{\vec r}{2}\right) \nonumber\\
&=\int dk_l d{\vec k}_T e^{ik_l(r_l-v_{\nu}\delta t)+i{\vec k}_T\cdot{\vec
 r}_T+ic_0({\vec k}_T^2)\delta t} |w(k_l,{\vec k}_T)|^2.
\end{align}
The wave function $w({\vec x}-{\vec v}t)$ that includes 
the spreading effect is expressed in the following form 
\begin{eqnarray}
& &w({\vec x}-{\vec v}t)=\int dk_l d{\vec k}_T \,e^{ik_l(x_l-v_{\nu}t)+i{\vec k}_T\cdot{\vec
 x}_T+iC_{ij}k_T^ik_T^jt} w(k_l,{\vec k}_T),\\
& &C_{ij}=C_0 \delta_{ij},~C_0={1 \over 2E},
\end{eqnarray}
instead of the Gaussian function of Eq. $(\ref{singular-correlation})$.
A  quadratic form in ${\vec
k}$ in an expansion of $E\left({\vec p}+{\vec k}\right)$ is included
and this makes the wave packet spread with time. The
coefficient $C_{ij}$ in the longitudinal direction is negligible for 
the neutrino and is neglected. Expanding the delta function in the form,
\begin{eqnarray}
\delta({\delta t}^2-r_l^2-{\vec r}_T^{\,2})=\sum_l{1 \over l!}(-{\vec r}_T^{\,2})^l\left({\partial \over
 \partial {\delta t}^2}\right)^l \delta (t^2-r_l^2),
\end{eqnarray}
we have  the correlation function  
 \begin{align}
&J_{\delta(\lambda)}=\int dr_l d{\vec r}_T e^{i\phi(\delta t,r_l)}
\tilde w( r_l -v_\nu\delta t,{\vec r}_T){1 \over
4 \pi} \left\{1+\sum_{n=1} {1 \over n!}(-{{\vec r}_T}^{\,2})^n \left(\frac{\partial}
{\partial (\delta t )^2}\right)^n\right\}\nonumber\\
&\  \times \delta(\delta t^2-r_l^2) 
\epsilon(\delta t) \nonumber\\
&=  \int dr_l  d{\vec r}_T dk_l d{\vec k}_T e^{i\phi(\delta
  t,r_l)+ik_l(r_l-v_{\nu} \delta t)+i{\vec k}_T\cdot{\vec r}_T+iC_0{\vec k}_T^2 \delta
  t}|w(k_l,{\vec k}_T)|^2 \nonumber \\
&\ \times {1 \over
4 \pi} \left\{1+\sum_{n=1} {1 \over n!}(-{{\vec r}_T}^{\,2})^n \left(\frac{\partial}
{\partial (\delta t )^2}\right)^n\right\}  \delta(\delta t^2-r_l^2) 
\epsilon(\delta t)
\nonumber \\
&=  \int dr_l dk_l  e^{i\phi(\delta
  t,r_l)+ik_l(r_l-v_{\nu} \delta t)} d{\vec r}_T  d{\vec k}_T
e^{+i{C_0{\vec k}_T^2 \delta
  t}}|w(k_l,{\vec k}_T)|^2 \nonumber \\
&\ \times{1 \over 4 \pi} \left\{1+\sum_{n=1} {1 \over n!}\left({\partial^2 \over
		    (\partial {\vec k}_T)^{\,2}}\right)^n \left(\frac{\partial}
{\partial (\delta t )^2}\right)^n\right\} e^{i{\vec k}_T\cdot{\vec r}_T}  \delta(\delta t^2-r_l^2) 
\epsilon(\delta t).\label{singular-correlation-centerG4}
\end{align}
The variable ${\vec r}_T$ is  integrated first and ${\vec k}_T$ is integrated
next. Then we have the expression 
\begin{align}
J_{\delta(\lambda)}&=  \int dr_l dk_l  e^{i{\phi}(\delta
  t,r_l)+ik_l(r_l-v_{\nu} \delta t)}  
|w(k_l,0)|^2 \nonumber \\
&\times  {1 \over 4 \pi} \left\{1+\sum_{n=1} {1 \over n!}(-2iC_0 \delta t)^n \left(\frac{\partial}
 {2 \delta t \partial \delta t }\right)^n\right\} (2\pi)^2 \delta(\delta t^2-r_l^2) 
\epsilon(\delta t).
\end{align}
Using the following identity 
\begin{eqnarray}
(2\delta t)^n\left({\partial \over 2\delta t \partial \delta t}\right)^n=\left({\partial
 \over \partial \delta t}\right)^n+O\left({1 \over \delta t}\right) \left(\frac{\partial}{\partial \delta t}\right)^{n-1},
\end{eqnarray}
and taking a leading term in ${1/\delta t}$, we have the final
expression of the correlation function at the long-distance region
\begin{eqnarray}
& &
\label{singular-correlation-centerF}
J_{\delta(\lambda)}= \pi e^{-C_0 p } \epsilon(\delta t){e^{i\bar{\phi}_c(\delta
  t)} \over 2\delta t} \int  dk_l  e^{k_l(i(1-v_{\nu}) \delta t +C_0)}  
|w(k_l,0)|^2.   
\end{eqnarray}

Hence $J_{\delta (\lambda)}$ in Eq. $(\ref{singular-correlation-centerF}  )$
becomes  almost the same form as 
Eq. $(\ref{lightcone-integration2-2})$ and the slow phase
$\bar \phi_c(\delta t)$ is modified slightly and the magnitude that is 
inversely proportional
to the time difference. $J_{\delta(\lambda)}$ has the universal form for
the general wave packets. By expanding the exponential factor and taking
the quadratic term of the exponent, the above   integral is written in 
the form 
\begin{align}
&\int  dk_l  (1 +{k_l(i(1-v_{\nu}) \delta
  t +C_0)}+{1 \over 2!}{(k_l(i(1-v_{\nu}) \delta
  t +C_0) )^2})  |w(k_l,0)|^2   \nonumber\\
&=w_0\left(1+C_0d_1+{d_2 \over
 2!}C_0^2-(1-v_{\nu})^2{\delta t}^2\right)+i(d_1(1-v_{\nu})\delta t
 +d_2C_0(1-v_{\nu}) \delta t ),\label{singular-correlation-centerFc}
\end{align}
where 
\begin{eqnarray}
& &\delta={d_1 \over E}+\frac{d_2}{2}{1 \over E^2},~
\gamma={d_1 \over 2E}+{d_2\over
 2!}\left({1 \over 2E}\right)^2-(1-v_{\nu})^2{\delta t}^2,\\
& &d_1= \frac{1}{w_0}\int d k_l k_l |w(k_l,0)|^2,~d_2= \frac{1}{w_0}\int d k_l
 k_l^2|w(k_l,0)|^2.
\end{eqnarray}
We substitute this expression into the correlation function and   have    
\begin{align}
J_{\delta(\lambda)}=\pi e^{-C_0 |\vec{p}| }\omega_0(1+\gamma) \epsilon(\delta t){e^{i\bar{\phi}_c(\delta
  t)(1+\delta)} \over 2\delta t}
, \ w_0= \int d k_l |w(k_l,0)|^2.
\end{align}

In  wave packets of time reversal invariance, $|w(k_l,0)|^2$ is the
even function of $k_l$. Hence $d_1$ vanishes 
\begin{eqnarray}
d_1=0,
\end{eqnarray}
 and the
correction are 
\begin{eqnarray}
\delta=\frac{d_2}{2}{1 \over E^2},\ 
\gamma={d_2 \over 2!}\left({1 \over 2E}\right)^2-(1-v_{\nu})^2{\delta t}^2.\ \square
\end{eqnarray}
%{\bf Q.E.D.}

The light-cone region ${\delta t}^2-|{\delta {\vec x}}|^2=0$ is so close 
to  neutrino orbits  that it gives a
finite   contribution to the integral
Eq. $(\ref{lightcone-integration1})$.  Since the light-cone singularity is
real, the integral is sensitive  only to the slow neutrino phase and shows  
interference of the neutrino. This theorem is applied to quite general
systems, where the neutrino interacts with a nucleus in a target.  
 
\subsubsection{Regular terms: short-range correlation}
Next, we study  regular terms of $\tilde \Delta_{\pi,\mu}(\delta
t,\delta{\vec x})$  in Eq. $(\ref{lightcone-integration1})$.  Regular
terms 
are  short-range and the
spreading effect is ignored and the Gaussian wave packet is
studied. First term is  $f_{short }$ in
$I_1$ and is composed of  Bessel functions. We  have   
\begin{align}
L_1=\int d{\vec x}_1 d{\vec x}_2\, e^{i\phi(\delta
 x)}e^{-\frac{1}{2\sigma_{\nu} } \sum_i \left({\vec x}_i-{\vec X}_{\nu} -{\vec
v}_\nu(t_i-\text{T}_{\nu})\right)^2}
f_{short}. 
\label{lightcone-integration2-1}
\end{align}
$L_1$ is evaluated at a large $|\delta t|$ in the form 
\begin{eqnarray}
L_1 = ({\pi \sigma_{\nu}})^{\frac{3}{2}}e^{iE_{\nu}\delta t}
\int d{\vec r} \, e^{-i\vec p_{\nu} \cdot \vec r-{1 \over 4\sigma_{\nu} }({\vec r} 
- {\vec v}_\nu\delta t)^2}  f_{short},~{\vec r}={\vec x}_1-{\vec x}_2.
\end{eqnarray}
Here the integration is made in the space-like region $\lambda <0$. We
write
\begin{eqnarray}
r_l=v_{\nu}\delta t +\tilde r_l,
\end{eqnarray}
and   rewrite $\lambda$ in the form 
\begin{eqnarray}
\lambda =\delta t^2-{r}_l^{\,2}-{\vec r_T^{\,2}} =\delta
 t^2-(v_{\nu}\delta t + \tilde r_l)^{2}-{\vec r_T^{\,2}}
\approx -2 v_{\nu}\tilde r_l \delta t -\tilde r_l^2-{\vec
 r_T^{\,2}}.
\end{eqnarray}
The $L_1$ for the large $|\delta t|$ is written
with these variables. Using  the asymptotic expression of the Bessel
functions, we  have 
\begin{align}
L_1&=({\pi
 \sigma_{\nu}})^{\frac{3}{2}}e^{i(E_{\nu}-|\vec{p}_{\nu}|v_{\nu})\delta t}
\int d{\vec r}_T d \tilde r_l \, e^{-i(  |\vec{p}_{\nu}| \tilde r_l)-\frac{1}{4\sigma_{\nu} }(
\tilde r_l^2+ {\vec r}_T^{\,2})} {i\tilde m \over 4\pi^2}\left({\pi \over
						       2\tilde m}\right)^{\frac{1}{2}}
\nonumber \\  
&\times\left({ 1  \over {2 v_{\nu}\tilde r_l |\delta t| +\tilde r_l^2+
{\vec r_T^{\,2}} }}\right)^{\frac{3}{4}}
e^{i\tilde m  \sqrt{2 v_{\nu}\tilde r_l |\delta t| +\tilde r_l^2+{\vec r_T^{\,2}}
}} . 
%\label{lightcone-integration2-1}
\end{align}
The Gaussian integration around ${\vec r}_T={\vec
0}$, $\tilde r_l=-2i\sigma_{\nu}|\vec{p}_{\nu}|$ give  the asymptotic expression of
$L_1$ at a large $|\delta t|$
\begin{eqnarray}
& &L_1=({\pi \sigma_{\nu}})^{\frac{3}{2}}\tilde L_1,\label{asymptotic-expansionL_1}\\
& &\tilde L_1=
e^{i(  E_{\nu}-|\vec{p}_{\nu}|
 v_{\nu})\delta t}  e^{-  \sigma_\nu |\vec{p}_{\nu}|^2}\frac{i \tilde m}{4\pi^2}\left(\frac{\pi}
 {2 \tilde m}\right)^{\frac{1}{2}}
 \left(\frac{ 1 }{{4 v_{\nu}\sigma_{\nu} |\vec{p}_{\nu}|
|\delta t| }}\right)^{{\frac{3}{4}}}
e^{i\tilde m \sqrt{2 v_{\nu}\sigma_{\nu} |\vec{p}_{\nu}||\delta t|}}.\nonumber
\end{eqnarray}
Obviously $L_1$ oscillates  fast as $e^{i\tilde mc_1|\delta t|^{\frac{1}{2}}}$ where $c_1$
 is determined by $|\vec{p}_{\nu}|$ and $\sigma_{\nu}$ and is short-range.
 The integration carried out with a different  stationary value of $r_l$ 
which takes into account the last term in the right-hand side gives  almost
 equivalent result. The integration  in the time-like region, 
$\lambda >0$,
 is carried in a similar manner and $L_1$ decreases  with time as
 $e^{-\tilde mc_1|\delta t|^{\frac{1}{2}}}$ and 
final  result is almost the same as that of 
the space-like region.
 It is noted that the long-range term which appeared 
from the isolated ${1/\lambda}$ singularity in 
Eq. $(\ref{lightcone-integration4-2}) $ does  not exist  in $L_1$ in
 fact. The reason for its absence is that the Bessel function decreases 
much faster in the  space-like region than ${1/\lambda}$ and
 oscillates much faster 
 than ${1/\lambda}$ in the time-like region. Hence the long-range
 correlation is not generated from the $L_1$  and 
 the light-cone singularity $\delta(\lambda) \epsilon(\delta t)$ and 
$1/{\lambda}$ are the
 only source of the long-range  correlation. 

Second term  of Eq. $(\ref{lightcone-integration1})$ is from $I_2$, of Eq. $(\ref{normal-term})$. We have, 
\begin{align}
L_2=&2 p_{\nu}\!\cdot\!(p_{\pi}-p_{\nu})(\pi\sigma_{\nu})^{\frac{3}
{2}}(4\pi\sigma_{\nu})^{\frac{3}{2}}\frac{1}{
  \left(2\pi\right)^3}\tilde L_2,\label{lightcone-integration5}\\
\tilde L_2=& \int {d\vec{q} \over 2\sqrt{\vec{q}^{\,2}+m_{\mu}^2}} 
 e^{-i\left(E_\pi-E_{\nu}-\sqrt{\vec{q}^{\,2}+m_{\mu}^2}-{\vec v}_{\nu}\cdot({\vec
 p}_{\pi}-{\vec q}-{\vec p}_{\nu})\right)\delta t}  \nonumber \\
&\times e^{ -{ \sigma_{\nu} ({\vec p}_{\pi}-{\vec q}-{\vec p}_{\nu})^2}} 
\theta \left(E_\pi-\sqrt{\vec{q}^{\,2}+m_{\mu}^2} \right).
\end{align}

 The angular velocity of the integrand  in $L_2$
 varies with ${\vec q}$ and the integral $L_2$  has  a short-range
 correlation of the length, $2 \sqrt{
 \sigma_{\nu}}$, in the time direction.
So the $L_2$'s contribution to the total probability comes from the small
 $|\delta t|$ region and   corresponds to the short-range component.

Thus the integral over the coordinates  is 
written in the form
\begin{eqnarray}
& &\int d{\vec x}_1 d{\vec x}_2\,e^{i\phi(\delta x)}e^{-{1 \over
 2\sigma_{\nu} } \sum_i
\left({\vec x}_i-{\vec X}_{\nu} -{\vec
v}_\nu(t_i-\text{T}_{\nu})\right)^2}
\tilde \Delta_{\pi,\mu}(\delta
t,\delta{\vec x})\nonumber \\
& &=2i  \frac{p_{\pi}\! \cdot\! p_{\nu}}{E_\pi} \left[ \left(1 +2p_{\pi}
 \!\cdot\! p_{\nu} \frac{\partial}{\partial {\tilde m}^2}+\cdots
 \right)e^{i\bar{\phi}(\delta t)}(J_{\delta(\lambda)}+L_1)+L_2\right]
\nonumber\\
& &\approx 2i (\pi\sigma_{\nu})^{\frac{3}{2}} \frac{ p_{\pi} \!\cdot\! p_{\nu}}{E_\pi} \left[\left(1
 +2p_{\pi} \!\cdot\! p_{\nu} 
{\partial  \over \partial {\tilde m}^2}+\cdots \right)\right. \nonumber \\ 
& &\left.~~~~\times\left(\frac{\sigma_{\nu}}{2} e^{i\bar \phi_c(\delta t)} 
{\epsilon(\delta t) \over |\delta t|}
 +\tilde L_1\right) -i\left( \frac{\sigma_\nu}{\pi}\right)^{\frac{3}{2}}\tilde L_2\right].\label{lightcone-integration1-1}
\end{eqnarray} 
In the above equation, $p_{\nu}^2=m_{\nu}^2$ is negligibly small
  compared to $\tilde{m}^2$, $p_{\pi}\!\cdot\! p_{\nu}$ and
${\sigma_{\nu}}^{-1}$, and  is  neglected   in most  places 
except the slow phase $\bar \phi(\delta t)$. The 
first term in the right-hand side of
Eq. $(\ref{lightcone-integration1-1})$ is  long-range  and 
the second term is short-range. The long-range term is  separated
from others in a clear manner.

\subsubsection{Convergence condition}

Now  we find  a condition for our method to be 
valid. In Eq. $(\ref{seperation-region})$, the integration region was
split into the one of finite
region $-p_{\pi}^0 \leq q^0 \leq 0$ and the region $0 \leq q^0
$. Accordingly, 
the correlation
function is written into a sum of the singular term and the regular 
term. The singular term is written with  the light-cone
singularity and the power series in 
Eq. $(\ref{singular-function})$. Hence this series must converge
 for the present  method of extracting the light-cone singularity 
to be applicable.

We study  the power series 
\begin{eqnarray} 
\label{power-series}
\sum_n (-2p_{\pi} \!\cdot\! p_{\nu} )^n {1 \over n!} 
\left({\partial  \over \partial {\tilde m}^2}\right)^n \tilde L_1,
\end{eqnarray}
 using the asymptotic expression of $\tilde
L_1$, Eq. $(\ref{asymptotic-expansionL_1})$, first.  The most
weakly converging  term
in $\tilde L_1$, is from ${\tilde m}^{\frac{1}{2}}$ and other terms converge
when this converges.  The series  
\begin{eqnarray}
S_1=\sum_n (-2p_{\pi} \!\cdot\! p_{\nu} )^n {1 \over n!} 
\left({\partial  \over \partial {\tilde m}^2}\right)^n ({\tilde m}^2)^{\frac{1}{4}}
\end{eqnarray}
 becomes  the form,
\begin{align}
S_1=&\sum_n \left({-2p_{\pi} \!\cdot\! p_{\nu} \over {\tilde m}^2 }\right)^n {1 \over
 n!} \left(n-\frac{1}{4}\right)!(-1)^n ({ \tilde m})^{\frac{1}{2}}  \nonumber\\
\approx &\sum_n \left(-{2p_{\pi} \!\cdot\! p_{\nu} \over {\tilde m}^2} \right)^n
 (-1)^n n^{-\frac{5}{4}}({ \tilde m})^{\frac{1}{2}} =\sum_n \left({2p_{\pi} \!\cdot\! p_{\nu}\over {\tilde m}^2} \right)^n n^{-\frac{5}{4}} 
({ \tilde
 m})^{\frac{1}{2}}.
\end{align} 
Hence the series converges if the geometric ration is less than 1. At $2p_{\pi} \cdot p_{\nu}={\tilde
m}^2 $ $S_1$ becomes finite, and the value is expressed by the zeta function,
\begin{eqnarray}
S_1=\sum_n n^{-\frac{5}{4}} ({ \tilde
 m})^{\frac{1}{2}}=\zeta\left(\frac{5}{4}\right)  ({
 \tilde m})^{\frac{1}{2}}.
\end{eqnarray}
Thus  in the region,
\begin{eqnarray}
{2p_{\pi}\!\cdot\! p_{\nu} \over {\tilde m}^2} \leq 1,
\label{convergence-condition-ratio}
\end{eqnarray}
 the series converges and the 
correlation function $\tilde I_1(p_{\pi},\delta x)$ has the singular terms.
Outside  this region, the power series diverges and the present  method
of extracting the light-cone singularity does not work. $I$ is evaluated 
directly and agree with the $I_2$.  

The power series Eq. $(\ref{power-series})$ is estimated with $\tilde L_1 \approx e^{i\tilde m \sqrt{2
 v_{\nu}\sigma_{\nu} |\vec{p}_{\nu}||\delta t| }} $ as
\begin{eqnarray}
S_2=\sum_n (-2p_{\pi} \!\cdot\! p_{\nu} )^n {1 \over n!} 
\left({\partial  \over \partial {\tilde m}^2}\right)^ne^{i\tilde m \sqrt{2
 v_{\nu}\sigma_{\nu} |\vec{p}_{\nu}||\delta t| }},
\end{eqnarray}
and becomes  oscillating  with $\sqrt{|\delta t|}$ of  the form,
 \begin{eqnarray}
S_2=
e^{i\tilde m\left|\sqrt{2
 v_{\nu}\sigma_{\nu} |\vec{p}_{\nu}||\delta t| }\right|\left(1-\frac{p_{\pi} \cdot p_{\nu}}
{{\tilde m}^2}\right)}.
\end{eqnarray}
 The present method of
separating the light-cone singularity  from the correlation function and
of evaluating  the finite-size correction of the probability is valid in
the kinematical region Eq. $(\ref{convergence-condition-ratio})$. 
%\newpage

%%%%%%%%%%%%%%%%%%%%%%%%%%%%%%%%%%%%%%%%%%%% Sub section Probability %%%%%%%%%%%%%%%%%%%%%%%%%%%%%%%%%%%%%%%%%%%%%%
%%%%%%%%%%%%%%%%%%%%%%%%%%%%%%%%%%%%%%%%%%%%%%%%%%%%%%%%%%%%%%%%%%%%%%%%%%%%%%%%%%%%%%%%%%%%%%%%%%%%%%%%%%%%%%%%%%%
\subsection{Time-dependent probability}

Substituting Eq. (\ref{lightcone-integration1-1}) into Eq. (\ref{probability-correlation1}), we have 
the probability of the event that  the neutrino is detected at
a space-time position $(\text{T}_{\nu},{\vec X}_{\nu})$, when the pion momentum distribution $\rho_{exp}({\vec
p}_{\pi})$ is known, in the following form  
\begin{align}
&\int d{\vec p}_{\pi} \rho_{exp}({\vec
p}_{\pi})  d{\vec p}_{\mu} \frac{d{\vec X}_{\nu}}{(2\pi)^3} d{\vec p}_\nu  \sum_{s_1,s_2}|\mathcal{M}|^2 
=g^2 m_{\mu}^2
 |N_{\pi\nu}|^2{(\sigma_{\nu}\pi)}^{\frac{3}{2}}\frac{\sigma_{\nu}}{(2\pi)^6} \int {d
 \vec{p}_{\pi} \over E_\pi}\rho_{exp} \label{total-probability2}({\vec p}_{\pi})\nonumber \\
&\times\int d{\vec X}_{\nu}{d{\vec
 p}_\nu  \over E_{\nu}}   p_{\pi} \!\cdot\! p_{\nu} \int dt_1 dt_2
\left[    e^{i {m_{\nu}^2  \over 2E_{\nu}}\delta t} 
\frac{\epsilon(\delta t)}{|\delta t|}
 +{2 \tilde L_1 \over \sigma_{\nu}}-i{2 \over \pi}\left( {\sigma_\nu \over \pi}\right)^{\frac{1}{2}}\tilde L_2\right]\nonumber\\
&\times  e^{-{1 \over 2 \sigma_{\pi} } \left({\vec X}_{\nu}
-{\vec X}_{\pi}+({\vec v}_{\nu}-\bar {\vec v}_{\pi})(t_1-\text{T}_{\nu}) + 
\bar{\vec{v}}_{\pi}(\text{T}_{\pi}- \text{T}_{\nu})\right)^2-{1 \over 2 \sigma_{\pi} }\left({\vec X}_{\nu}-{\vec X}_{\pi}+({\vec
v}_{\nu}-\bar{{\vec v}}_{\pi})(t_2-\text{T}_{\nu}) + \bar {\vec{v}}_{\pi}(\text{T}_{\pi}
- \text{T}_{\nu})\right)^2}.
\end{align}
From a pion mean free path obtained in the Appendix, the
coherence condition, Eq. $(\ref{coherence-condition})$, is satisfied
and  the pion Gaussian parts are regarded as constant in $t_1$ and $t_2$,
 \begin{eqnarray}
& &e^{-\frac{1}{2\sigma_{\pi}}\left(\vec{X}_{\nu} - \vec{X}_{\pi} +
			     (\vec{v}_{\nu} - \bar {\vec{v}}_{\pi})(t_1 -
			     \text{T}_{\nu}) + \bar {\vec{v}}_{\pi}(\text{T}_{\pi} -
			     \text{T}_{\nu})\right)^2}\approx \text{constant in
}t_1,\\
& &e^{-\frac{1}{2\sigma_{\pi}}\left(\vec{X}_{\nu} - \vec{X}_{\pi} +
			     (\vec{v}_{\nu} - \bar {\vec{v}}_{\pi})(t_2 -
			     \text{T}_{\nu}) + \bar {\vec{v}}_{\pi}(\text{T}_{\pi} -
			     \text{T}_{\nu})\right)^2}\approx \text{constant in
}t_2,\label{coherence-conditions}
 \end{eqnarray}
when an integration over  $t_1$ and $t_2$ are made in a distance of 
our interest which is of the order of a few $100$ [m].  
The integration over  $t_1$ and $t_2$ will be made in the next section.

 When the above  conditions Eq. $(\ref{coherence-conditions})$ are
 fulfilled, an area where the neutrino is produced is inside of a
 same pion and  neutrino waves are treated coherently and are capable
 of showing interference. In a much larger distance where this 
condition is not satisfied,  two positions can not be in the same pion
 and the interference disappears.

\subsubsection{Integrations over times}
Integrations over  the times $t_1$ and $t_2$ are carried
and a probability at a finite  T is obtained here. The  following integral  of 
the slowly decreasing   term is  
\begin{eqnarray}
& &i  \int_0^{\text{T}} dt_1 dt_2  {e^{i {\omega_{\nu}}\delta t }
 \over |\delta t|}\epsilon(\delta t)   
= \text{T} \left\{\tilde g(\text{T},\omega_{\nu})+g(\infty,\omega_{\nu})\right\},\ 
\omega_{\nu}={m_{\nu}^2 \over 2E_{\nu}},\label{probability1} 
\end{eqnarray}
where $\tilde g(\text T,\omega_{\nu})$ vanishes at $\text T \rightarrow \infty$.
We understand that the short-range part $L_1$ cancels with
$g(\infty,\omega_\nu) $ and write the total probability with 
$\tilde g(\text{T},\omega_\nu) $ and the short-range term from $ I_2$.

The integral  of the short-range term, $ \tilde L_2$, is  
\begin{eqnarray}
& &\frac{2}{\pi}\sqrt{\sigma_\nu \over \pi}\int dt_1 dt_2 \tilde
 L_2(\delta t) \nonumber\\
& &= \frac{2}{\pi}\sqrt{\frac{\sigma_\nu}{\pi}} \int_0^{\text{T}} dt_1 dt_2 
 \int \frac{d\vec{q}}{2\sqrt{\vec{q}\,^{2}+m_{\mu}^2}} 
 e^{-i\left(E_{\pi}-E_{\nu}-\sqrt{\vec{q}^{\,2}+m_{\mu}^2}-{\vec v}_{\nu}\cdot({\vec
 p}_{\pi}-{\vec q}-{\vec p}_{\nu})\right)\delta t}  \nonumber \\
& &\ \ \ \times e^{ -{ \sigma_{\nu} ({\vec p}_{\pi}-{\vec q}-{\vec p}_{\nu})^2}}
 \theta\left(E_{\pi}-\sqrt{\vec{q}\,^{2}+m_{\mu}^2}\right)
    \nonumber \\
& &=\text{T} G_0,
\end{eqnarray}
where the constant $G_0$ is given in the integral   
\begin{eqnarray}
& &G_0=2 \sqrt{\sigma_\nu \over \pi}  \int \frac{d\vec{q}}{\sqrt{\vec{q}^{\,2}+m_{\mu}^2}} \delta \left(E_{\pi}-E_{\nu}-\sqrt{\vec{q}^{\,2}+m_{\mu}^2}-{\vec v}_{\nu}\cdot({\vec
 p}_{\pi}-{\vec q}-{\vec p}_{\nu})\right)  \nonumber
 \\
& &\times e^{ -{ \sigma_{\nu} ({\vec p}_{\pi}-{\vec q}-{\vec p}_{\nu})^2}}
 \theta\left(E_{\pi}-\sqrt{\vec{q}^{\,2}+m_{\mu}^2}\right), 
\end{eqnarray}
and is estimated numerically. Due to the rapid oscillation in $\delta t$,
$ \tilde L_2$'s contribution  to the probability comes from the 
small $|\delta t|$ region and the integrations over the time becomes 
constant in T.  Hence this has no finite-size  correction. The regular 
term $\tilde L_1$ is also the same.  
  %%%%%%%%%%%%%%%%%%%%%%%% 800MeV 1eV Pi 4GeV  %%%%%%%%%%%%%%%%%%%%%%%%%%%%%% 
\begin{figure}[t]%
% \begin{minipage}{0.5\hsize}
\begin{center}
\includegraphics[angle=-90,scale=.6]{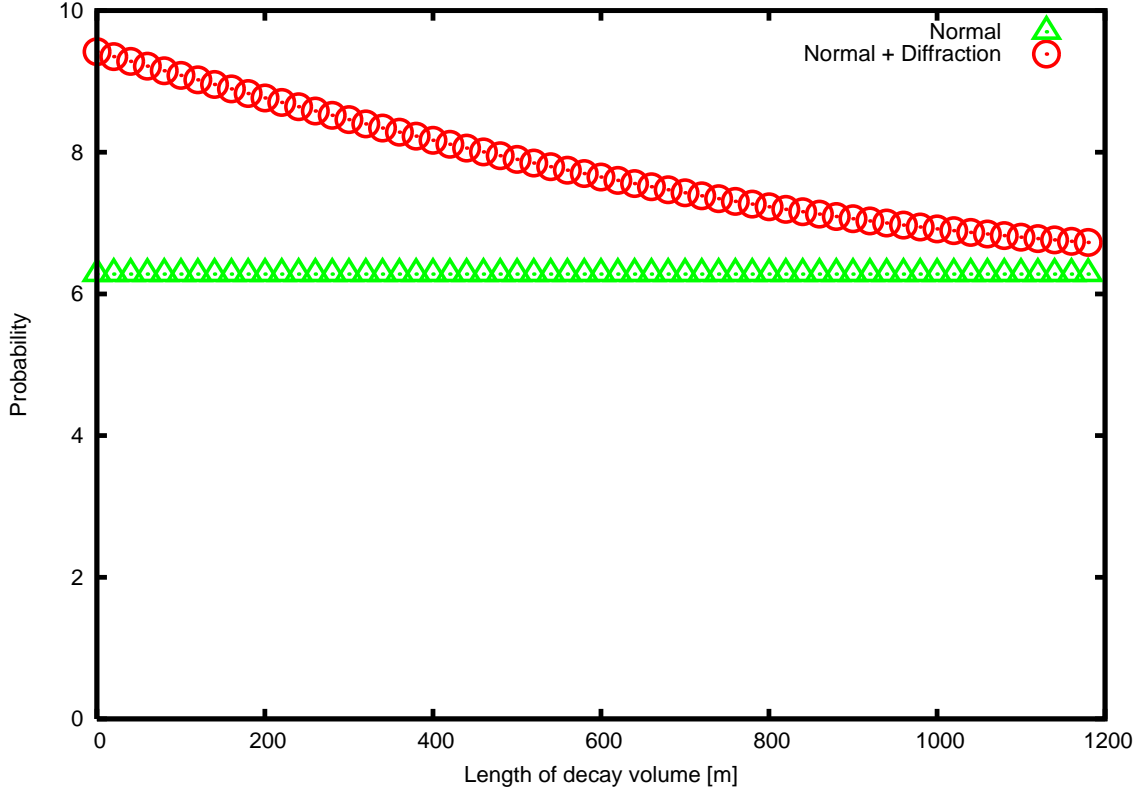}
   \end{center}
\caption{The  probability per unit time of the event that  the neutrino 
is detected 
 in the 
forward direction
 at a distance L is 
given. The constant shows the short-range normal term and   the
 long-range diffraction term is written on top of the normal term.  The horizontal axis 
shows the distance in~[m] and the probability  of the normal term is
 normalized to 2$\pi$.  Clear excess of more than $2/5$ of the normal
 term
 is seen in the 
 distance below 1200~[m]. The neutrino mass, pion energy, neutrino energy are
  1~[eV/$c^2$], 4~[GeV], and 800~[MeV]. A target nucleus with which
the neutrino interacts in a detector is ${}^{16}O$.}
\label{fig:virtual-pi-singular}
\end{figure}%
%%%%%%%%%%%%%%%%%%%%%%%%%%%%%%% %%%%%%%%%%%%%%%%%%%

\subsubsection{Total transition probability    }

Adding   the slowly decreasing part  and the short-range part,  we have 
the final expression of the total 
probability. The neutrino coordinate ${\vec X}_{\nu}$  is integrated in Eq. $(\ref{total-probability2})$ and a factor  $({\sigma_{\pi}\pi})^{\frac{3}{2}}$ emerges. 
This factor is cancelled  with   $({4\pi/\sigma_{\pi}})^{\frac{3}{2}}$
of   the
normalization in Eq. $(\ref{probability})$ and a final result is 
independent of
$\sigma_{\pi}$.  The total transition 
probability is expressed in the form,  
\begin{eqnarray}
& &P=\text{T}g^2 m_{\mu}^2
 D_0  \sigma_{\nu}
\int {d\vec{p}_{\pi} \over E_{\pi}}\rho_{exp}({\vec
p}_{\pi})\int {d\vec{p}_{\nu} \over E_{\nu}}  (p_{\pi}\! \cdot\! p_{\nu}) 
 [\tilde g(\text{T},\omega_{\nu}) 
 +G_0 ], 
\label{probability-total}\\
& &
D_0=|N_{\pi\nu}|^2{(\sigma_{\nu}\pi)}^{\frac{3}{2}}
{(\sigma_{\pi}\pi)}^{\frac{3}{2}}{1 \over (2\pi)^6}={1 \over (2\pi)^3},
\label{probability-31}
\end{eqnarray}
where $\text{L} = c\text{T}$ is the length of decay region. 
The first term in the right-hand side of Eq. $(\ref{probability-total})$
depends on the time interval T, and the neutrino wave packet 
size $\sigma_{\nu}$, but the second
term does not.

At  a finite $\text{T}$, the first term does not vanish and
the probability Eq. $(\ref{probability-total})$ has the finite-size
correction. Its  relative ratio over  $G_0$ is  independent   of
detection process. So 
we compute  
$\tilde g(\text{T},\omega_{\nu})$ and $G_0$ of 
Eq. ($\ref{probability-31}$) at the forward direction $\theta=0$ and the 
energy dependent total probability that is integrated over the neutrino
angle in the following.  

The probabilities per unit  time  in the forward direction 
are plotted in 
Fig.\,\ref{fig:virtual-pi-singular} for
 $m_{\nu}=1\,[\text{eV}/c^2]$,   $E_\pi = 4$\,[GeV], and the neutrino
 energy $E_\nu = 800$\,[MeV]. For the wave packet size of the neutrino, the size
 of the nucleus of the mass number $A$, $\sigma_{\nu}=
 A^{\frac{2}{3}}/m_{\pi}^2$ 
is used. The value becomes  $\sigma_{\nu}= 6.4/m_{\pi}^2$ for the ${}^{16}O$ nucleus and this is used for the following
 evaluations.   From 
this figure it is seen   that there
 is an excess of the flux at short distance region $\text{L}<600$ [m] and the
 maximal excess is about $0.4$ at $\text{L}=0$. The slope at  $\text{L}=0$ is
 determined by $\omega_{\nu}$.
The slowly decreasing   term has  the  finite magnitude and 
 the finite-size correction is large. 
%%%%%%%%%%%%%%%%%%%%%%%%%%%%%%%%%%%%%%%%%%%%%%%%%%%%%%%%%%%%%%%%%%%
\section{Neutrino spectrum }
%%%%%%%%%%%%%%%%%%%%%%%%%%%%%%%%%%%%%%%%%%%%%%%%%%%%%%%%%%%%%%%
\subsection{Integration over neutrino angle}
In Eq. ($\ref{probability-31} $), $\tilde g(\text
T,\omega_{\nu})$  has an angle   dependence different from 
that  of  $G_0$.
In  $G_0$,  the cosine of
neutrino angle $\theta$ is
determined approximately from a mass-shell condition,
\begin{eqnarray}
(p_{\pi}-p_{\nu})^2=p_{\mu}^2=m_{\mu}^2,
\end{eqnarray}
because the energy and momentum conservation is approximately well satisfied.  
Hence the product of the momenta is expressed with the masses
\begin{eqnarray}
p_{\pi}\!\cdot\! p_{\nu}={m_{\pi}^2-m_{\mu}^2 \over 2}, 
\label{on-shell-angle}
\end{eqnarray}
and  the cosine of the angle satisfies 
\begin{eqnarray}
1-\cos \theta= {m_{\pi}^2-m_{\mu}^2 \over 2|\vec{p}_{\pi}||\vec{p}_{\nu}|}-{m_{\pi}^2
 \over 2|\vec{p}_{\pi}|^2}.
\label{angle-energy-relation}
\end{eqnarray}
The $\cos \theta $ is very close to 1 in a high-energy region.
On the other hand,   $\tilde
g(\text{T},\omega_{\nu})$ of Eq. ($\ref{probability-31} $), 
is present in the  domain of the momenta
Eq. $(\ref{convergence-condition-ratio})$ i.e., in the kinematical region,  
\begin{eqnarray}
\label{long-kinematical}
|\vec{p}_{\nu}|(E_{\pi}-|\vec{p}_{\pi}|)\leq p_{\pi}\!\cdot\! p_{\nu} \leq {m_{\pi}^2-m_{\mu}^2 \over 2}. 
\end{eqnarray}
Since  the angular region of Eq. ($\ref{long-kinematical}$) is 
slightly different from Eq. $(\ref{on-shell-angle})$ and it is impossible
to distinguish the latter  from the former region
experimentally, the neutrino angle is integrated.   
We integrate over the neutrino angle of both terms separately.
We have  the normal term, $G_0$, in the form
\begin{align}
&\int \frac{d\vec{p}_\nu}{E_\nu} 
(p_\pi\cdot p_\nu) G_0  \nonumber\\
 &\simeq \int \frac{d\vec{p}_\nu}{E_\nu}
 (p_\pi\!\cdot\! p_\nu
 )2\sqrt{\frac{\sigma_\nu}{\pi}}\left(\frac{\pi}{\sigma_\nu}\right)^{\frac{3}{2}}\int \frac{d\vec{q}}{\sqrt{\vec{q}^{\,2} + m_\mu}} \nonumber \\
&\times \delta\left(E_\pi - E_\nu - \sqrt{\vec{q}^{\,2} + m_\mu^2}\right)\delta^{(3)}\left(\vec{p}_\pi - \vec{p}_\nu -
 \vec{q}\right)\theta\left(E_\pi - \sqrt{\vec{q}^{\,2} + m_\mu^2}\right) \nonumber\\
 &=\frac{(2\pi)^2}{\sigma_\nu}\left({m_{\pi}^2-m_{\mu}^2 \over
 2}\right){1 \over |\vec{p}_{\pi}|}\int_{E_{\nu,min}}^{E_{\nu},max} dE_{\nu},
\end{align}
where 
\begin{eqnarray}
E_{\nu,min}={m_{\pi}^2-m_{\mu}^2 \over
 2(E_{\pi}+|\vec{p}_{\pi}|)},\ E_{\nu,max}={m_{\pi}^2-m_{\mu}^2 \over
 2(E_{\pi}-|\vec{p}_{\pi}|)},
\end{eqnarray}
and  the Gaussian function is approximated by the delta function for
the computational convenience. The angle is determined uniquely. 

 We compute the correction  term next. There are two cases depending on
 the  minimum angle of satisfying the convergence  condition Eq. $(\ref{convergence-condition-ratio})$,
\begin{eqnarray}
\cos \theta_c={\frac{E_\pi E_\nu - \frac{1}{2}(m_\pi^2 - m_\mu^2)}{|\vec{p}_\pi||\vec{
 p}_\nu|}}.
\end{eqnarray}

 In the
 first energy region, 
\begin{eqnarray}
-1 \leq \cos \theta_c ,
\end{eqnarray}
the convergence  condition is satisfied  in $\cos \theta_c \leq \cos \theta $,
 and we have the integral  
\begin{align}
&\int \frac{d\vec{p}_\nu}{E_\nu} (p_\pi\!\cdot\! p_\nu)
 \tilde{g}(\text{T},\omega_\nu) \nonumber\\
&= 2\pi\int \frac{|\vec{p}_\nu|^2d|\vec{p}_\nu|}{E_\nu} \int_{\cos \theta_c}^{1} d\cos\theta(E_\pi E_\nu
 - |\vec{p}_\pi||\vec{ p}_\nu|\cos\theta)\tilde{g}(\text{T},\omega_\nu) \nonumber\\
&= 2\pi\int_{E_{\nu,min}}^{E_{\nu},max}  \frac{dE_\nu}{2|\vec{p}_\pi|}\left\{\frac{1}{4}\left(m_\pi^2 -
 m_\mu^2\right)^2 - (E_\pi E_\nu -
 |\vec{p}_\pi||\vec{p}_\nu|)^2\right\}\tilde{g}(\text{T},\omega_\nu).
\end{align}
Here  the angle is very close to the former value but is not unique.

In the second  
 region, 
\begin{eqnarray}
\cos \theta_c \leq -1,
\end{eqnarray}
the convergence condition is satisfied in arbitrary angle, 
and we have  the integral
\begin{eqnarray}
& &\int \frac{d\vec{p}_\nu}{E_\nu} (p_\pi\!\cdot\! p_\nu)
 \tilde{g}(\text{T},\omega_\nu) \nonumber\\
& &= 2\pi\int \frac{|\vec{p}_\nu|^2d|\vec{p}_\nu|}{E_\nu} 
\int_{-1}^{1}
d\cos\theta(E_\pi E_\nu
 - |\vec{p}_\pi||\vec{ p}_\nu|\cos\theta)\tilde{g}(\text{T},\omega_\nu) \nonumber\\
 & &= 4\pi\int_{0}^{E_{\nu},min} dE_\nu  E_\pi E_{\nu}^2
\tilde{g}(\text{T},\omega_\nu).
\end{eqnarray}
\begin{figure}[t]%
%\label{figure:angle-energy}
\centering{\includegraphics[scale=.6,angle=-90]{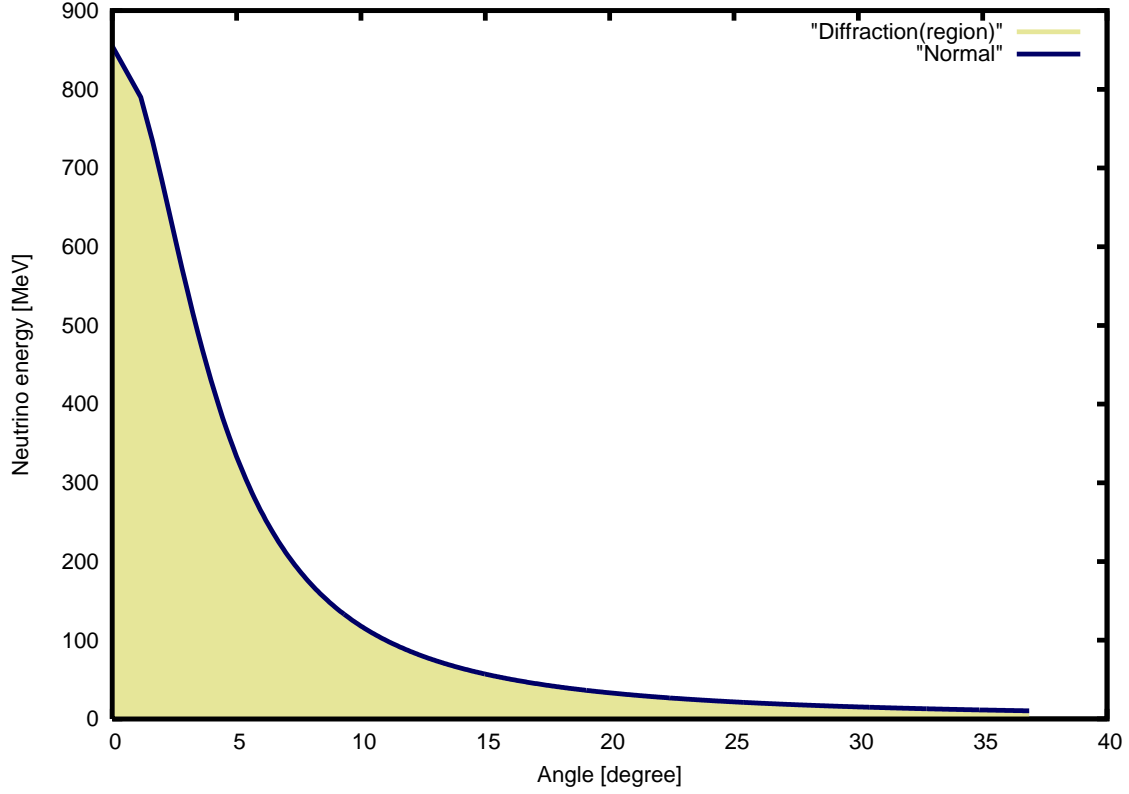}
\caption{The relation between the neutrino angle and energy is
 shown. The kinetic-energy is determined uniquely as Eq. $(\ref{angle-energy-relation})$ in the normal term and 
is not in the diffraction term.  
 The value  in the normal component is on the line and that in the diffraction component  is under the line.  The horizontal axis shows the angle and the vertical axis shows the energy.
    The
 energy of the pion is $E_\pi=2$ [GeV].}}
\label{figure:angle-energy}
\end{figure}%
The angle is fixed to one value, Eq. $(\ref{angle-energy-relation})$, in 
the normal asymptotic term and is
 in a continuous range in the correction  term. 
  The angle  
 dependences of the energies    are given in
  Fig.\,3. 
 Finally we have 
\begin{eqnarray}
\frac{dP}{dE_\nu}=\frac {dP^{(0)}}{dE_\nu}+\frac {dP^{(d)}}{dE_\nu}\label{total-probability-energy}, 
\end{eqnarray}
where
\begin{eqnarray}
& &\frac {dP^{(0)}}{dE_\nu}=\text{T}g^2 m_{\mu}^2
 D_0  \int {d
 \vec{p}_{\pi} \over E_{\pi}}\rho_{exp}({\vec
p}_{\pi}) 
 \frac{2\pi}{|\vec{p}_\pi|}\times {\pi}( m_\pi^2
 - m_\mu^2),  \\
& &\frac {dP^{(d)}}{dE_\nu} =\text{T}g^2 m_{\mu}^2
 D_0  \int {d
 \vec{p}_{\pi} \over E_{\pi}}\rho_{exp}({\vec
p}_{\pi}) 
 \frac{2\pi}{|\vec{p}_\pi|} \biggl(\theta(E_{\nu}-E_{\nu,min})
\Bigl\{\frac{1}{4}\left(m_\pi^2 -
 m_\mu^2\right)^2 \nonumber \\
& &~- (E_\pi E_\nu - |\vec{p}_\pi| |\vec{p}_\nu|)^2 \Bigr\} 
 +\theta(E_{\nu,min}-E_{\nu})
2p_{\pi} E_{\pi}E_{\nu}^2\biggr) \frac{ \sigma_{\nu}}{2}\tilde{g}(\text{T},\omega_\nu). 
\end{eqnarray}
 $\frac {dP^{(0)}}{dE_\nu}$ is  proportional to  T  and 
$\frac {dP^{(d)}}{dE_\nu}$ is constant. 
%%%%%%%%%%%%%%%%%%%%%%%%%%%%%%%%%%%%%%%%%%%%%%%%%%%%%%%%%%%%%%%%%%%%%%%%%
%%%%%%%%
\subsection{Neutrino spectrum }
%%%%%%%%%%%%%%%%%%%%%%%%%%%%%%%%%%%%%%%%%%%%%%%%%%%%%%%%%%%%%%%%%%%%%%
\subsubsection{Sharp pion momentum}
%%%%%%%%%%%%%%%%%%%%%%%%%%%%%%%%%%%%%%%%%%%%%%%%%%%%
For the initial pion of a discrete momentum ${\vec P}_{\pi}$,   
\begin{eqnarray}
\rho_{exp}({\vec p}_{\pi})=\delta({\vec p}_{\pi}-{\vec P}_{\pi}),
\end{eqnarray}
 the rates are expressed in the form,
\begin{eqnarray}
& &\frac{1}{\text{T}}\frac{dP^{(0)}}{ dE_\nu}=g^2 m_{\mu}^2
 D_0   {1 \over E_{\pi}} 
 \frac{2\pi}{|\vec{P}_\pi|} {\pi}( m_\pi^2
 - m_\mu^2) \label{total-probability-energy2n},  \\
& &\frac{1}{\text{T}} \frac{dP^{(d)}}{dE_\nu}=g^2 m_{\mu}^2
 D_0   \frac{1}{E_{\pi}} 
 \frac{2\pi}{|\vec{P}_\pi|}\biggl(\theta(E_{\nu}-E_{\nu,min}) \left\{\frac{1}{4}(m_\mu^2 -
 m_\pi^2)^2 - (E_\pi E_\nu - |\vec{P}_\pi|
 |\vec{p}_\nu|)^2\right\}\nonumber\\
& &+\theta(E_{\nu,min}-E_{\nu})
 2p_{\pi}E_{\pi}E_{\nu}^2\biggr){ \sigma_{\nu} \over 2}  \tilde{g}(\text{T},\omega_\nu)
.\label{total-probability-energy2d}
\end{eqnarray}
Equations $(\ref{total-probability-energy2n})$ and $(\ref{total-probability-energy2d})$ are independent of the position
${\vec X}_{\pi}$ and an average over  ${\vec X}_{\pi}$ is easily made and
the results are  obviously the same.

The   rate  $P^{(0)}$  is independent of  
$\sigma_{\nu}$ \cite{Stodolsky}, and   agrees with the standard value,
 \begin{eqnarray}
\Gamma&=&g^2 m_{\mu}^2
 D_0   {1 \over E_{\pi}}
 \frac{2\pi}{|\vec{P}_\pi|} {\pi}( m_\pi^2
 - m_\mu^2) \int_{E_{\nu,min}}^{E_{\nu,max}} dE_{\nu}\nonumber\\
&=&g^2 m_{\mu}^2 {1 \over 4\pi}\frac{m_{\pi}^2}{E_{\pi}}\left(1-{m_{\mu}^2
 \over m_{\pi}^2}\right)^2.
\end{eqnarray} 
$\tilde{g}(\text{T},\omega_\nu)$ behaves as ${2}/({\omega_{\nu} \text{T}})$ at a large T, 
where $\text{T}=\text{T}_{\nu}-\text{T}_{\pi}$, and the correction 
 term at high energy pion becomes, 
\begin{eqnarray}
\Gamma^{(d)} = \frac{m_{\pi}^2 \sigma_{\nu}}{8\pi}
 \left(1-\frac{m_{\mu}^2}{m_{\pi}^2}\right)^2{E_{\pi} \over m_{\nu}^2 \text T}\Gamma.
\label{diffraction-magnitude}
\end{eqnarray}

%\subsubsection{Integration over neutrino angle}

%%%%%%%%%%%%%%%%%%%%%%%%%%%%%%% cos integral Enu 800 mnu 1.0 %%%%%%%%%%%%%%%%%%%
 \begin{figure}[t]%
% {0.5\hsize}
\begin{center}
  \includegraphics[angle=-90,scale=.6]{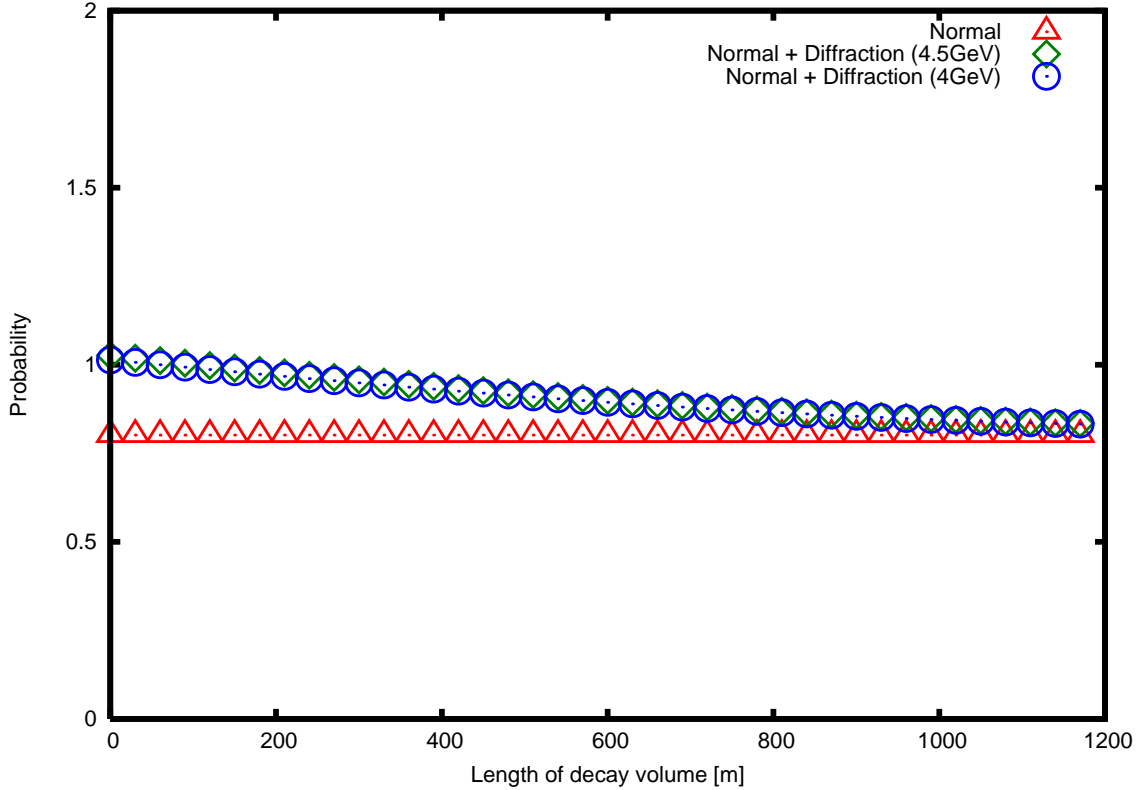}
   \end{center}
  \caption{The  total probability per unit time  of the event that 
the neutrino is detected  in any   
  angle  at  L is 
given. The constant shows the  normal term and   the diffraction term is 
written on top of the normal term.  The horizontal axis 
shows the distance in~[m] and the total probability  is normalized to a 
unity at $\text{L}=0$. The excess becomes less clear than the forward
  direction, but is seen in the distance below 1200\,[m]. The neutrino mass, pion energy, neutrino energy are
  1.0~[eV/$c^2$], 4~[GeV] and 4.5~[GeV] , and 800~[MeV]. A target nucleus with which
the neutrino interacts in a detector is ${}^{16}O$.}
 \label{fig:total-int-1}
%%%\end{minipage}
\end{figure}%
%%%%%%%%%%%%%%%%%%%%%%%%%%%%%%% virtual pi subtraction %%%%%%%%%%%%%%%%%%%
 %%%%%%%%%%%%%%%%%%%%%%%%%%%%%%% cos integral Enu 800 mnu 0.6 %%%%%%%%%%%%%%%%%%%
 \begin{figure}[t]%
% {0.5\hsize}
\begin{center}
  \includegraphics[angle=-90,scale=.6]{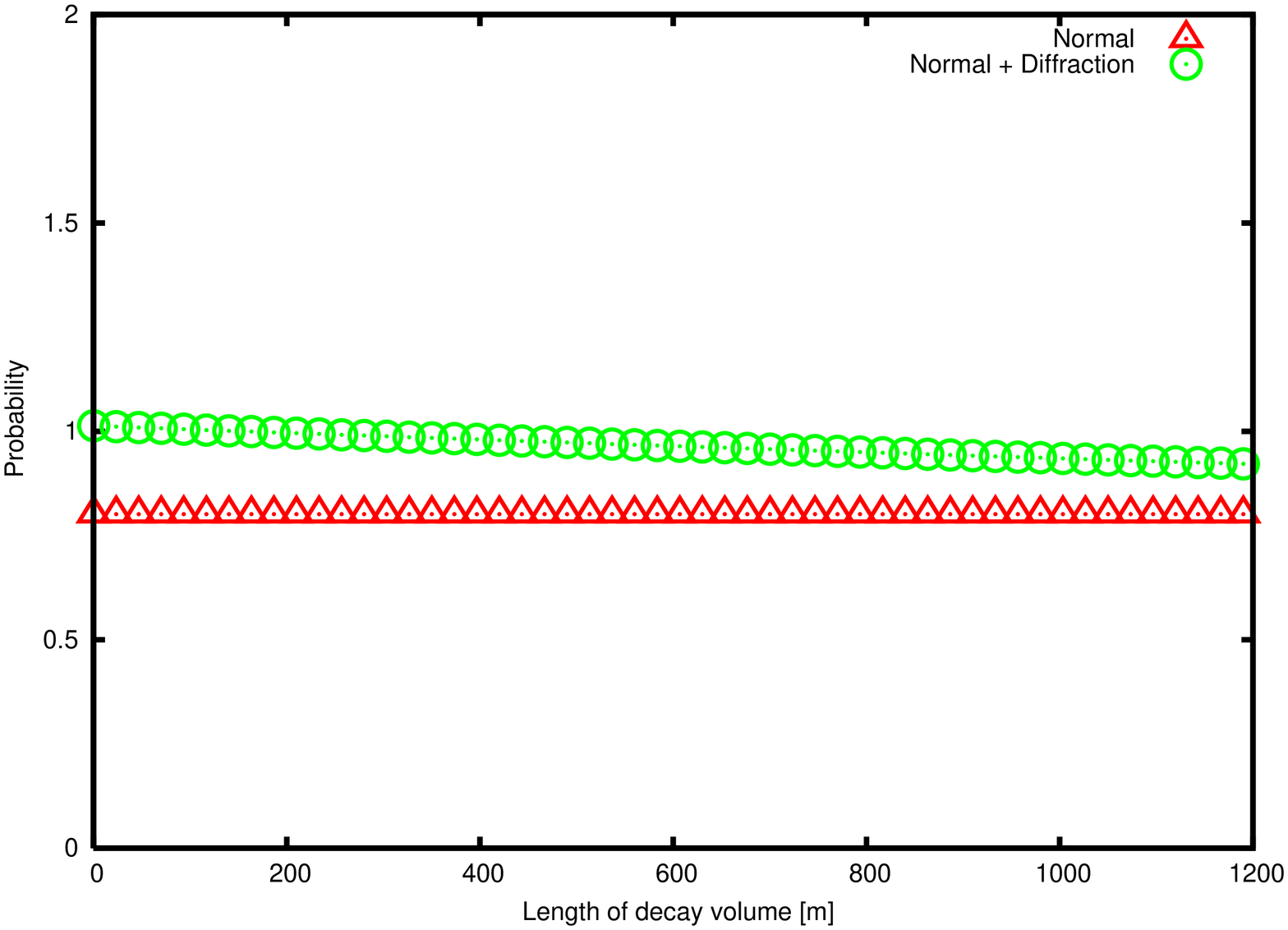}
   \end{center}
  \caption{The  total   probability per unit time of the event that
  the  neutrino is detected in any angle  at  L is 
given. The constant shows the  normal term and   the diffraction  term is written on top of the normal term.  The horizontal axis 
shows the distance in~[m] and the probability of the normal term  is
  normalized to 0.8. Clear uniform excess is seen in the
 distance below 1200\,[m]. The neutrino mass, pion energy, neutrino energy are
  0.6~[eV/$c^2$], 4~[GeV], and 800~[MeV]. A target nucleus with which
the neutrino interacts in a detector is ${}^{16}O$.}
 \label{fig:total-int-.6}
%%%\end{minipage}
\end{figure}%
%%%%%%%%%%%%%%%%%%%%%%%%%%%%%%% virtual pi subtraction %%%%%%%%%%%%%%%%%%%
\subsubsection{Position dependence}
\begin{figure}[t]
% \begin{minipage}{0.5\hsize}
   \begin{center}
   \includegraphics[angle=-90,scale=.6]{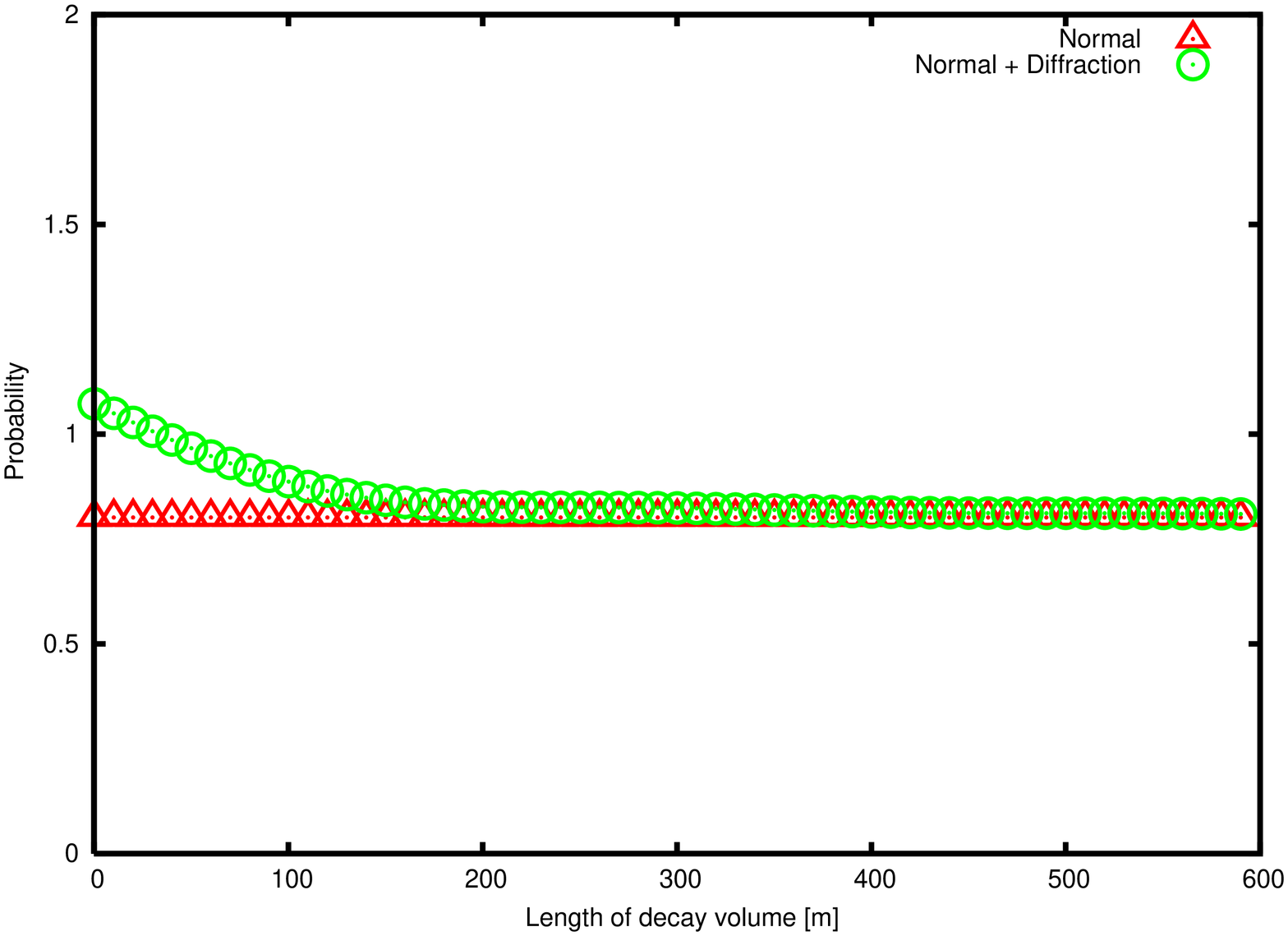}
   \end{center}
\caption{The  total  probability per unit time of the event that the neutrino is
 detected in any angle  at  L is 
given. The constant shows the  normal term and   the diffraction 
 term is written on top of the normal term.  The horizontal axis 
shows the distance in~[m] and the probability of the normal term  is
 normalized to 0.8.  Clear excess and decreasing behavior are  seen in the
 distance below 600~[m]. The neutrino mass, pion energy, neutrino energy are
  1~[eV/$c^2$], 4~[GeV], and 100~[MeV]. A target nucleus with which
the neutrino interacts in a detector is ${}^{16}O$.}
 \label{fig:total-int-1-100}
\end{figure}

%%%%%%%%%%%%%%%%%%%%%%%%%%%%%%%%%%%%%%%%%%%%%%%%%%%%%%%%%%%%%%%%%
In  $P/\text{T}$,  $\tilde g(\text T,\omega_{\nu})$ varies with the
 distance L defined by $\text{L}=c\text T$.   $P/\text{T}$ 
for  $m_{\nu}=1.0\,[\text{eV}/c^2]$ and  $E_{\pi}=4$\,[GeV] and
 $4.5$\,[GeV] are given in  Fig.\,$\ref{fig:total-int-1}$, and for
$m_{\nu}=0.6\,[\text{eV}/c^2]$ are  given in
 Fig.\,$\ref{fig:total-int-.6}$. 
The rate decreases extremely slowly for a light neutrino 
and a longer distance is necessary
to observe the non-uniform behavior for smaller neutrino mass.
For the detection of the muon neutrino, the neutrino energy
 should be larger than the muon mass.  For smaller  energy, the electron 
neutrino is observed.   The probability for $m_{\nu}=1.0\,[\text{eV}/c^2]$ with the energy  $100$\,[MeV]
is given in  Fig.\,$\ref{fig:total-int-1-100}$.
The slowly decreasing component of the probability becomes more
 prominent with lower values.  Hence to observe this component, the 
experiment of the lower neutrino energy is more convenient.   
%%%%%%%%%%%%%%%%%%%%%%%% cos integral 100MeV 0.1eV %%%%%%%%%%%%%%%%%%%%%%%%%%%%%% 
%%%%%%%%%%%%%%%%%%%%%%%% virtual pi probability E-plot %%%%%%%%%%%%%%%%%%%%%%%%%%%%%% 
%%%%%%%%%%%%%%%%%%%%%%%% cos integral 10MeV 1.0eV %%%%%%%%%%%%%%%%%%%%%%%%%%%%%% 
 \begin{figure}[t]
  \begin{center}
   \includegraphics[angle=-90,scale=.6]{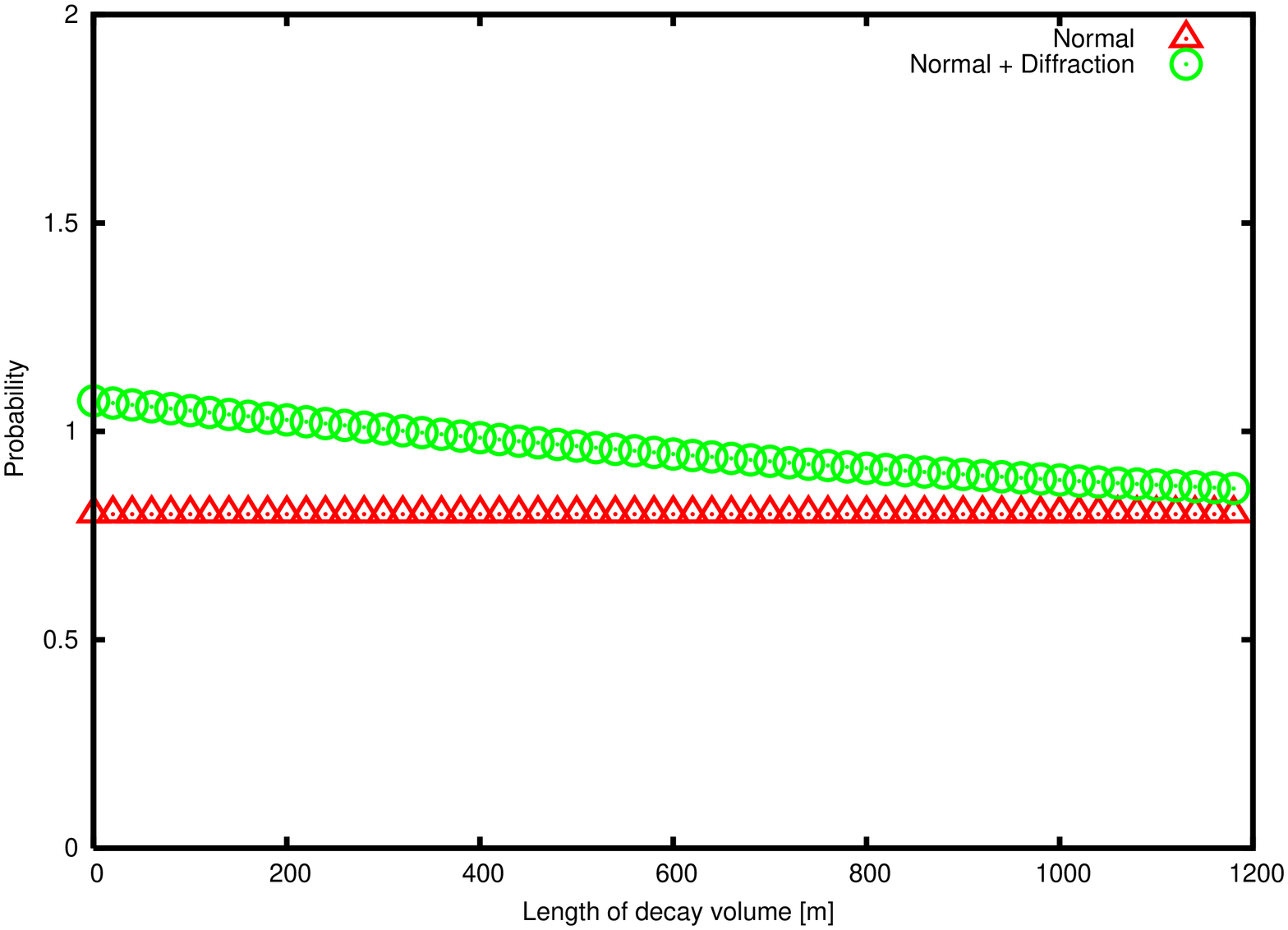}
  \end{center}
  \caption{The  probability per unit time of the event that  the
  neutrino is detected in any angle  at  L is 
given. The constant shows the normal term and   the diffraction 
 term is written on top of the normal term.  The horizontal axis 
shows the distance in~[m] and the probability  is normalized to
  0.8. Clear 
excess is seen in the
 distance below 1200~[m]. The neutrino mass, pion energy, neutrino energy are
  0.1~[eV/$c^2$], 4~[GeV], and 10~[MeV]. A target nucleus with which
the neutrino interacts in a detector is ${}^{16}O$.}
 \label{fig:total-int-.1-10}
%\end{minipage}
\end{figure}%
%%%%%%%%%%%%%%%%%%%%%%%% virtual pi subtraction E-plot
%%%%%%%%%%%%%%%%%%%%%%%% %%%%%%%%%%%%%%%%%%%%%%%%%%%%%% 
We plot $P/{\text T}$ for $m_{\nu}=0.1
\,[\text{eV}/c^2]$, $E_{\nu}=10\,[\text{MeV}]$ in
Fig.\,$\ref{fig:total-int-.1-10}$.
$P/{\text T}$  decreases more slowly  than before.  So
in order to  observe    the T-dependent behavior for the small
neutrino  mass less than or about the same as $0.1\,[\text{eV}/c^2]$, the electron
neutrino should be used. The decay of the muon and others will be
studied in a forthcoming paper.

From Eq. $(\ref{probability-31})$, and $\tilde g(\text T,\omega_{\nu})=\frac{c\omega_{\nu}}{
2\text{T}}$ at a large T, the typical length $l_0$ for the decreasing
behavior  is  
\begin{eqnarray}
l_0~[\text{m}] ={2E_{\nu} \hbar c \over m_{\nu}^2 }= 400{E_{\nu}[\text{GeV}] \over
 m_{\nu}^2[\text{eV}^2/c^4]}.
\end{eqnarray}
By observing  this behavior, the neutrino absolute mass   would be
determined. The neutrino's 
energy is measured with uncertainty $\Delta E_{\nu}$, which is of the 
order of $0.1 \times E_{\nu}$. This uncertainty is $100$\,[MeV] for the energy
$1$\,[GeV] and is accidentally same order as that of the minimum uncertainty 
$\hbar/|\delta \vec{x}|$ derived from  the nuclear size $|\delta \vec{x}|$. 
The total probability for a larger
value of energy uncertainty is easily computed using
Eq. (\ref{total-probability-energy}).   Figures 4-7 show the
distance dependence of the probability. If the mass is around $1\,[\text{eV}/c^2]$ the
excess of the neutrino flux of
about $20$ percent at the distance less than a few hundred meters is
found. 
%In the  long-baseline neutrino oscillation experiments, the
%neutrino flux at the near detectors has observed excesses of about $10-20$
%percent
%\cite{excess-near-detectorK2K,excess-near-detectorMini,
%excess-near-detectorMino}. We 
%believe this is connected with the excesses found in this
%paper. 
We use mainly $m_{\nu}=1\,[\text{eV}/c^2]$ throughout this section.
Because the rate  has a constant  term and the T-dependent 
term, the T-dependent   term is extracted  easily 
by subtracting the constant term from the total rate. The slowly 
decreasing  component decreases with the scale $l_0$ determined by the
neutrino's mass and the energy.

%%%%%%%%%%%%%%%%%%%%%%%% cos integral Energy dependence L=100m 1.0eV 4GeV %%%%%%%%%%%%%%%%%%%%%%%%%%%%%% 
 \begin{figure}[t]
  \begin{center}
   \includegraphics[angle=-90,scale=.6]{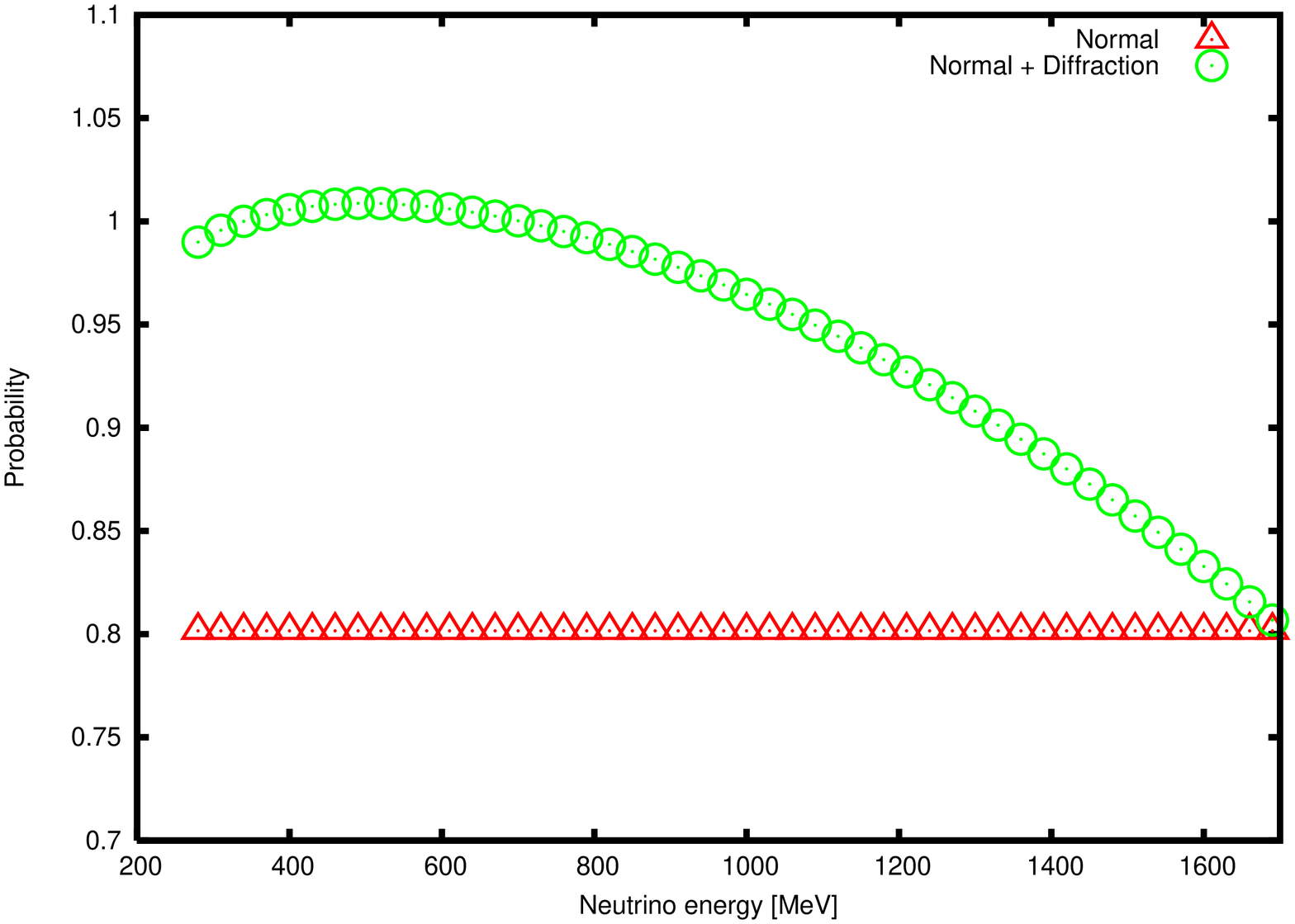}
  \end{center}
  \caption{The energy dependence of probability of the event that the neutrino 
is detected  in any angle   at distance $\text{L} = 100$ [m] is 
given. The lower curve  shows the  normal term and   the correction 
  term is added  on top of the normal term.  The horizontal axis 
shows the neutrino energy in~[MeV] and the probability of the normal
  term  is
 normalized to 0.8.  The neutrino mass and pion energy are
  1.0~[eV/$c^2$] and 4~[GeV]. A target nucleus with which
the neutrino interacts in a detector is ${}^{16}O$.}
 \label{fig:total-ene-1}
%\end{minipage}
\end{figure}% 
%%%%%%%%%%%%%%%%%%%%%%%%%%%%%%%%%%%%%%%%%%%%%%%%%%%%%%%%%%%%%%%
\subsubsection{Energy  dependence}
%%%%%%%%%%%%%%%%%%%%%%%%%%%%%%%%%%%%%%%%%%%%%%%%%%%%%%%%%%%%%%%
The energy spectrum of the neutrino is
  studied next. Since the correction  term has the origin in the final
  states that do not conserve  the kinetic energy, that 
  shows unusual behavior. 
   In    Fig.\,$\ref{fig:total-ene-1}$, the spectrum for the neutrino mass 
and pion energy, 1.0~[eV/$c^2$] and 4~[GeV], are given. The  spectrum
  of the normal term is flat, whereas  that 
of the  diffraction is not flat and  has a maximum at the energy 
$E_{\nu}\approx { E_{\nu,max}/3}$. The former is caused by the fact that
  the neutrino energy in the rest system is 
fixed to one value from the energy-momentum conservation, whereas  that 
of the  diffraction is not fixed to one value from the non-conservation
  of the kinetic energy. 
  A unique property of the correction term for neutrino  is identified by
 its  energy spectrum. 

\begin{figure}[t]
\begin{center}
\includegraphics[scale=.6,angle=-90]{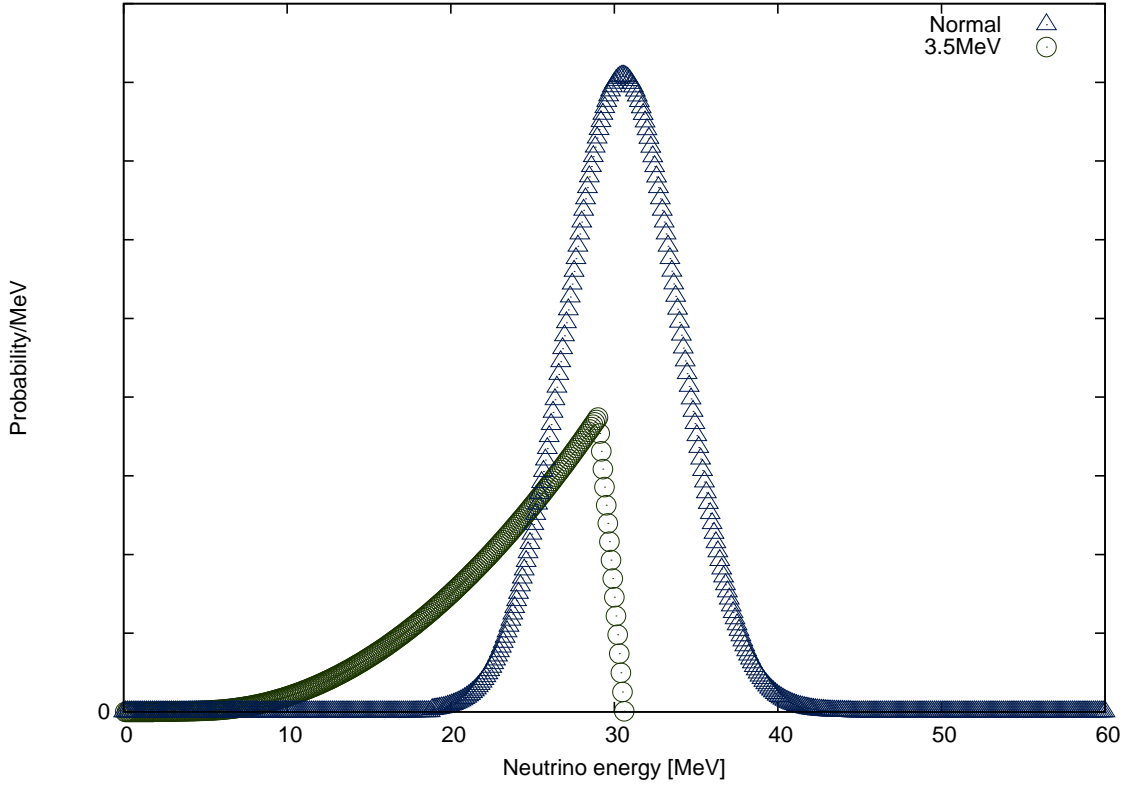}
\end{center}
\caption{The energy dependence of probability of the event that the
 neutrino  is detected  in the rest
 system of the pion are  given for the wave packet size of $5$ [MeV]. The 
normal component  becomes  wide due to the
 wave packet effect. The diffraction  component has lower energies and
 is   wider  than the
 normal component.  The magnitude of the diffraction 
term is arbitrary.  The neutrino mass is 1.0~[eV/$c^2$]. The length is
 $\text{L}=10$ [m].
  }
\label{figure:rest-pion1}
\end{figure}
The energy spectrum of the normal and correction  terms from a
pion at rest  for the wave
packet size of the momentum width $5$ [MeV] is
given in  Fig.\,\ref{figure:rest-pion1}. The spectrum has 
a  peak at the value  derived from the  energy-momentum conservation,
\begin{eqnarray}
m_{\pi}=E_{\nu}+E_{\mu},\ {\vec p}_{\nu}+{\vec p}_{\mu}=0
\end{eqnarray} 
of  the two-body decay.
The neutrino energy is uniquely determined to the
value 
\begin{eqnarray}
E_{\nu}={m_{\pi}^2-m_{\mu}^2 \over 2m_{\pi}},
\label{energy-rest}
\end{eqnarray}
for plane waves, but the  spectrum becomes broad due to the finite wave
packet effect.  
The correction term   derived from the leading singularity is
shown  in the low energy region at the  length is $\text{L}=10$ [m].

\begin{figure}[t]
\begin{center}
\includegraphics[scale=.60,angle=-90]{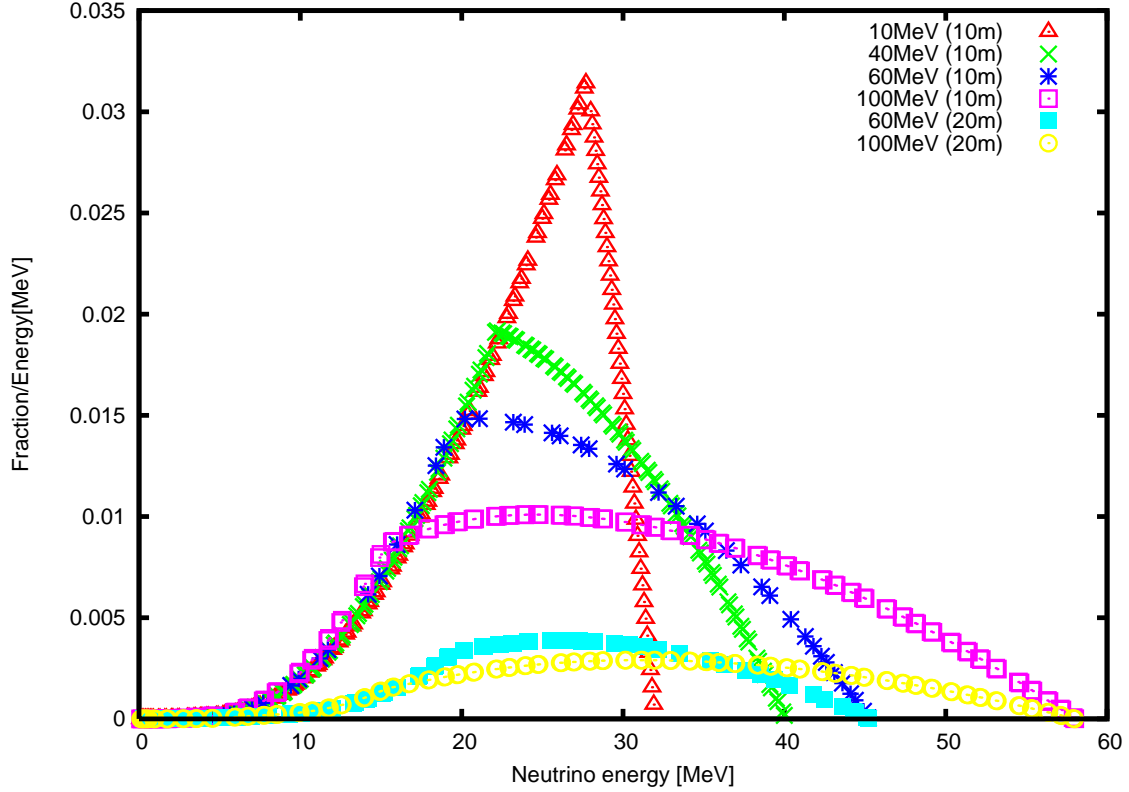}
\end{center}
\caption{The fraction  of the correction term over the normal term 
of the event that the neutrino of certain
 energy  is detected. 
The fractions  are given for the energy of the pion 10, 40, 100 [MeV/c]
 and the length $\text{L}=10$ [m], and for the energy of the pion 60,
 100 [MeV/c]
 and the length $\text{L}=20$ [m]. Target is ${}^{56}$Fe. The fraction is small in lower energy and  
 larger in higher  energy.  The correction  term may be observable in
 these energy regions too. The neutrino mass is 1.0~[eV/$c^2$].  }
\label{figure-intermediate-pion}
\end{figure}

 Figure $\ref{figure-intermediate-pion}$  shows the energy spectrum of 
the fraction of the correction  term over the normal term for various
parameters of the pion energy and distance,
which   are
  computed with the $(V-A)\times(V-A)$  interaction and is 
represented in a  latter section (Sec.7.3) for ${}^{56}Fe$. The neutrino  
spectrum varies depending on the pion energy.    Only the leading 
  term was  taken into account. 
Because the neutrino has the energy different   from that of the kinetic-energy
conserving term, this component would have been  misunderstood  as  a 
background  that  is  not connected  with the system. This component  is 
derived from the Schr\"odiger
equation, and appears at all times. 
  The finite-size correction is not invariant under the Lorentz
  transformation and its  magnitude  becomes 
  larger in  higher energy. 
Thus the fraction of the electron mode varies with the pion's energy. 
{\bf This unusual
behavior is a characteristic feature of the correction  component.} Our 
result, in fact, shows that this  background becomes  larger as  the 
pion's energy becomes larger but  has
   the universal property.

\subsubsection{Wide distribution of pion momentum  }
When a momentum distribution $\rho_{exp}({\vec p}_{\pi})$ of 
initial pions is known,  an energy-dependent probability 
is computed  using  the expression   Eq. $(\ref{total-probability-energy})$.
 Equation $(\ref{total-probability-energy})$ is also independent of the position
${\vec X}_{\pi}$ and depends upon  a pion momentum and 
a neutrino momentum  and the time interval  $\text{T}=\text{T}_{\nu}-\text{T}_{\pi}$. In
experiments, a position of a pion is not measured and an average over 
a position is made. 
An  average probability  agrees with  
Eq. $(\ref{total-probability-energy})$. This  probability varies slowly 
with  the pion's momentum and is regarded constant in the energy range 
of the order of $100$\,[MeV]. So the experimental observation of the
correction  
term is quite easy.
   
\subsection{On the universality of the finite-size correction  }
 The finite-size correction of the 
probability of the event that  the neutrino is detected has various
unique properties. 

This component is constant  in  T, hence the total 
probability is not
proportional to T in this region. In classical particle's decay, the 
decay process occurs randomly and follows Markov process. Hence 
an average number of   the event  is necessary proportional
to T. Now due to the finite-size correction, this property does not
hold. This is not surprising in  $\text{L} \leq l_0$,
because the quantum mechanical interference  modifies the
probability.

The finite-size correction  is expressed with the universal function $\tilde
g(\text {T},\omega_{\nu})$, where $\omega_{\nu}={m_{\nu}^2}/{(2E_{\nu})}$. This  
is  determined  only with  the mass and energy of 
the neutrino and is  independent of details of other parameters of 
the system such as the size, shape, and position  of the wave packets
and others. Hence the correction  has    
the genuine property of the wave function 
$|\text {muon,~neutrino}(t)\rangle$, of Eq. $(\ref{state-vector} )$, and is capable of  
experimental measurements.

\subsubsection{Violation of kinetic-energy conservation }
The probability computed with $S[\text {T}]$  reflects the wave
function at a finite time $t$, and  the states of 
non-conserving  kinetic energy lead  the 
finite-size correction.  
  So conservation  laws derived from the space-time symmetry 
get  modified and various probabilities become different from those of
$\text {T} \rightarrow \infty$.
The leading finite-size corrections have, nevertheless, universal forms
that are proportional to $\tilde g(\text{T},\omega)$. 
\subsubsection{Comparisons on the neutrino finite-size correction  with
   diffraction of 
classical waves through a hole}

{\bf a.  Inelastic channel}. 

The correction for neutrino  emerges as a macroscopic  quantum
phenomenon. The neutrino  of varying kinetic energies  is expressed with
the many-body 
wave function composed of the pion, muon, and neutrino. Accordingly,  the
 probability of the event  that  the neutrino is detected  becomes 
very 
different from  that of 
the free isolated neutrino, and its  probability receives  the large finite-size
correction of the universal behavior. Its  magnitude is determined by
the overlap of wave functions, and  depends
on the wave packet size.  So the finite-size correction is determined
by both of initial and  final states.
We should
note that quantum mechanical probability is determined with  the overlap 
of the in-coming  waves  with the out-going  waves  and depends on the
 both  states.

In a classical wave phenomenon, on the other hand, an intensity of the
wave is 
determined uniquely with  the in-coming wave.   Its magnitude  is
directly observed. 
Hence the finite-size correction and  the interference
pattern  are determined only by  the  in-coming
wave. Thus interference of the quantum mechanical wave is different
from that of the classical wave. 

{\bf b. Pattern in longitudinal  direction}

The  finite-size correction 
results from the wave natures at a finite time $t$,
and is  generated by the states  orthogonal to the states at $t
\rightarrow \infty$. Hence  its
magnitude is positive semi-definite  and 
depends on the time interval
T. Consequently  the neutrino flux at a finite t has the excess that decreases with the
distance in the  direction to the neutrino momentum and vanishes at the
infinite distance. 

The diffraction pattern of  light through a hole or the interference 
pattern of light  in a double slit experiment are those of classical waves and 
 different. The
intensity has modulations in the perpendicular direction to the wave vector. The 
interference term is a product of two waves of different phases and 
so oscillates.   
 Integrating the  intensity
 over the whole screen, the oscillating interference terms cancel and
 the total intensity  is constant. 

Thus  the pattern of the
 neutrino is very different from that of light.

{\bf c.   $\omega_{\nu}$ is  Einstein minus  de Broglie  
frequencies  }

The pattern of the finite-size correction for neutrino is determined by  
the angular velocity $\omega_{\nu,diff}=\omega_{\nu,E}-\omega_{\nu,dB}$. Since $\omega_{\nu,E}$ and 
$\omega_{\nu,dB}$ are almost the same, they are almost
cancelled and $\omega_{\nu,diff}$ becomes extremely small and stable in $E_{\nu}$.

The interference pattern of the light on the screen of the double slit 
experiment, on the other hand,  
  is determined with the angular
velocity $\omega_{\gamma,dB}$. Since  $\omega_{\gamma,dB}$ is large and proportional
to the energy, the pattern varies rapidly with the position in the
screen and with the energy.   The diffraction
pattern of the light passing through the hole varies   rapidly.
\subsubsection{Muon in the pion decay}
In experiments of observing  the muon in the pion decays,  the neutrino
is not observed.
In this situation, the finite-size correction for the muon   has a 
magnitude that  is determined by  the ratio of the  mass and energy,
${m_{\mu}^2/(2E_{\mu})}$.
%%%%%%%%%%%%%%%%%%%%%%%% cos theta  dependence of
%%%%%%%%%%%%%%%%%%%%%%%% the diffraction  %%%%%%%%%%%%%%%%%%%%%%%%%%%%%% 
 \begin{figure}[t]
  \begin{center}
   \includegraphics[angle=-90,scale=.60]{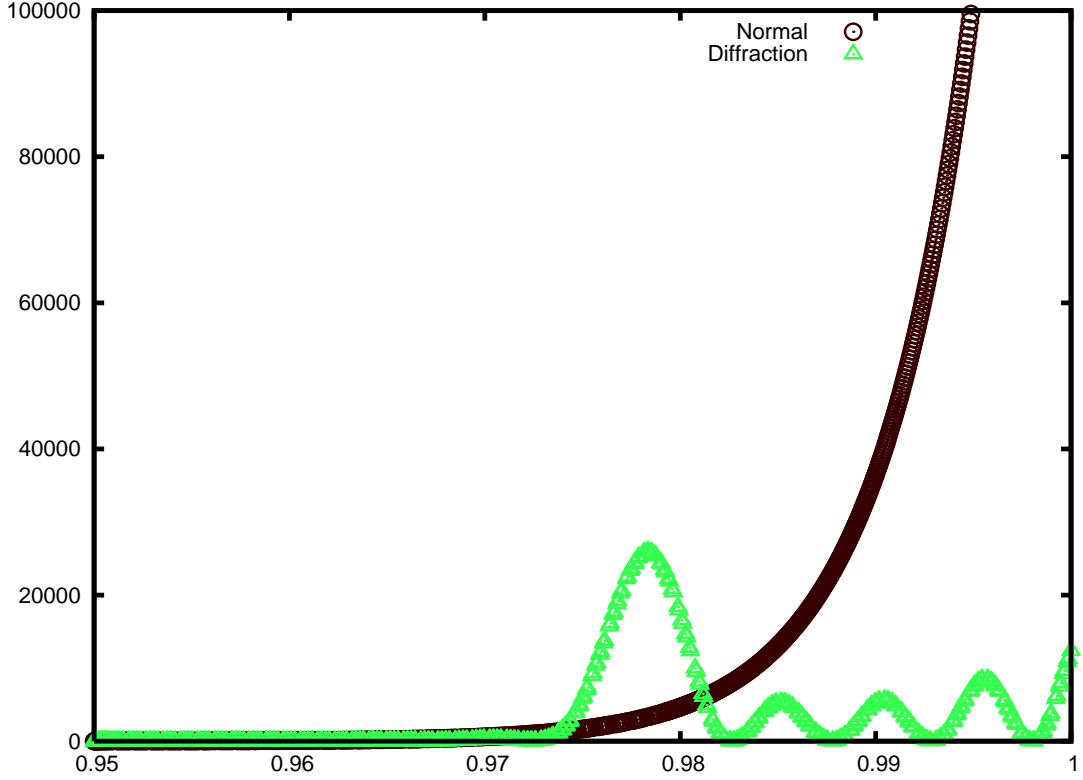}
  \end{center}
  \caption{The probabilities of the events that the neutrino are
  detected at certain angle in  the normal and correction terms  are
  given. The large peak toward $\cos \theta =1$ shows  the
  normal component and the small peaks at the tail of the previous peak 
 show   the correction  component.  The horizontal axis 
shows the cosine of the angle between ${\vec p}_{\nu}$ and ${\vec
  p}_{\pi}-{\vec p}_{\mu}$ and the vertical axis shows the
  probability.   The  pion energy,  muon energy, and time interval 
  are  250~$m_{\nu}$,  210~$m_{\nu}$, and  30 $m_{\nu}^{-1}$.}
 \label{fig:omega}
%\end{minipage}
\end{figure}% 
Since  the muon mass is  larger than the neutrino mass
by $10^8$,  the value $m_{\mu}^2/(2E_{\mu})$ for the muon is much 
larger  than that of 
the neutrino by $10^{16}$. For
the muon of energy 1\,$[\text{GeV}]$, the length is of the order of $l_0=10^{-14}$\,[m].
This value is a microscopic size and $\tilde g(\text{T},\omega_{\mu})$ vanishes
at a macroscopic T. 
Hence the probability of the event for  the muon at the
macroscopic distance 
becomes constant.  The
muon from the pion decay has no  finite-size correction.  This  
probability agrees with the production
probability.  The muon and neutrino behave  differently at the finite 
distance.

If  the muon is observed under  a condition that the neutrino
is  detected at the finite $\text {T}$, $S[{\text{T}}]$ is applied and
the probability of the event that  the muon is detected has the contribution from the
neutrino
diffraction. The diffraction component gives a wide energy spectrum for
the muon since that comes from the tail of the
distribution function.  Fig.\,${\ref{fig:omega}}$ shows   a 
probability integrated over  the neutrino energy in this condition 
that both the muon and neutrino are detected,  
which is obtained from Eq. $(\ref{integrated-amplitude-honbun})$. In this
figure, we use units $c=1,\hbar=1$  and express the energy and time with
the neutrino mass $m_{\nu}$. 
Energy of the pion is $250$ $m_{\nu}$ and the muon has the energy $210$
$m_{\nu}$ 
and has an angle with the pion of $\cos \theta =0.95-1$. The  cosine of 
the angle between ${\vec p}_{\pi}-{\vec p}_{\mu}$ and ${\vec p}_{\nu}$ 
is in the  horizontal axis. T is $30$  $m_{\nu}^{-1}$. The neutrino 
mass of an unphysical magnitude  of the order of MeV
 and the value of T are chosen in such manner  that the numerical 
computation of diffraction component is easily made. Qualitative
features  of  Fig.\,$\ref{fig:omega}$ are that there exist  a large peak at 
$\cos \theta \approx 1$ and small peaks at the tail region.
 The former is the peak from the root of $\omega=0$ at $\delta {\vec p} \approx 0$ and
$\delta E =0$ and the latter are the peaks  from  the roots of
$\omega=0$ of $\delta {\vec p} \neq 0$ and $\delta E \neq 0$. The latter
 peaks, which  do not exist in the
probability of detecting only the muon,  show the feature of the 
diffraction component of the probability when  the neutrino is
detected. Thus the  diffraction component is observed in the muon also 
when the neutrino is detected simultaneously.   Experimental 
verification of the diffraction term of this situation using the muon   
may be made in future.

As was shown in Section 3, the production rate is common to the muon and
neutrino, since they  are produced in the same decay process.  However because 
they propagate differently, they   are detected independently with the  
apparatus. The rates are then  different. The probability of the event that the
neutrino is detected is affected by the large $P^{(d)}$, but   that for
the muon is negligibly small.    

  If both particles are measured
simultaneously, the rates for both are the same. 

Thus the wave-zone for neutrino
is quite  wide,
and  the transition amplitude for  a neutrino  at a
finite distance detected by a nucleus becomes different from that of the infinite
distance. The neutrino wave is 
 a superposition of those waves that are  produced at 
different space-time positions  and the  
probability  is modified by the diffraction term. The
overlap between the 
neutrino wave that is detected with a nucleus in a detector and those
that are produced  from  a pion decay shows the neutrino diffraction
of  unique properties.
So
the neutrino flux measured with  its collisions with a nucleus in targets is 
different from that defined from the norm of wave function. 
%\newpage

%\newpage
\section{Implications}
In this section, various physical quantities of neutrino processes which 
are modified by the  finite-size correction  are studied. Particularly 
 neutrino nucleon total cross sections,  quasi-elastic cross sections,
electron-neutrino production anomaly, a proton target enhancement, and
an anomaly in atmospheric neutrino  are such processes that 
have significant contributions from the neutrino diffraction.

\subsection{Total cross sections of $\nu_{\mu}N$ scattering}

 Neutrino collisions with hadrons in high-energy regions  are understood 
well with the  quark-parton model. A total cross section of a  high-energy 
neutrino is proportional to
the  energy and  is written in the form 
\begin{eqnarray}
\label{total-crossection}
& &\sigma^{\nu}={M_N E_{\nu}G_F^2 \over \pi }(Q+ \bar{Q}/3),
%& &\sigma^{\bar \nu}={M_N E_{\nu} G_F^2 \over \pi }(\bar Q+1/3 Q)\nonumber 
\end{eqnarray}
using integrals of quark-parton distribution functions $q(x)$ and $\bar
q(x)$ and  $Q=\int_0^1 dx xq(x),\bar Q=\int_0^1 dx x\bar q(x)$. The cross
section is
proportional to the neutrino energy and a current value is
$\sigma_{\nu}/E= 0.67 \times 10^{-38}[\text{cm}^2/\text{GeV}]$. 

Now the rate of process of the neutrino  produced in a decay of a pion and 
reacts    with a nucleus at a finite distance has   a finite-size correction. It
modifies  the probability of the event that the neutrino collides with the nucleus in 
target. We estimate its effect
hereafter. Including 
the diffraction term, the effective neutrino flux becomes     
%Thus  experiments seem 
%consistent with  Eq.\,$(\ref{total-crossection})$.  However  recent 
%experiments of NOMAD and MINOS gave the total cross sections in wide 
%energy ranges with small uncertainties. They  have slight energy dependences
% and might be inconsistent with Eq.\,$(\ref{total-crossection} )$. They 
%are compared with our theoretical calculations. 
%The diffraction term  was  not  
%included in the neutrino flux used in NOMAD and MINOS.  
%Here  we include 
%the diffraction term  into the neutrino flux
%and have the flux in 
the sum of the normal and
diffraction terms  
\begin{eqnarray}
f=f^{(0)}(1+r^{(d)}),
\end{eqnarray}
where $f^{(0)}$ is a flux derived from the normal component, and 
$r^{(d)}$ is the rate of the diffraction component over the
normal component and is a function of the combination  
$({m_{\nu}^2 \over 2cE_{\nu}}\text{L})$,
\begin{eqnarray}
r^{(d)}=d_{0} \tilde g(\frac{m_{\nu}^2}{2cE_{\nu}}\text{L}),
\label{E-depedent-correction}
\end{eqnarray}
where  L is a length of the decay volume and the coefficient $d_0$
is determined from  geometries of experiments. 

When a detector is
located at the end of the decay volume, the correction
factor Eq. $(\ref{E-depedent-correction})$ is used. In actual case, the
detector is located in a distant region from the decay volume. There are 
material or soil between them and pions are stopped in beam dump. The neutrino
is produced in the decay region  and  propagates freely afterward. Since 
the wave packets of one $\sigma_{\nu}$ form the complete
set \cite{Ishikawa-Shimomura}, the wave packet of the size at the 
decay volume  is the $\sigma_{\nu}$
determined with  the detector.
The 
neutrino flux at the end of the decay volume is computed with the
diffraction term of the decay volume 's length L and the wave packet
size of the detector. 
Wave packets of this $\sigma_{\nu}$  propagate
 freely from the end of decay volume to the detector. The final value of 
neutrino flux  at the detector is found combining both effects.
%using the factor Eq.\,$(\ref{E-depedent-correction})$. 
When neutrino changes its flavor  in
this period, the final probability for each flavor is written with  a
usual formula of flavor oscillation. 
 
The true  neutrino events   in experiment is converted to the cross
 section $\sigma^{exp}(E_\nu)$ that includes 
 the diffraction component and is connected
with the cross section computed  with only the normal component $\sigma^{the}(E_\nu)$ by  
the rate   
\begin{eqnarray}
%& &
\sigma^{the}(E)
%={N_{event} \over f}={N_{event} \over f_{normal}}{f_{normal}
% \over f_{normal}(1+ r_{diff})} \\
%& &
=\sigma^{exp}(E_\nu) \frac{ 1 }{1+ r^{(d)} }.
\end{eqnarray}
Conversely the experimental  cross section is written as 
\begin{eqnarray}
\sigma^{exp}(E_\nu)/E_\nu= ( 1+ r^{(d)} )(\sigma^{the}(E_\nu)/E_\nu).
\label{energy-dependece}
\end{eqnarray}
$\sigma^{the}(E_\nu)/E_\nu$ is  constant from Eq. ($\ref{total-crossection}$) so  the E-dependence of 
$\sigma^{exp}(E_\nu)/E_\nu$ is due to E-dependence of $ r^{(d)}$,
Eq. $(\ref{E-depedent-correction})$.  

The correction $ r^{(d)}$ depends  on the geometry of 
experiments and the material of the detector.  
We compute $ r^{(d)}$ using the experimental conditions 
of  MINOS \cite{excess-total-detectorMino}  and NOMAD \cite{excess-total-detectorNOMAD} and the total cross sections. 
%%%%%MINOS%%%%%%%%%%%
\begin{figure}[t]
\includegraphics[scale=.45,angle=-90]{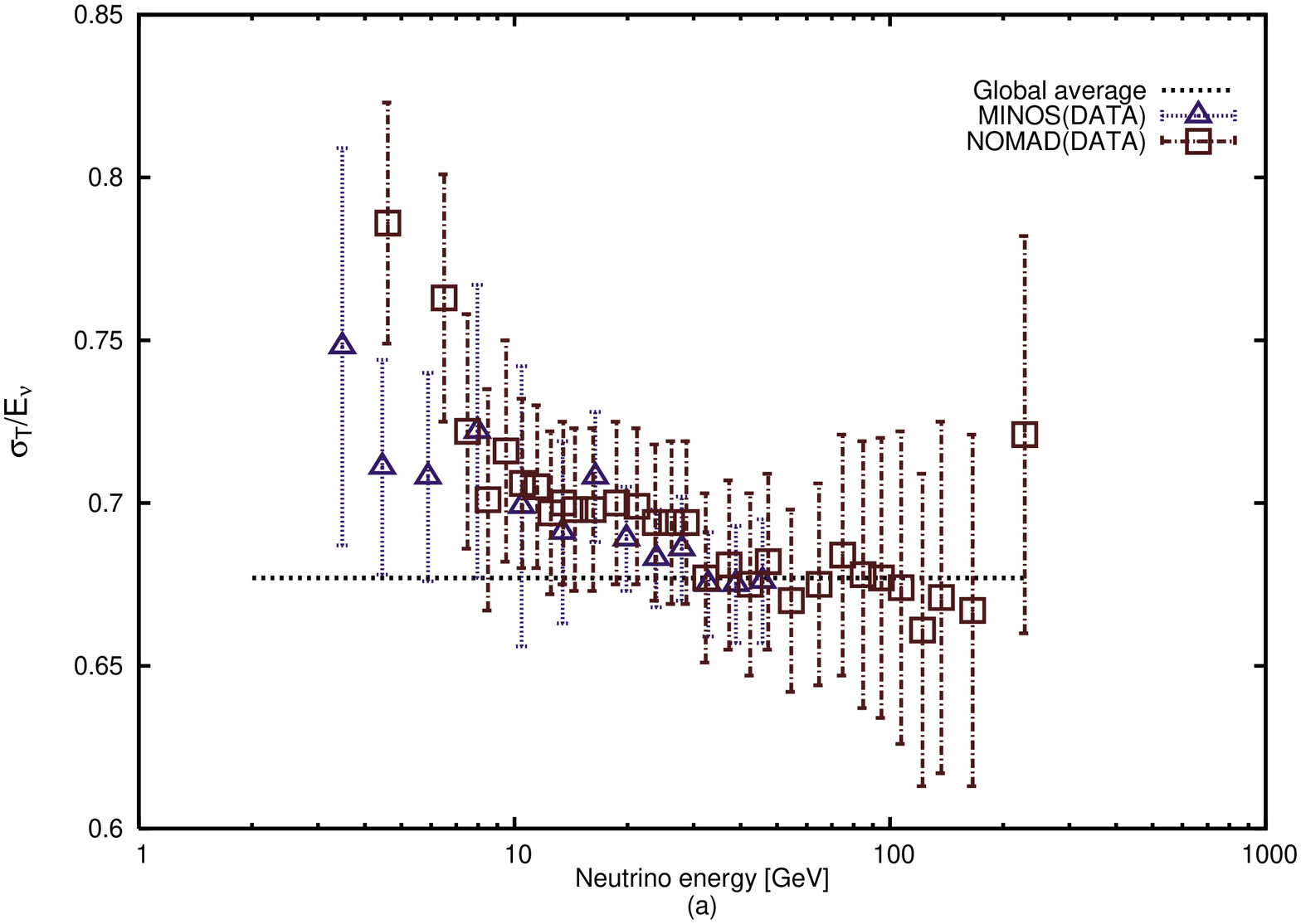}\\
%\caption{Total cross section of MINOS and NOMAD are given.}
%\label{MINOS:fig}
%\end{figure}
%%%%%%MINOS%%%%%%%%%%%%%
%%%%%MINOS%%%%%%%%%%%
%\begin{figure}[t]
\includegraphics[scale=.45,angle=-90]{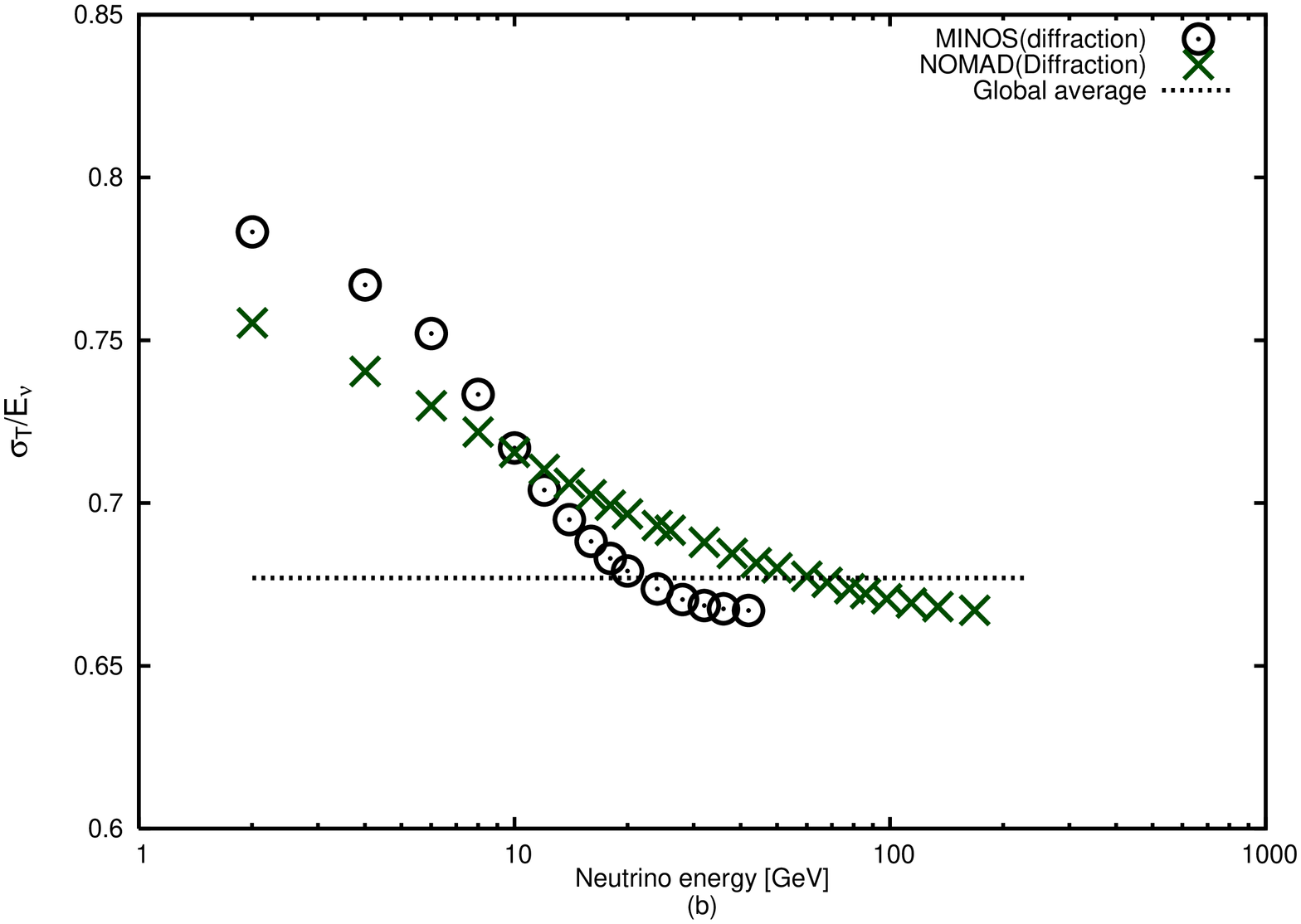}
\caption{Neutrino-Nucleon total cross section of MINOS and NOMAD  (a) and total cross sections of the sums of normal
 and correction  terms in geometries  of MINOS and NOMAD (b) 
are given. The horizontal axis shows the neutrino energies  in [GeV] and the vertical
 axis shows the ratio of the cross section over the energy.}
\label{NOMAD and NOMAD:fig}
\end{figure}
%%%%%%NOMAD%%%%%%%%%%%%%
The geometry of MINOS and  NOMAD are the following. The lengths between the
pion source and the neutrino detector, $\text{L}_{det-so}$, and those  
of the decay region, $\text{L}_{decay-reg}$, 
 are:
\begin{eqnarray}
%& &NuteV~:  L_{detector-source}=30M, L_{decay-region}=20M
%\label{short-baseline}\\
NOMAD~&:&  \text{L}_{det-so}=835\,[\text{m}],\ \text{L}_{decay-reg}=290\,[\text{m}], \\
%& &MiniBoone~: L_{detector-source}=500M, L_{decay-region}=M
%\label{long-baseline} \\
MINOS~&:&\text{L}_{det-so}=1040\,[\text{m}],\ \text{L}_{decay-reg}=675\,[\text{m}].
%& &CDHS~:L_{detector-source}=600M, L_{decay-region}=M, \nonumber 
\end{eqnarray} 
%$.7$M of $L_{decay-region}$ is the length of decay region in air, and
%whole decay region is $1.8$ M. 
%Since lengths between the detectors and the sources 
%Eq.\,$(\ref{long-baseline})$ of MiniBoone and CDHS are  much longer 
%we apply the diffraction term to the decays in
%flight in air of  short baseline
%experiments Eq.\,$(\ref{short-baseline})$.  
Also  pion beam spreading was included from  
 angle of initial pion;   $0$ to $10\,[\text{mrad}]$ for NOMAD and
 $0$ to $15\,[\text{mrad}]$ for MINOS.

The wave packet size is estimated  with  the size of target nucleus. 
From the size of the nucleus of the mass number $A$, we have  
$\sigma_{\nu}= A^{\frac{2}{3}}/m_{\pi}^2$. 
For various material   the value are
\begin{eqnarray}
& &  \sigma_{\nu}= 5.2/m_{\pi}^2;\ {}^{12}C~ nucleus,
\label{wave-packet-size}\\
%& & \sigma_{\nu}= 5.2/m_{\pi}^2; {}^{16}O ~nucleus.\nonumber \\
& & \sigma_{\nu}= 14.6/m_{\pi}^2;\ {}^{56}Fe ~nucleus. 
\end{eqnarray}

Including the geometries, beam spreadings, and wave packet sizes, we
   computed the total cross sections and compared with the experiments
   in Fig.\,$\ref{NOMAD and NOMAD:fig}$. 
   These cross sections  computed theoretically  slowly 
decrease  with the energy in the geometry dependent manner 
   and agree with the experiments. Since the experimental parameters
   such as the neutrino energy and others are different in two
   experiments, the agreements of the theory with the experiments are
   highly non-trivial.
So the large cross sections at low-energy regions may be attributed to 
   the diffraction component. 

We have compared only NOMAD and MINOS here. Many experiments are listed 
in particle data \cite{particle-data} and most of them have similar
energy dependences and agree qualitatively with the presence of the 
diffraction components. It is important to notice 
that the magnitude of diffraction component is sensitive to geometry.
Furthermore, if a kinematical constraint
Eq. ($\ref{angle-energy-relation}$) on 
the angle between ${\vec p}_{\pi}$ and  ${\vec p}_{\nu}$ was required,
only the events of the normal term was selected. Then  the cross section 
should agree with that of the normal term.

\subsection{Quasi-elastic  cross sections}

Quasi-elastic or one pion production
processes are understood relatively well theoretically. The diffraction modifies 
the total events of these processes also.

The cross sections for  
\begin{eqnarray}
& &\nu+n \rightarrow \mu^{-}+p(+\pi^0),\\
& &\nu+p \rightarrow \mu^{-}+p+\pi^{+},\\
& & \bar \nu+p \rightarrow \mu^{+}+n(+\pi^0) ,
\end{eqnarray}
and the neutral current
 process  
\begin{eqnarray}
\nu+N \rightarrow \nu+N(+\pi^0),
\end{eqnarray}
are known well  using CVC, PCAC, and 
vector dominance and are studied recently by MiniBooNE \cite{excess-qenear-detectorMini}. The parameter is the axial vector meson $M_A$ and
higher mass contributions. So these cross section are used to study the 
diffraction terms.

\subsection{Electron neutrino anomaly}
\begin{figure}[t]
\begin{center}
\includegraphics[scale=.6,angle=-90]{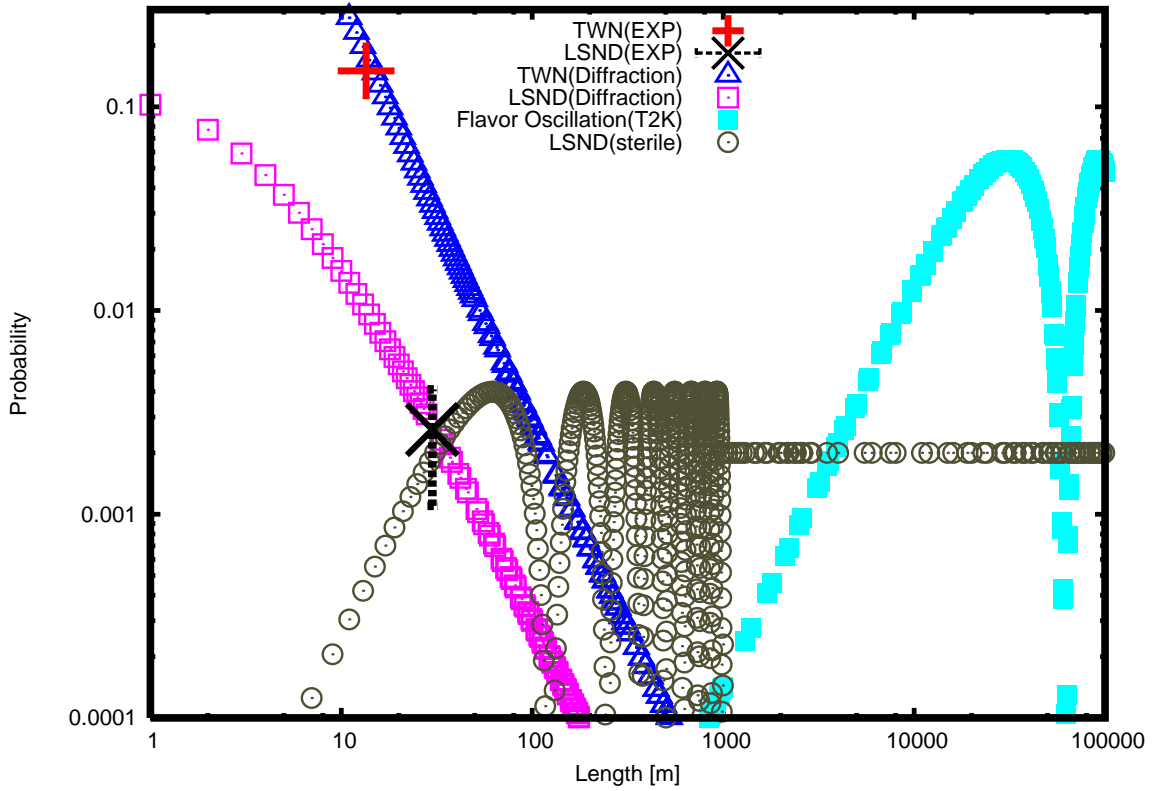}
\end{center}
\caption{Experiments of LSND and TWN are compared with the theoretical
  values of the correction  terms. TWN(EXP) and  LSND(EXP) show the
  experimental values and  TWN(Diffraction) is computed with the
  parameters $m_\nu=0.2\,[\text{eV}/c^2]$,
	    $E_\nu=250[\text{MeV}]$, $P_\pi=2[\text{GeV}/c]$,
	    LSND(Diffraction) is computed with  $m_\nu=0.2\,[\text{eV}/c^2]$, $E_\nu=60\,[\text{MeV}]$,
	    $P_\pi=300\,[\text{MeV}/c]$. Flavor oscillation
  oscillation(T2K) \cite{T2K-nue} shows the values for 
	    $\sin^2\theta_{13}=0.11$, $\delta m^2_{23} =
	    2.4\times10^{-3}\,[\text{eV}^2/c^4]$, $E_\nu = 60\,
	    [\text{MeV}]$, and LSND(sterile) shows with
  $\sin^2\theta=0.004$, $\delta
	    m^2 = 1.2\, [\text{eV}^2/c^4]$, $E_\nu = 60\,[\text{MeV}]$.}
\label{LSND :fig}
\end{figure}
In pion decays, a branching ratio  of an electron mode is smaller
than that of a muon mode  by 
factor $10^{-4}$ due to the helicity suppression of the decay of a
pseudo-scalar particle caused by the charged current
interaction.  \cite{Sakai-1949,Jack,Ruderman,Anderson}. 
This behavior of the total rates  has been  confirmed by the 
observations of charged leptons.

Now the probability  of the event that  a neutrino is detected inside the  
coherence length,
where the neutrino retains the wave natures,  is affected by the 
finite-size correction. Because this correction  comes from the states
that have different kinetic-energy from the initial value,  the
neutrino in this region  
does not follow the conservation law satisfied in the asymptotic 
region $t \rightarrow \infty$. The rate that electron neutrino  is
detected  is not suppressed. The  ratio of the probability of the event that 
the electron neutrino  is detected over that of the muon neutrino event 
becomes   substantially larger in near-detector regions. 

\begin{figure}[t]
\begin{center}
\includegraphics[scale=.6,angle=-90]{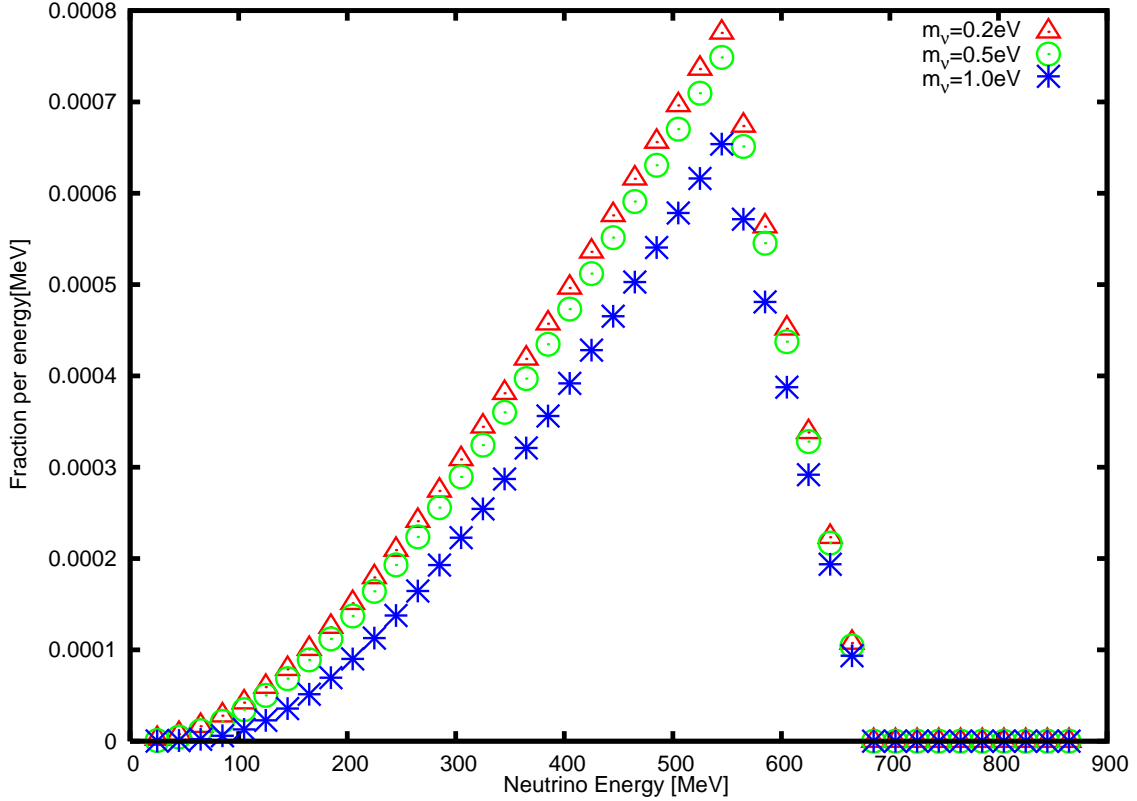}
\end{center}
\caption{Fraction of the electron neutrino of the mass 0.2
  $[\text{eV}/c^2]$, 0.5 $[\text{eV}/c^2]$ and 1$[\text{eV}/c^2
]$
  at L=110\,[m], distance=170\,[m] of T2K geometry and $P_\pi=2$ [GeV/$c$].}
\label{T2K-confuguration}
\end{figure}

To compute  the transition probability and the spectra of electron 
and muon neutrinos, we  start from 
the $(V-A)\times(V-A)$ interaction Lagrangian
(Hamiltonian). The result of the probability is almost the same as 
that of Eq. $(\ref{weak-hamiltonian})$ in the muon mode but is different 
in the electron mode since the diffraction component does not satisfy 
the rigorous conservation of the kinetic-energy and momentum.
   In I, it was found that the  initial pion is described by  a wave
   packet of a large size. Hence the initial pion of the plane wave is
   studied here.  The 
amplitude $\mathcal{M}$   is
written with the hadronic $V-A$ current and  Dirac spinors  in the form
%\begin{widetext}
\begin{eqnarray}
%\label{amplitude}
\mathcal{M} = \int d^4xd{\vec k}_{\nu}
\,N\langle 0 |J_{V-A}^{\mu}(x)|\pi \rangle 
\bar{u}({\vec p}_l)\gamma_{\mu} (1 - \gamma_5)\nu({\vec k}_{\nu})\nonumber\\
\times e^{ip_l\cdot x + 
ik_\nu\cdot(x - \text{X}_\nu)
 -\frac{\sigma_{\nu}}{2}({\vec k}_{\nu}-{\vec p}_{\nu})^2}\label{V-A-interaction},  
\end{eqnarray}
%\end{widetext}
where 
$N=ig \left({\sigma_\nu/\pi}\right)^{\frac{4}{3}}\left({m_l m_{\nu}}/{
 E_l E_{\nu}}\right)^{\frac{1}{2}}$, and  the time $t$ is
 integrated in the region $\text{T}_{\pi} \leq t \leq \text {T}_{\nu}$. 
$\delta L_{int}={\partial \over \partial x_{\mu}}G^{\mu}$ in
Eq. $(\ref{total-derivative})$ is included in the amplitude
Eq. $(\ref{V-A-interaction})$, hence $P^{(d)}$  is 
neither  proportional to $m_{electron}^2$ nor suppressed.  
  The transition
probability 
to this final state is
written, after the spin summations are made, with the correlation
function and the neutrino wave function in the form 
 \begin{eqnarray}
\int  \frac{d{\vec
 p}_l}{(2\pi)^3} \sum_{s_1,s_2}|\mathcal{M}|^2 =   \frac{N_1}{E_\nu}\int d^4x_1 d^4x_2 
e^{-\frac{1}{2\sigma_\nu}\sum_i ({\vec x}_i-\vec{x}_i^{\,0})^2} \Delta_{\pi,l}(\delta x)
e^{i \phi(\delta x)},
\label{probability-correlation} 
\end{eqnarray}
where $N_1=g^2
\left({4\pi}/{\sigma_{\nu}}\right)^{\frac{3}{2}}V^{-1}$, $V$ is
a normalization volume for the initial pion, $\vec{x}_i^{\,0} = \vec{\text{X}}_{\nu} + {\vec
v}_\nu(t_i-\text{T}_{\nu})$, $\delta x
=x_1-x_2$, $\phi(\delta x)=p_{\nu}\!\cdot\!\delta x $
  and 
\begin{align}
\Delta_{\pi,l} (\delta x)=
 {\frac{1}{(2\pi)^3}}\int
 \frac{d {\vec p}_l}{E({\vec p}_l)}\left(2(p_{\pi}\cdot p_{\nu})( p_{\pi}\cdot p_l)-m_{\pi}^2 (p_l\cdot p_{\nu})\right)  
 e^{-i(p_{\pi}-p_l)\cdot\delta x }. 
\label{pi-mucorrelation}
%& &\delta t=t^1-t^2,\delta {\vec x}={\vec x}^1-{\vec x}^2.\nonumber
\end{align}

\begin{figure}[t]
\begin{center}
\includegraphics[scale=.4,angle=-90]{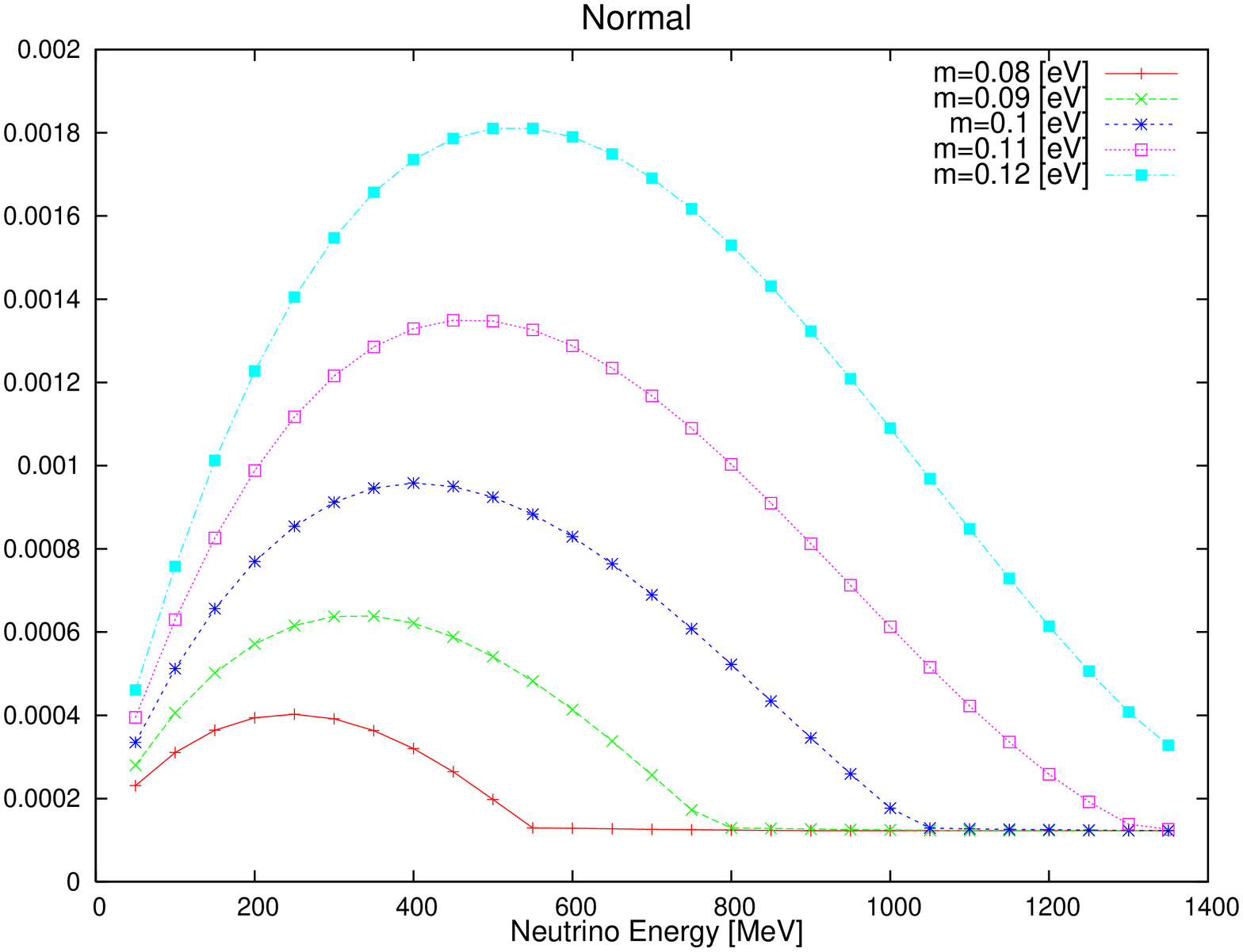}\\
\ \\
\includegraphics[scale=.4,angle=-90]{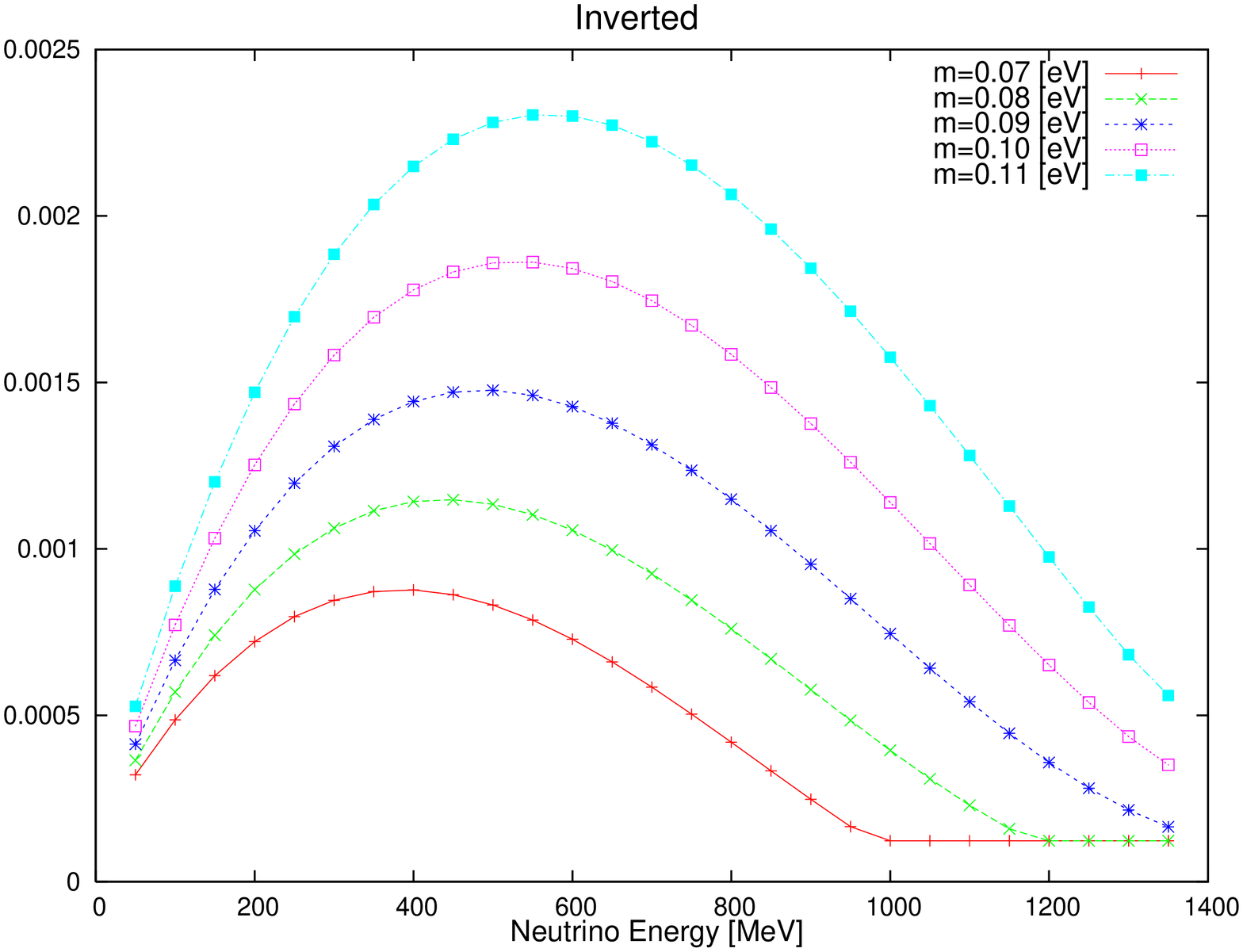}
\end{center}
\caption{Fraction of electron neutrino  from  a pion of 4 $[\text{GeV}/c]$ 
 in normal and inverted hierarchies at L=110\,[m], distance=170\,[m]. 
Pion's life time is included. 
The neutrino mass is 0.08(red-solid), 0.09(green-cross), 0.10(purple-cross), 0.11(pink-box), 
0.12(blue-box) $[\text{eV}/c^2]$, for normal hierarchy, 
and  0.07(red-solid), 0.08(green-cross), 0.09(purple-cross), 0.10(pink-box), 
0.11(blue-box) $[\text{eV}/c^2]$, for inverted  hierarchy.}
\label{normal 4GeV 1:figure}
\end{figure}

The 
probability of the event that  a neutrino of $p_{\nu}$ is detected at
${\vec X}_{\nu}$  is expressed 
as the sum of the normal term $G_0$ and the
diffraction term $\tilde g(\text{T},\omega_\nu)$, 
\begin{eqnarray}
%\label{probability-3}
P=N_2\int \frac{d^3 p_{\nu}}{(2\pi)^3}
\frac{p_{\pi}\! \cdot\! p_{\nu}(m_{\pi}^2-2p_{\pi}\! \cdot\! p_{\nu}) }{E_\nu}
 \left[\tilde g(\text{T},\omega_{\nu}) 
 +G_0 \right]
\label{probability-leptons}, 
\end{eqnarray}
where $N_2 = 8\text{T}g^2 \sigma_\nu$ and $\text{L} = c\text{T}$ is the
length of decay region. In $G_0$, the energy and momentum are conserved
approximately well and 
\begin{eqnarray}
p_l \approx p_{\pi}-p_{\nu}
\end{eqnarray}
is satisfied. Hence from a  square of the both hand sides, the mass shell
condition  
\begin{eqnarray}
m_l^2 \approx m_{\pi}^2-2  p_{\pi} \cdot p_{\nu}
\end{eqnarray}
is obtained. Thus the normal terms are proportional to the square of
lepton masses and the electron mode is suppressed \cite{Sakai-1949,Jack,Ruderman,Anderson}. 
 In $\tilde{g}(\text{T},\omega_{\nu})$, on the other hand, momenta satisfy
\begin{eqnarray}
p_l \neq p_{\pi}-p_{\nu},
\end{eqnarray}
and  the diffraction  terms are not proportional to the square of
lepton masses and the electron mode is not suppressed.

The total probability  of the events that    a neutrino or a charged lepton is detected  in the 
pion decay at macroscopic distance  is written 
 in  the form, 
%\cite{Ishikawa-Tobita-prl1,Ishikawa-Tobita-prl2}
\begin{eqnarray}
P= P^{(0)}+P^{(d)}_{lepton}.
\label{probability-lepton}
\end{eqnarray}
In Eq. $(\ref{probability-lepton})$,  $P^{(0)}$ is  the normal
term  that is obtained  from the decay probability $G_0$ in
Eq. $(\ref{probability-leptons})$ and $P^{(d)}_{lepton}$ is  the diffraction  term that
is determined from $\tilde g$ in Eq. $(\ref{probability-leptons})$. 
The former probability agrees to that obtained using the plane waves and
the  latter one has not been  included before and its effect is
estimated here.
The diffraction term  at  $\text{T}$ is described with  its mass
 and energy   in the universal form

\begin{eqnarray}
P^{(d)}_{lepton}/\text T=\frac{8}{15}g^2m_{\pi}^4\left(1-\frac{m_{\mu}^2}{m_{\pi}^2}\right)^4
\left(1+\frac{3m_{\mu}^2}{m_{\pi}^2}\right)
\frac{m_{\pi}^2
 \sigma_{\nu}p_{\pi}}{\text{T} m_{\nu}^2},
\label{diffraction} 
\end{eqnarray}
which  decreases slowly with a 
distance $\text{L}$ and  vanishes at infinite  distance. Hence at $\text{L}=\infty$,
the probability agrees with  the normal component, 
\begin{eqnarray}
P= P^{(0)} = \text{T}\Gamma.
\label{probability-lepton-as}
\end{eqnarray}
 The magnitude of
 $\tilde g(\text{T},\omega_{lepton})$ at the macroscopic distance is given in 
Fig.\,2. At $\text{L}=100\,[\text{m}]$,
 $E=1\,[\text{GeV}]$ for the mass
 $1\,[\text{eV}/{c^2}]\,(\nu)$, $0.5\,[\text{MeV}/{c^2}]\,(e)$, and $100\,[\text{MeV}/{c^2}]\,(\mu)$, the
 values are, 
\begin{eqnarray}
& &\tilde g(\text{T},\omega_{\nu} ) \approx 3, \nonumber\\
& &\tilde g(\text{T},\omega_e ) \approx 0, \nonumber \\
& &\tilde g(\text{T},\omega_{\mu} ) \approx 0.
%\label{fig:gtilde}
\end{eqnarray}
%%%%%%%%%%%%%%%% figure of \tilde{g} %%%%%%%%%%%%
\begin{figure}[t]
\begin{center}
\includegraphics[scale=.4,angle=-90]{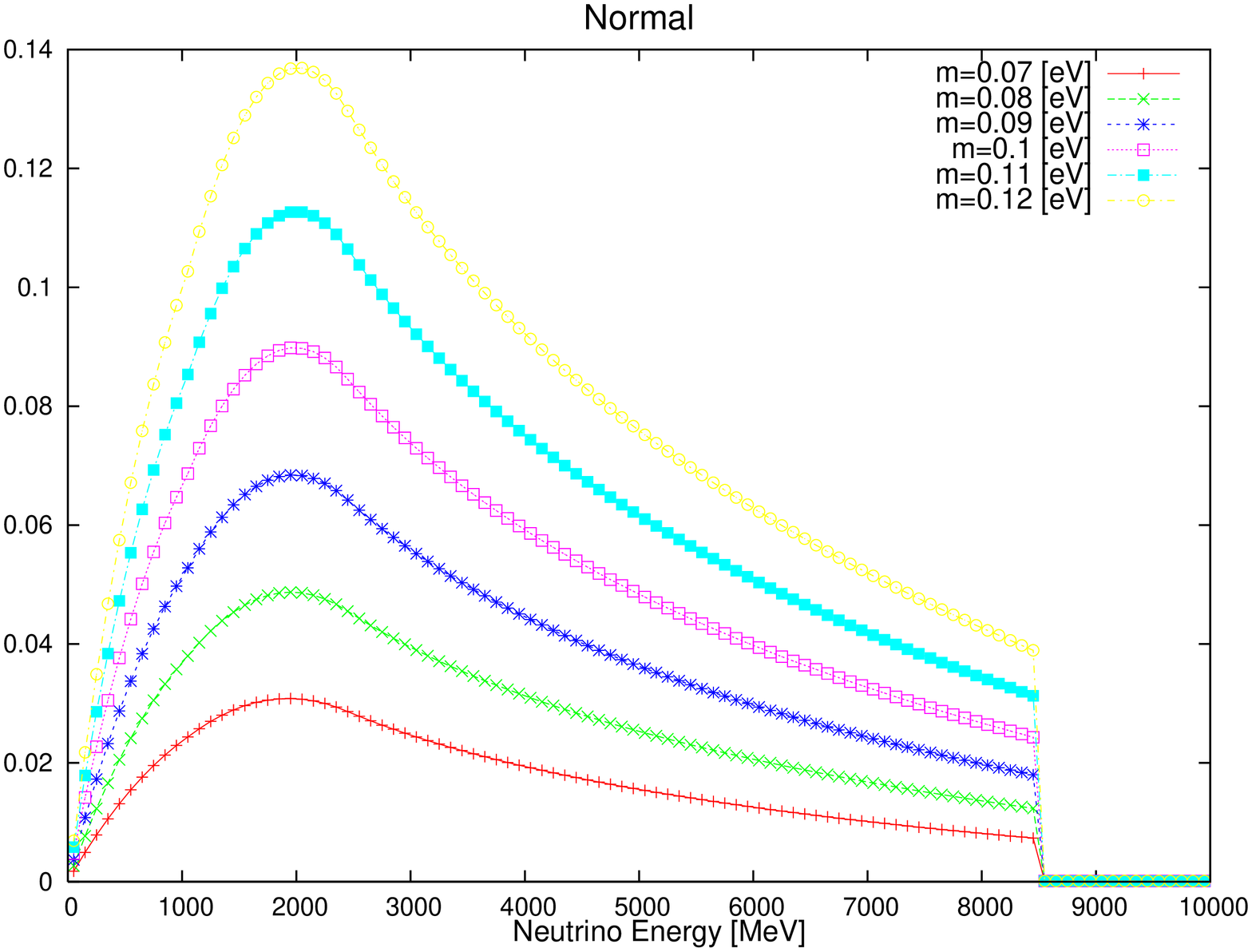}\\
\ \\
\includegraphics[scale=.4,angle=-90]{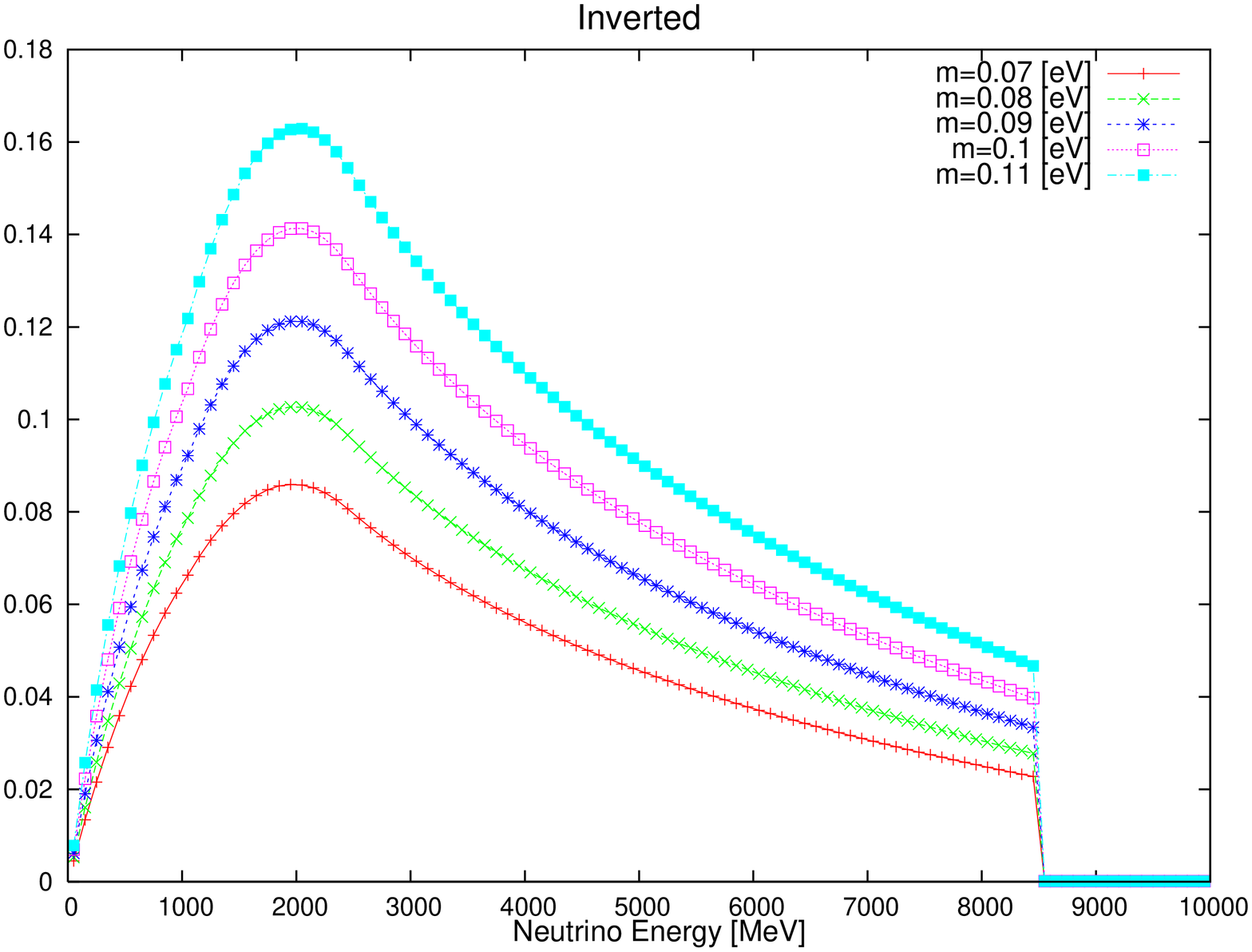}
\end{center}
\caption{Fraction of electron neutrino from  a pion of 20 $[\text{GeV}/c]$  in 
normal and inverted hierarchies at L=100\,[m], distance=200\,[m]. 
Pion's life time is included. 
The neutrino mass is 0.07(red-solid),
0.08(green cross),
0.09(purple cross),
0.10(pink-box),
0.11(bluebox),
0.12(yellow-cross) $[\text{eV}/c^2]$ \cite{ICARUS-NEAR}. }
\label{20GeV normal hyierlchy}
\end{figure}
In this region, they satisfy
\begin{eqnarray}
\tilde g(\text{T},\omega_l ) \approx {m_{\nu}^2 \over m_l^2} \tilde g(\text{T},\omega_{\nu} ), 
\end{eqnarray}
hence  the diffraction component at a macroscopic distance is finite in 
the neutrino and negligibly small in others. 
It
is striking that  the  probability of the event that the neutrino is detected  has an
additional 
term and  is not equivalent to  that of the charged lepton even though they are
 produced in the same decay process.  

 The conservation law of the kinetic energy is violated in $S[\text T]$, and 
results to  the unique finite-size correction expressed as $P^{(d)}$ to 
the electron mode. Pion does not decay to  massless Fermion and anti-Fermion 
in the weak $(V-A)\times(V-A)$ interaction when the kinetic energy, momentum, 
and angular momentum are conserved, and cause the suppression of the electron mode. This conditions 
hold  in $S[\infty]$ and $P^{(0)}$, but does not in $S[\text T]$ 
and $P^{(d)}$.
  Consequently the electron mode 
in  $P^{(d)}$ is not suppressed.   
    We study non-suppression of the electron mode and other implications 
derived from   $P^{(d)}$,   here.
In Fig. $\ref{LSND :fig}$, experiments of LSND \cite{excess-LSND} and the
two neutrino experiment( TWN) \cite{excess-two-neutrino} are compared with
the diffraction components and the flavor oscillations.  Theoretical
values are obtained including geometries of the experiments. Since those 
of LSND and TWN are different, the theoretical value for the LSND are
smaller than that for TWN. The experimental values plotted with crosses 
 agree with the theoretical values. The  values  from the flavor
 oscillations expected from  the current parameters are also shown. 
The mass-squared differences and mixing angles from the recent ground 
experiments lead negligible values for both experiments. A sterile 
neutrino of the  mass around $1$ $[{\text{eV}}/c^2]$ is necessary to fit 
the data of  LSND with the flavor oscillation. The agreements of 
the values from the neutrino diffraction in  LSND and TWN suggest that 
it is unnecessary to introduce  additional parameters.    

\begin{figure}[t]
\begin{center}
\includegraphics[scale=.4,angle=-90]{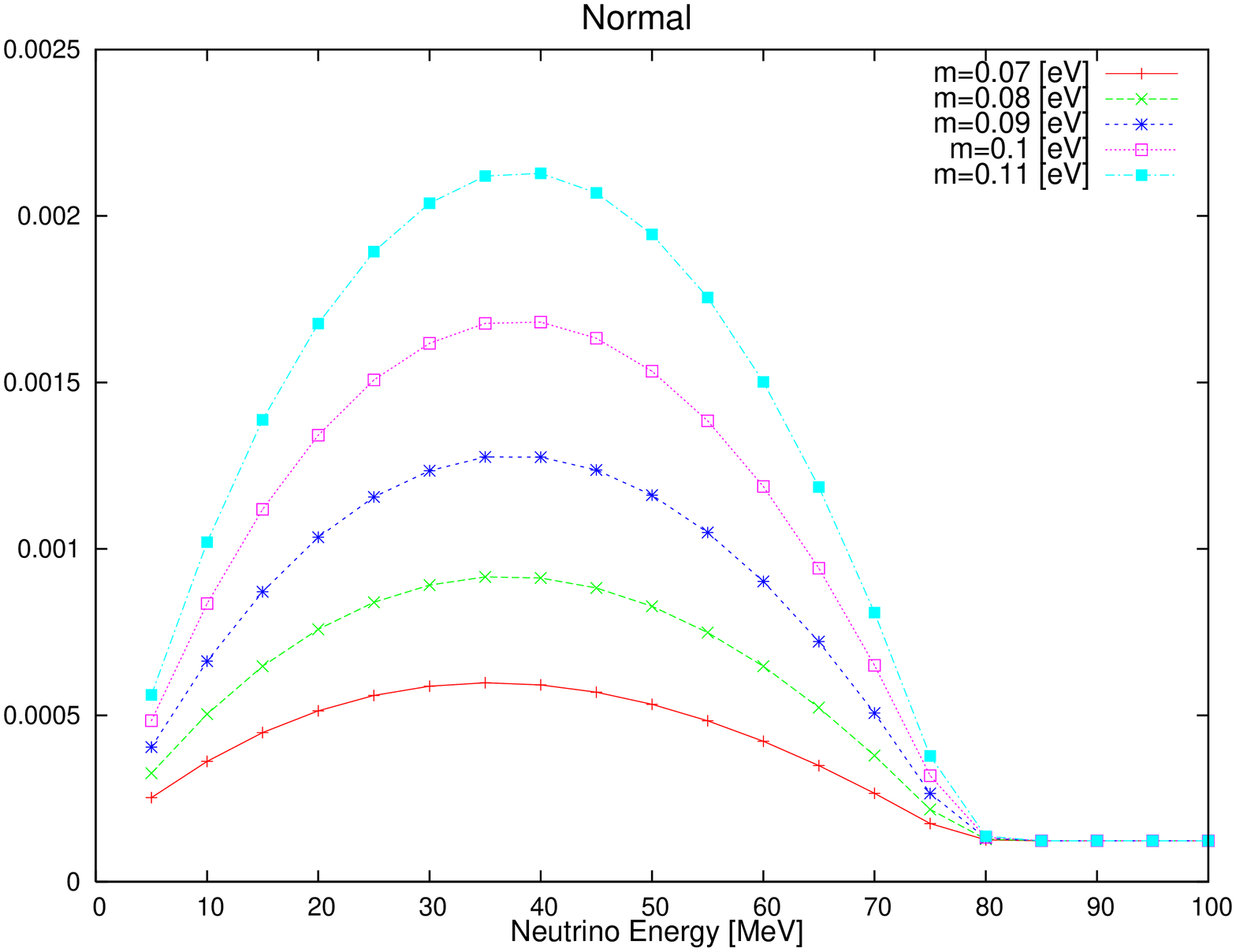}\\
\ \\
\includegraphics[scale=.4,angle=-90]{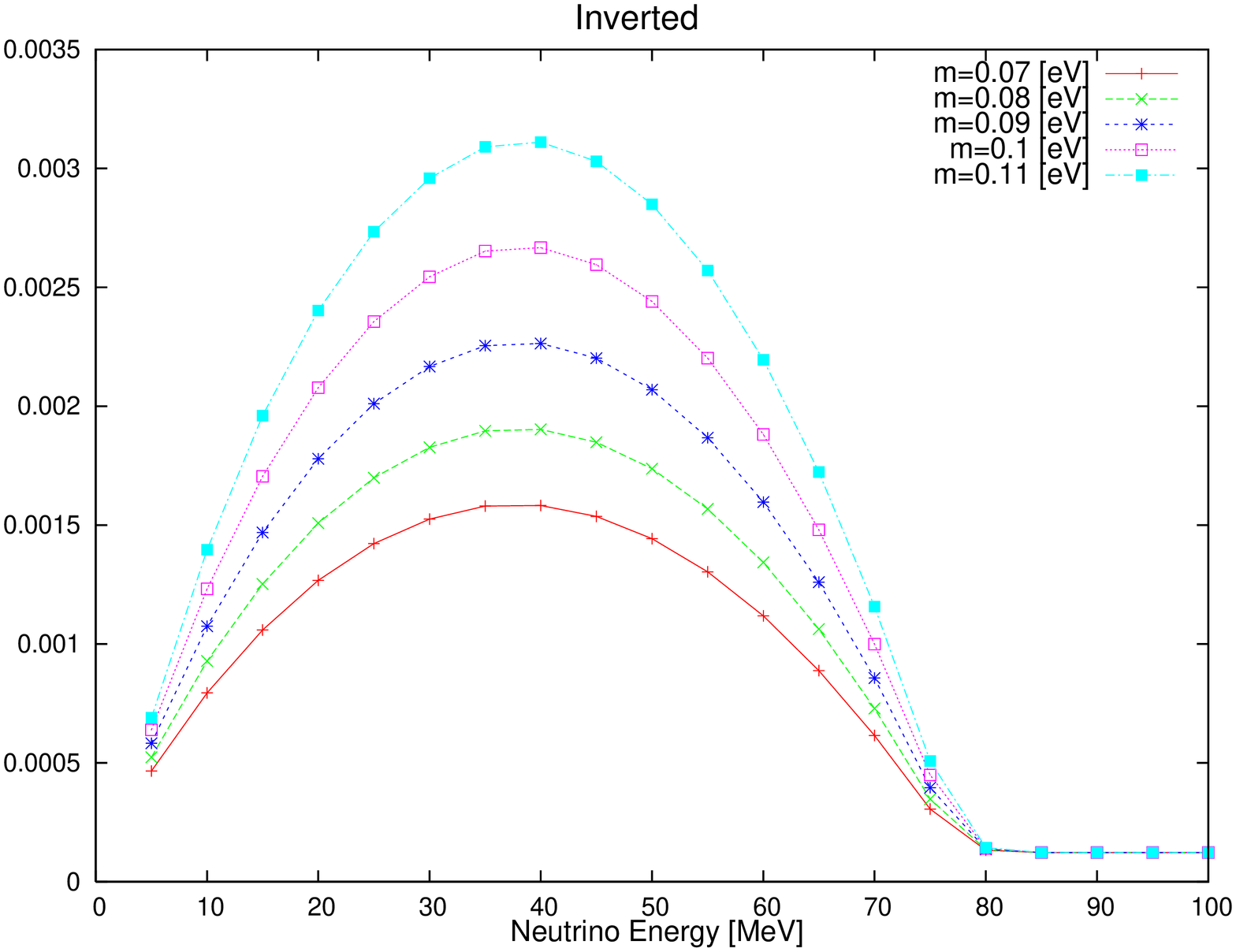}
\end{center}
\caption{Electron neutrino spectrum from  a pion of 2 $[\text{GeV}/c]$  in 
normal and inverted hierarchies at L=110\,[m], distance=50\,[m]. 
Pion's life time is included. 
The neutrino mass is 0.07(red-solid),
0.08(green cross),
0.09(purple cross),
0.10(pink-box),
0.11(bluebox),
0.12(yellow-cross) $[\text{eV}/c^2]$ \cite{ICARUS-NEAR}. }
\label{t2k-ground-50m}
\end{figure}

%\centering
In Fig. $\ref{T2K-confuguration}$, the maximum possible fraction of 
the electron neutrino in a geometry of T2K experiment is shown. The
 spreading of the pion beam is ignored in this Figure. Since the    
diffraction is sensitive to the pion beam spreading, the real value may
becomes smaller than this figure. In lower energy region of the pion, the fraction becomes extremely small.  
%\centering
%\centering
%\begin{figure}[t]
%\begin{center}
%\includegraphics[scale=.35,angle=-90]{t2k-4GeV-inv1.eps}
%\end{center}
%\caption{Elctron neutrino spctrum from  a pion of 4 $[\text{eV}/c]$  in 
%inverted hierarchy. Pion's life-time is included. The neutrino mass is 0.07-0.11 $[\text{eV}/c^2]$
% }
%\label{4GeV inverted hierarchy1:figure}
%\end{figure}
 Figure $\ref{normal 4GeV 1:figure}$ 
shows  electron neutrino spectra from  the pion of  life-time of this energy for the neutrino mass 0.08-0.12 of normal  hierarchy and for the mass  0.07-0.11 of inverted   hierarchy . The 
spectrum is sensitive to the absolute neutrino mass in this parameter regions. In lower energy region of the pion, the fraction becomes extremely small.      
%\centering
Figure $(\ref{20GeV normal hyierlchy})$ 
shows the electron neutrino spectra including the pion's life-time at higher pion 
energy for the neutrino mass 0.08-0.12 [eV/$c^2$] of normal  hierarchy and for 
the mass  0.07-0.11 [eV/$c^2$] of inverted   hierarchy. The fraction becomes larger and 
the spectrum is sensitive to the absolute neutrino mass. Figure $(\ref{t2k-ground-50m})$ shows the electron neutrino spectrum at lower energy. 
%%%%%%%%%%%%%%%%%%%%%%%%%%%%%%%%%%%%%%%%%%%%%%%%%%%%%%%%%%%%%%%%
\subsubsection{Electron(positron) enhancement}
%%%%%%%%%%%%%%%%%%%%%%%%%%%%%%%%%%%%%%%%%%%%%%%%%%%%%%%%%%%%%%
The finite-size correction  is small  in the event that an electron or positron
is detected if their wave packets  are of nuclear sizes. That  would become 
larger and non-negligible, if the wave packets for them  of much larger than 
nuclear sizes are used.
For a detector of having $\sigma_{e} \approx (10^{-8})^2\, [\text{m}^2]$, the electron from 
a  decay of pion has the finite size correction
\begin{eqnarray}   
\Gamma^{(d)} = {m_{\pi}^2 \sigma_{e} \over 4\pi} \left(1-{m_{\mu}^2 \over m_{\pi}^2}\right)^2
{E_{\pi} \over m_{e}^2 \text T}\Gamma,
\end{eqnarray}
in the high-energy region where the pion's life time is ignorable. 
In the  energy region for  the pion 
\begin{eqnarray}
\frac{E_{\pi}}{m_{\pi}c^2} > 10^{5},
\end{eqnarray} 
the life time is negligible at $c\text{T} \leq 4\times 10^5$\,[m],
and $P^{(d)} $ gives a significant effect. An excess of positron from high energy positive pion
 would be observed.  

\subsection{Proton target  anomaly}
Magnitude of  diffraction component depends upon the size of the nucleus
which neutrino interacts with and is expressed by the wave packet size
$\sigma_{\nu}$. It becomes larger with the larger target. It is known
and used in the text that nuclear size is proportional to $A^\frac{2}{3}$ and 
the large $A$ nuclear gives a large diffraction component, generally. Proton has a 
smallest intrinsic size. However    a proton is expressed by a wave function of 
its  position in matter. So  the wave packet size is determined by a size
of this wave function. Since a proton is the lightest nucleus, it has
the largest size. We estimate this size using center of mass gravity effect
between proton and electron.
For the proton's mass $m_{p}$ and the electron mass $m_{e}$, an  
electron's coordinate
${\vec x}_{electron}$ and the proton  coordinate ${\vec x}_{p}$
are expressed as,
\begin{eqnarray}
& & {\vec x}_{electron}={\vec X}+{m_{p} \over
 m_{e}+m_{p}}{\vec r}\approx {\vec X}+\left(1 +{1 \over 2000}\right){\vec r},\\
& &{\vec x}_{p}={\vec X}-{m_{e} \over
 m_{e}+m_{p}}{\vec r}\approx{\vec X}-{1 \over 2000}{\vec r}.
\end{eqnarray}
If the wave function of the atom is 
\begin{eqnarray}
\Psi({\vec R}) \varphi({\vec r}),
\end{eqnarray}
and the function of the relative coordinate, $\varphi({\vec
r}\,)$, is extended
by an amount $R_{atom}$ which is about $10^{-10} [\text{m}]$ then the proton 
is extended with a radius  
\begin{eqnarray}
R_{p}={1 \over 2000}R_{atom}\approx 5\times 10^{-14} \,[\text{m}].
\label{center-mass-gravity}
\end{eqnarray} 
This  value is much shorter than the atomic scale and is larger than one 
nucleon's size  $1\,[\text{fm}]=10^{-15}\,[\text{m}]$ by factor $50$. 
Thus proton  in  solid  is  extended to  the size ${1 \over
2000} $ of the atomic wave
  function, which is larger than  the nuclear size  of ${}^{16}O$. 
Hence proton  gives the important role in the neutrino diffraction. 
Its size may be  in the range 
\begin{eqnarray}
l_{proton}(U)=5\times 10^{-14}-10^{-13} \,[\text{m}].  
\label{proton-size}
\end{eqnarray}

An enhancement of  diffraction contribution due to  the proton 
is expected in 
\begin{eqnarray}
\bar \nu+p \rightarrow \mu^{+} +\mathbf{X},\ \mathbf{X}=n,\ p\pi^{-},\ n\pi^{0}\text{ and }others.
\end{eqnarray}
           
%%%%%%%%%%%%%%%%%%%%%%%%%%%%%%%%%%%%%%%%%%%%%%%%%%%%%%%%%%%%%
\subsection{Atmospheric neutrino}
%%%%%%%%%%%%%%%%%%%%%%%%%%%%%%%%%%%%%%%%%%%%%%%%%%%%%%%%%%%%
The neutrino flavor oscillation was  found first with  an atmospheric
neutrino. Neutrinos are produced from decays of  charged pions and muons 
in secondary cosmic rays.  Since the matter density is low in
atmosphere, these 
charged particles travel freely long distance. Thus  neutrinos produced in 
decays of pions or muons   show the diffraction phenomenon and the 
diffraction components  are added to
the neutrino fluxes. These neutrino events  may be  observed  in 
detectors set in the ground, such as Super-KamiokaNDE(SK) if the absolute 
mass is a reasonable value. The minimum mass  allowed 
 from the mass-squared difference is about the value, 
$\sqrt {\delta m^2} \approx 10^{-2} \,[\text{eV}/c^2]$. Then  the length that the
diffraction component is observed becomes  
$\text{L}_0={2E_{\nu} c \over m_{\nu}^2 } \approx 20\,[\text{km}]$ for $E_{\nu}=1 \,[\text{GeV}] $, which is longer than
the height of troposphere. Hence the diffraction 
component  could  be observed  with  the angle-dependent excess of  the 
electron and muon neutrino fluxes. Since the diffraction components from 
pion decays are common to  both  neutrinos,
their ratio is not sensitive to the diffraction.  Instead of this ratio, 
a ratio of the neutrino flux to the flux of charged leptons is good to
see the signal of the neutrino diffraction. 
%%%%%%%%%%%%%%%%%%%%%%%%%%%%%%%%%%%%%%%%%%%%%%%%%%%%%%%%%%%%%%%%%%%%%
\section{  Unusual properties from $S[\text T]$  }
%%%%%%%%%%%%%%%%%%%%%%%%%%%%%%%%%%%%%%%%%%%%%%%%%%%%%%%%%%%%%%%%%%
Unique properties of $S[\text T]$ expressed in Section 2, lead  various 
unusual properties to observables. 
%%%%%%%%%%%%%%%%%%%%%%%%%%%%%%%%%%%%%%%%%%%%%%%%%%%%%%%%
\subsection{ Unitarity}
%%%%%%%%%%%%%%%%%%%%%%%%%%%%%%%%%%%%%%%%%%%%%%%%%%%%%%
Probability of the event that  the neutrino is detected per unit time  $P(\text{L})$ decreases 
with the distance L. This behavior   appears to  suggest   
that the probability is not preserved and is inconsistent with the
unitarity, if $S[\infty]$ is applied. However at T, the probability is derived 
from  $S[\text T]$  that satisfies $S^{\dagger}[\text{T}]S[\text{T}]=1$, 
Section 2.4, and is consistent with the unitarity.  The probability at L 
is determined with 
S-matrix $S[\text T],\ \text{L}=c\text{T}$ and 
 has two components
$P=P^{(0)}+P^{(d)}(\text{L})$. Both terms are positive semi-definite and 
the latter is the finite-size correction that decrease  with L.  This  
behavior is a natural consequence of
the wave nature of the states at finite T  and  is consistent 
with the unitarity.  The unitarity leads that the life time
of the pion becomes larger if the neutrino is detected at a finite T.
%%%%%%%%%%%%%%%%%%%%%%%%%%%%%%%%%%%%%%%%%%%%%%%%%%%%%%%%%%%%%%%%%%%
\subsection{ Lepton number non-conservation}
%%%%%%%%%%%%%%%%%%%%%%%%%%%%%%%%%%%%%%%%%%%%%%%%%%%%%%%%%%%%%%%%%%
The probability of the event that the neutrino is detected is different from 
that of  the charged lepton even though they are produced in pair.  They propagate 
with different velocities and different wave lengths along the 
light-cone. Consequently they  have different retarded effects and are 
 detected with different probabilities at finite T. The probabilities  from 
$S[\text T]$ depend on boundary conditions, which differ  in both cases, 
Eq. $(\ref{symmetry-breaking})$, and the two probabilities are different.

If the neutrino and charged lepton are observed simultaneously, they are 
expressed by $S[\text T]$ of one boundary condition, and the 
charged lepton shows the same behavior.  Such an experiment is not easy 
and has not been made. 

The charged lepton has small finite-size corrections  if the sizes of 
wave packets are almost the same.  The sizes in detectors of ordinary 
experiments belong to this and  the finite-size corrections are negligible, and 
the   probability is computed with $S[\infty]$ using the plane waves. 
This situation has been studied well experimentally and
agrees with the theoretical calculations obtained with $S[\infty]$. 

Now the boundary conditions
of the above two cases are different. 
One  boundary condition  leads 
unique  probability   and the different boundary conditions may lead
different probabilities. The fact that those of the neutrino from $S[\text T]$ is different from those of the charged leptons from $S[\infty]$, is a natural consequence. 
It is meaningless to compare the probability for neutrino  
in the first case  with that for the charged lepton in the
second case, because they follow the different boundary conditions.

Decay probabilities  computed at $\text{T}=\infty$  agree with  the 
probability of the event that the decay products are detected at $\text{T}=\infty$. 
If the finite-size correction is finite in the neutrino and vanishes in the 
charged lepton,
the probabilities of the events that they are detected   become asymmetric, even 
though they are produced in pair.  
The fact that the probability for the neutrino at T is larger  
than  that of the charged lepton  does not mean the violation of the 
lepton number conservation.
Because  the neutrino propagates with almost the light speed, 
the probability of the event is enhanced in a similar manner as the retarded 
electric potential of a moving charged body.

%%%%%%%%%%%%%%%%%%%%%%%%%%%%%%%%%%%%%%%%%%%%%%%%%%%%%%%%%%%%%%%%%%%
\subsection{ Dependence on wave packet size}
%%%%%%%%%%%%%%%%%%%%%%%%%%%%%%%%%%%%%%%%%%%%%%%%%%%%%%%%%%%%%%%%%%
It is known that the total probability at $\text{T}=\infty$ does not
depend on the
wave packet size \cite{Stodolsky}. The result of the present paper
Eq. $(\ref{total-probability-energy})$ in 
fact shows that  the  asymptotic value, the first  term in the 
right-hand side,   is independent 
of the wave packet size. Now the finite-size
correction, the second term in
Eq. $(\ref{total-probability-energy})$,   is proportional 
to $\sigma_{\nu}$.   Since $S[\text T]$ is
determined with the boundary condition  Eqs. $(\ref{expansion-wave-packets}),(\ref{bounday-LSZT1}),\text{ and}~ (\ref{bounday-LSZT2})$  that depend on 
$\sigma_{\nu}$,  the finite-size correction 
  depends on $\sigma_{\nu}$. That  increases with 
$\sigma_{\nu}$   and diverges at $\sigma_{\nu}=\infty$. The 
diverging correction  at $\sigma_{\nu}=\infty$ is consistent, in fact, 
with the divergence  of Eq. $(\ref{diverging-integral})$,
and  the fact that the total cross section diverges for
the plane waves \cite{Asahara} obtained without the damping factor 
$e^{-\epsilon |t|}$.  The latter divergence occurs because the denominator of 
the neutrino propagator vanishes, which is connected with the boundary 
condition. The finite-size correction   has the 
universal properties,  despite of the $\sigma_{\nu}$ dependent behavior. 

%%%%%%%%%%%%%%%%%%%%%%%%%%%%%%%%%%%%%%%%%%%%%%%%%%%%%%%%%%%%%%%%%%%%
\subsection{Non-conservation of kinetic energy} 
%%%%%%%%%%%%%%%%%%%%%%%%%%%%%%%%%%%%%%%%%%%%%%%%%%%%%%%%%%%%%%%%%%%%%
\subsubsection{Violation of symmetries}
 $S[\text T]$ does not commute with the free Hamiltonian $H_0$,
and satisfies 
 Eqs. $(\ref{commutation-relation-S(T)  })$ and $(\ref{momentum-non-conservation1})$. Particularly if the parent and 
  decay products overlap,   $H_{int}$ has a finite expectation value and 
the kinetic energy  is different from that of $H$. 
The kinetic energy is not  conserved, despite the fact that the total 
energy  is conserved. The transition probability from these  states 
was computed  analytically, and  exhibits  the  diffraction 
in   the finite-size correction. 

The finite-size correction is not invariant under Lorentz transformation
  and the magnitude of the finite-size corrections depends on the systems.
  
\subsubsection{Helicity suppression}
 
Suppression of the branching ratio of electron mode over that of 
the muon mode is due to  the helicity suppression, which is caused by 
conservation law of the kinetic energy and angular momentum. The 
helicity suppression hold in $\Gamma$ of the decay of a pseudo-scalar 
particle to a neutrino and charged lepton caused by $V-A$ weak
interaction. Now $S[\text T]$ violates the conservation law  of the kinetic energy  and the angular
momentum. $P^{(d)}$ comes  from the kinetic-energy non-conserving
states, and is not suppressed in the electron mode. 
The finite-size corrections to the probabilities  of the events that  the 
electron neutrino  are almost the same as those of the muon mode, 
and give the  dominant contribution at small T.
Thus when the neutrino is observed in suitable near-detector region, the
electron neutrino is substantially enhanced.

\subsubsection{Large finite-size correction} 

The finite-size correction vanishes with the use of $S[\infty]$  but 
becomes finite with $S[\text T]$. We study the amplitudes in tree level and 
identify the reason why  $S[\text T]$ gives finite corrections.

$S[\infty]$ is Poincar\'{e} invariant
and  is expressed with Feynman diagrams in perturbative expansions.
The energy and momentum of initial states are given and those of final 
and  intermediate states are limited from exact conservation at  each 
vertex. In  $S[\infty]$,   the states of infinite  energies do not couple.  
The two point functions are short range and the light-cone singularity 
does not couple.

Now  $S[\text T]$ is   Poincar\'{e} non-invariant
and   final and  intermediate states of unlimited kinetic-energy and momentum
can couple and produce the  light-cone singularity.
The
light-cone singularity is the real function and extended to a large area and gives 
the universal finite-size  correction to light particles. 
It
is  remarkable that the states of the ultraviolet region give the
observable effect to the probability of  the tree diagram.

The decay rate of the pion becomes different if 
the neutrino is detected in the region of the finite-size correction. The 
life time  becomes shorter   than the 
normal value. 
This phenomenon that the life time is modified by  its 
interaction with matter is known in  the literature as  quantum Zeno
 effect. 
Neutrinos actually interact extremely weakly 
with matter,  and  a majority of  neutrinos are passing
 freely without any interaction  and are not  affected by this effect.  
Consequently  the majority of the pions are  not affected by  the  
finite-size  effect and its life
 time is not modified and has the normal life time. Although the detected
 neutrino receives the large finite-size correction, its effect is
 negligibly small for  observables of the pions.

%%%%%%%%%%%%%%%%%%%%%%%%%%%%%%%%%%%%%%%%%%%%%%%%%%%%%%%%%%%%%%%%%%%%
\subsection{Overlap of wave functions }
%%%%%%%%%%%%%%%%%%%%%%%%%%%%%%%%%%%%%%%%%%%%%%%%%%%%%%%%%%%%%%%%%%
\begin{figure}[t]
\begin{center}
\includegraphics[scale=1.2,angle=0]{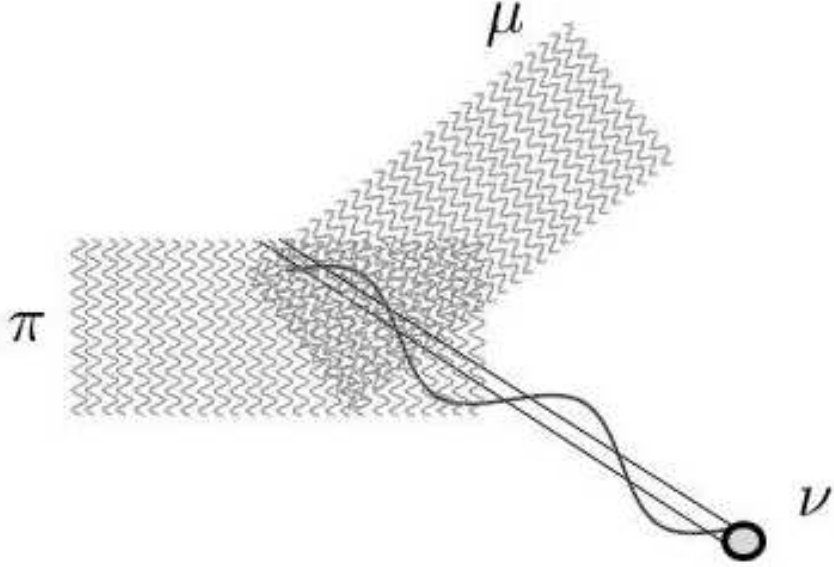}
\end{center}
\caption{Overlap of large waves  of pion and muon  and a small 
wave  of a neutrino. They overlap in long  area, and the rate is 
computed with  $S[\text T]$ and has  
the large finite-size correction. The neutrino wave along the light-cone 
has a large wave length Eq. (\ref{neutrino-light-cone}), and give the finite-size correction of macroscopic size.  
  }
\label{overlap-confuguration1}
\end{figure}

%\centering
\begin{figure}[t]
\begin{center}
\includegraphics[scale=.7,angle=0]{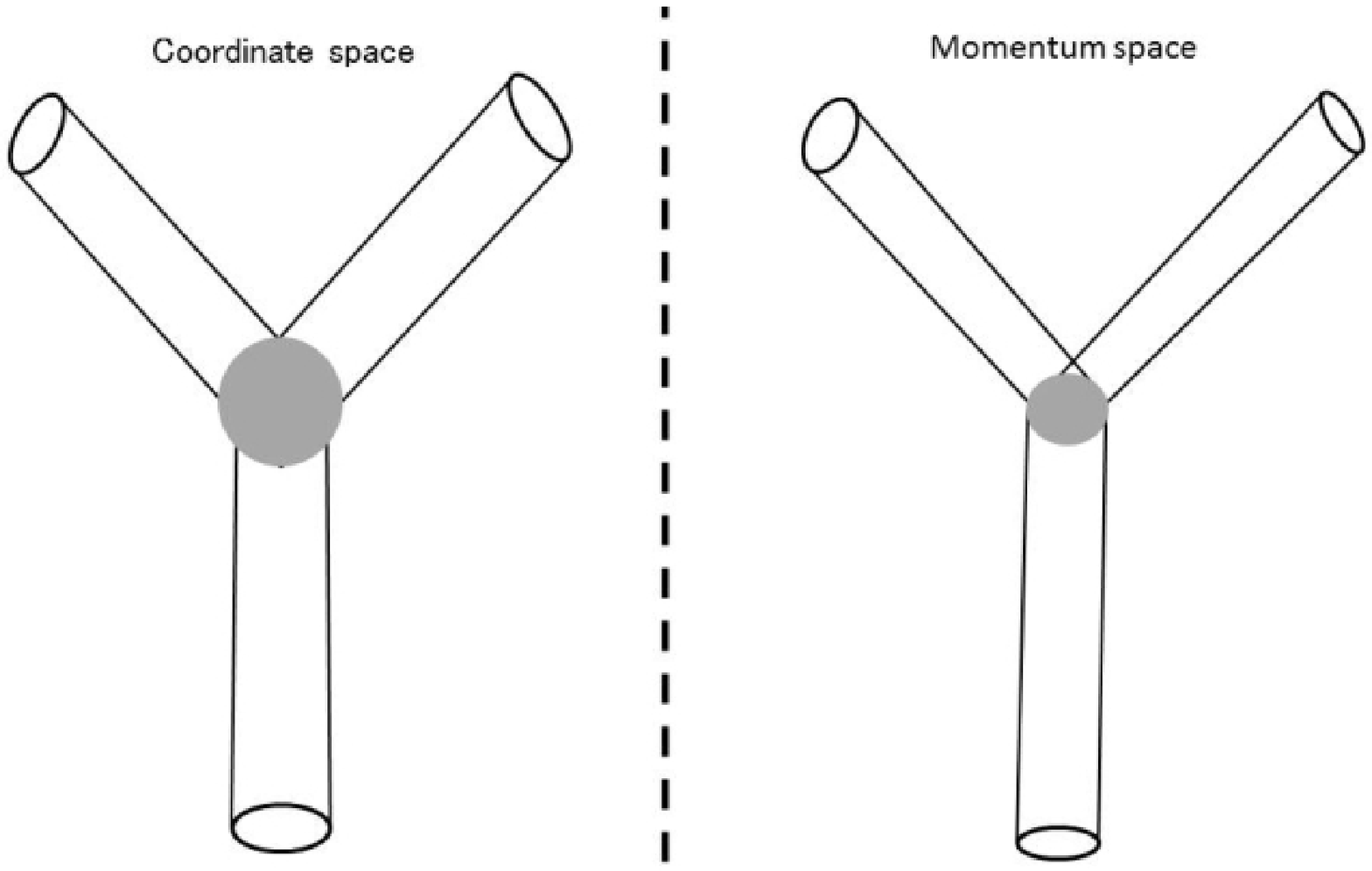}
\end{center}
\caption{Overlap of wave functions of finite  sizes in momentum and coordinate. 
 Wave functions overlap in small regions and the finite-size correction 
vanishes. $S[\text T]$ agrees with $S[\infty]$. }
\label{overlap-confuguration2}
\end{figure}

%\centering

The  present diffraction of  transition amplitude appears  when  
  the wave of parent overlap with the waves of  daughters of varying kinetic 
energy. The  neutrino  
is detected by the final states that are produced in the neutrino collision in the 
detector and its wave function is determined by   the apparatus. 
Accordingly  the diffraction pattern and  the finite-size correction depend on the 
wave function that the neutrino interacts. This is a unique 
property of quantum physics.
 In classical physics,  physical variables are observable and 
interference patterns do not depend on the apparatus.

The present phenomena appear always when the wave functions that retain the 
coherence overlap in wide area.  
Overlap of wave functions  of various spatial sizes  are presented in the 
following figures.  The 
plane waves and small wave packet in  particle decays  are shown  
in Fig. (\ref{overlap-confuguration1}). 
Short range fluctuations of the correlation functions overlap in the 
microscopic  region  and  give the constant 
probabilities which agree with those at  the asymptotic regions. Now the long 
range fluctuation 
expressed by the light-cone singularity  in $S[\text T]$ 
 extends to macroscopic area, and gives the long distance 
  effects to  light particles and  gives the short distance effect to massive 
particle. The angular velocity $\omega$ along the light-cone is given as 
\begin{eqnarray}
\omega t =(E({\vec p}\,)-c|\vec{p}\,|)t=\frac{m^2}{2|\vec{p}\,|}t,\label{neutrino-light-cone}
\end{eqnarray}
and becomes extremely small for neutrino. Hence the probability decreases 
slowly as $\text{T}_0/\text{T}$, $\text{T}_0={2|\vec{p}\,|}/{m^2}$.  
 The light-cone singularity appears always in the many body correlation 
functions in tree levels, hence the finite-size corrections always  
appear in the tree levels. But the magnitude is inversely proportional to $m^2$,
and the corrections become significant only in light particles  

\begin{figure}[t]
\begin{center}
\includegraphics[scale=.7,angle=0]{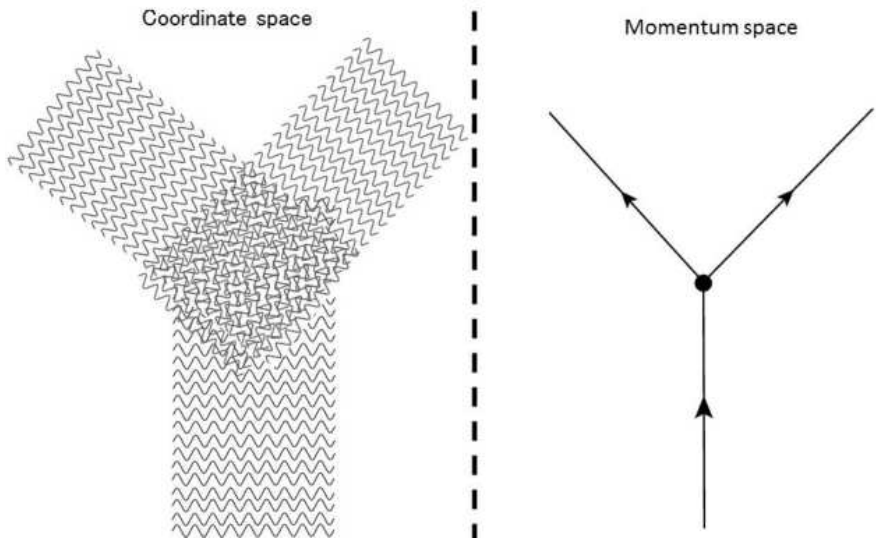}
\end{center}
\caption{Overlap of wave functions of large sizes in coordinate space becomes those of small sizes in momentum space. Wave functions overlap in small regions and $S[\infty]$ with the $i \epsilon$ prescription is applied.
  }
\label{overlap-confuguration3}
\end{figure}

Overlap of wave functions of other sizes are shown  in Figs.\,(\ref{overlap-confuguration2}) 
and  (\ref{overlap-confuguration3}), in configuration  and momentum space. In Fig.\,(\ref{overlap-confuguration2}) they overlap in small region and get 
contributions from short-range fluctuations, and  are studied  with either 
$S[\text T]$ or  $S[\infty]$ of the $e^{-\epsilon |t|}$ prescription. They give equivalent probabilities.
In Fig.\,(\ref{overlap-confuguration3}), the parent and daughters overlap 
in wide area in   configuration space but they overlap in small region 
in momentum space. Accordingly the asymptotic values  are treated with  
$S[\infty]$ with the damping factor $e^{-\epsilon |t|}$ 
or the value $ \displaystyle{\lim_{\sigma \rightarrow \infty} \left[\lim_{\text T
\rightarrow \infty} P(\text T,\sigma)\right]}$ obtained from the configuration
of Fig. 22.
%%%%%%%%%%%%%%%%%%%%%%%%%%%%%%%%%%%%%%%%%%%%%%%%%%%%%%%%%%%%%%%%%
\section{Summary and implications}
%%%%%%%%%%%%%%%%%%%%%%%%%%%%%%%%%%%%%%%%%%%%%%%%%%%%%%%%%%%%%%%%
The probability of the events that the final states are
detected at T  is computed with $S[\text T]$ that 
satisfies the boundary  conditions at T.  $S[\text T]$ was formulated and 
applied to the  pion in nucleon collision and the neutrino in pion
decay. The probability is modified  
from the asymptotic value $\Gamma \text T$ and has the correction  
$P^{(d)}$. The 
correction violates the 
Poincar\'e symmetry of the    Lagrangian and gives  the large
enhancement in the 
electron neutrino, which satisfies  $\Gamma \approx 0$  by the helicity 
suppression. 
Due to $P^{(d)}$, the probability  of the events for  the neutrino is 
different from that   for the charged particle. 
This deviation is caused by the mass difference between the neutrino and
charged lepton, and by  the large distance between the position 
of detection and   that of  production. The waves of light particle have
the same velocity and cause  the interference of waves to be constructive,
but those of   massive particle have varying  velocity  that depends on
the momentum and cause the interference not to be  constructive.

The probability  amplitude at  $\text T \leq \tau$ reveals 
the wave nature  similar to  but different from classical waves around a 
disorder or obstacle. The states   in the overlapping region have   
non-constant  kinetic energies, and show diffraction of waves. They modify 
the transition  amplitude and probability and result  in the finite-size 
corrections.
The probability of the event that  decay products  are detected  is    
decomposed to the normal term
$\Gamma T$ and the finite-size correction $ P^{(d)}(\propto C_0(\sigma) \tilde g(\text{T},\omega))$
 that gives the  1/T correction. 
The  correction has   a universal form $\tilde g(\text{T},\omega)$ of 
the magnitude  proportional to ${E_{\nu} \over m_{\nu}^{2} \text T}$, which  
  is extremely large  and becomes observable with  
macroscopic experiments  for a light particle.  

  $S[\text T]$ are formulated with wave packets. $\Gamma$ does not
  depend on the wave packets but   $P^{(d)}$
  depends.  Hence the wave packets can not be replaced with plane
  waves.    Constructively added waves of the pion and lepton in the 
overlapping region  form    
the  light-cone singularity in the correlation function, and its overlap   
with the neutrino wave gives the large 
finite-size correction to the probability. 
  $P^{(d)}$  appears in vacuum and is  determined by the 
fundamental physical quantities of the Lagrangian.  The origin, mechanism, 
characteristic features, and 
implications are presented. 

The modified rates in the pion decays   were  compared with  previous neutrino 
experiments in Section 7. 
First, the slight energy dependence of the total $\nu N$ cross
sections at high-energy regions,
which is hard to understand in the standard theory,  was
shown to agree with the excess of the effective neutrino flux due to the 
diffraction.
 The excesses of neutrino events  will be observed in other reactions as
well at  macroscopic short distance regions. 
Theoretical calculations 
at distances of the order of a few hundred meters were 
computed and shown in
Figures of Section 6.  From these figures, the excesses are  not 
large but are sizable magnitudes. Hence  these excesses  shall 
be observed in these  distances. 
Actually fluxes measured in the near detectors of the long-baseline 
experiments of K2K \cite{excess-near-detectorK2K} and MiniBooNE
\cite{excess-near-detectorMini}  may show  excesses of about
$10-20$ percent of the Monte Carlo estimations. Monte Carlo estimations of
the fluxes are obtained using naive decay probabilities and do not have
the interference effects we presented in the present work. So the excesses of
these experiments may be related with the excesses due to
interferences. The excess is not clear in MINOS 
\cite{excess-near-detectorMino}. With more
statistics, quantitative  analysis  might become  possible to test the new 
universal term on the neutrino  flux at the finite distance.   
 If the mass is in the range from $0.1\,[\text{eV}/c^2]$ to $2\,[\text{eV}/c^2]$, the 
near detectors  at T2K, MiniBooNE, MINOS and other experiments  might 
be able to measure these signatures. 

Second, the fraction of electron  mode  modified by $P^{(d)}$  was compared. Since the exact conservation law of kinetic-energy 
and momentum  does not hold in the overlapping region, $P^{(d)}$
 violates   the  helicity suppression. Thus the corrections 
preserve the universality of weak interaction, and  are about the same for 
both modes. Consequently,    the electron mode is enhanced drastically at a finite T.  The theoretical value of  the fraction of  electron mode 
   was
compared with LSND and TWN, and agreements were obtained.  Further
confirmation of the diffraction component by observing the electron 
neutrino in pion decay will be made using modern version of LSND or similar
experiments. T2K near detector is a possible place for that. 

Third, anomalies  
in proton target and atmospheric neutrino were pointed out. They would 
supply also specific signature of the neutrino diffraction.   
  The neutrino diffraction is sensitive to the absolute neutrino
mass but is not so to other parameters such as pion and neutrino
energies. Hence the observations of the neutrino diffraction is easy, and 
may give the  absolute neutrino mass.

We summarize the reasons why the interference term of the 
long-distance behavior emerges in the neutrino but not in the charged lepton. 
The probability of the event that the decay product are detected at T, 
 Eqs. $(\ref{pion-probability})$ and $( \ref{probability-correlation1})$, shows that their behaviors at large T are determined by 
 the light-cone singularity 
of  the correlation function $\Delta(\delta t, \delta\vec{x})$ and the
wave function of detected particle.
  From relativistic invariance,  the particle's momentum is unlimited and 
   the singularity near  $\lambda{=c^2\delta t}^2-{\delta \vec x}^2=0$,
which is extended to  large distance $|\delta {\vec x|} \rightarrow
\infty$, emerges in  $\Delta(\delta t, \delta\vec{x})$.  
The neutrino wave function  along the light-cone region behaves as 
\begin{eqnarray}
& &\psi_{\nu}(\delta t,{\delta \vec x})=\mathcal{F}\frac{e^{i(E_{\nu}\delta t-{\vec p}_{\nu}\cdot{\delta \vec x})}}{x}
= \mathcal{F}{e^{i{m_{\nu}^2 \over
2E_{\nu}}\delta t}\over c\delta t},
\end{eqnarray}
where $\mathcal{F}$ has no dependence on the distance $|\vec{x}|$.
Consequently the
 relation between the energy width $\delta E$ and the time
interval $\delta t$ becomes
\begin{eqnarray}
\delta t \delta{m_{\nu}^2 \over 2E_{\nu}} =\delta t \delta E 
\times{1 \over 2} \left({m_{\nu} \over E_{\nu}}\right)^2 \approx \hbar,\end{eqnarray}
and 
\begin{eqnarray}
\delta t \delta E \approx 2 \left(\frac{E_{\nu}}{m_{\nu}}\right)^2 \hbar k.
\end{eqnarray}
The ratio $({E_{\nu}/m_{\nu}})^2$ is of order $10^{18}$ 
and $\delta t$ becomes macroscopic  even for the energy width
$\delta E$  of 100 [MeV].  Then $c \delta
t$ becomes 
\begin{eqnarray}
c \delta t  \approx 10^3 \,[\text{m}],
\end{eqnarray} 
 which is about the distance between the pion source and the near 
detector in fact. So  the interference effect 
appears in this distance and is observable using the apparatus of much
smaller size.     Now the lepton wave function behaves in the same region as,
\begin{eqnarray}
& &\psi_{l}(\delta t,{\delta \vec x})=\mathcal{F}\frac{e^{i(E_{l}\delta t-{\vec p}_{le}\cdot{\delta \vec x})}}{x}
= \mathcal{F}{e^{i{m_{l}^2 \over
2E_{l}}\delta t}\over c\delta t},
\end{eqnarray}
where $m_l$ is much larger than $m_{\nu}$ and $c\delta t$ is a microscopic size for charged leptons.
 
In  time interval ${\text T} \leq l_0/c$,  $P^{(d)}$  
gives  finite corrections, and the  probability of the event that  the
neutrino is detected  deviates from the 
probability of the event that the neutrino is produced.  
  In another region  $l_0/c \leq {\text T}, \tau_{\pi}$,  the parent 
disappears  and does not overlap with  daughters. 
 The decay products have the initial energy, and the  neutrino behaves 
like a free isolated particle. 
The diffraction term vanishes and  the probability of the event for the 
detection  is computable with $S[\infty]$ and agrees with the probability of the event  for the production. 
 At $T \gg \tau_{\pi}$, the corrections  vanish, and  the  normal term remains.  In this region, the flavor 
oscillations appear    among the isolated 
neutrinos, and   are detected 
at the position  of much longer distance than 
those of the finite-size corrections.
Thus both effects  separate clearly, Fig.\ $(\ref{LSND :fig})$, and 
independent, and the flavor 
oscillation was not discussed in the present paper.

The amplitude and 
probability in the lowest
order of interactions were studied, except some of the pion' life time,  and  
 effect of electroweak gauge theory was not included.  
 They do not modify the long-distance effect of the paper that is due to  
the overlap of wave functions.  The effects would appear to other light 
particles around  overlapping regions, and give new insights to 
our understandings.
We will study these   problems  and other large scale 
physical phenomena  in subsequent papers.

%\section{References}

%\section{References}
    
  %%%%%%%%%%%%%%%%%%%%%%%%%%%%%%%%%%%%%%%%%%%%%%%%%%%%%%%%%%%%%%%%%%%%%%%%%%%
%\newpage
\section*{Acknowledgement}
This work was partially supported by a 
Grant-in-Aid for Scientific Research ( Grant No. 24340043).
Authors  thank Dr. Kobayashi, Dr. Nishikawa, Dr. Nakaya,
and Dr. Maruyama for useful discussions on 
the near detector of T2K experiment, Dr. Asai,
Dr. Kobayashi, Dr. Kawamoto, Dr. Komamiya, Dr. Minowa, Dr. Mori, and Dr. Yamada
for useful discussions on interferences. 

%\newpage
\appendix
\def\thesubsection{\Alph{section}-\Roman{subsection}}

%{\bf Appendix D: light-cone singularity }  

%%%%%%%%%%%%%%%%%%%%%%%%%%%%%%%%%%%%%%%%%%%%%%%%%%%%%%%%%%%%%%%%%%%%%
\section{Wave packets sizes}
%%%%%%%%%%%%%%%%%%%%%%%%%%%%%%%%%%%%%%%%%%%%%%%%%%%%%%%%%%%%%%%%%%%%
%%%%%%%%%%%%%%%%%%%%%%%%%%%%%%%%%%%%%%%%%%%%%%%%%%%%%%%%%%%%%%%%%%%
\subsection{Proton  mean free path}
%%%%%%%%%%%%%%%%%%%%%%%%%%%%%%%%%%%%%%%%%%%%%%%%%%%%%%%%%%%%%%%%%%%%
We estimate the wave packet size of a proton first and those of a pion
and a neutrino next, 
following a method of our previous works
\cite{Ishikawa-Shimomura,Ishikawa-Tobita-ptp}. 
 
A mean free path of a  charged particle in matter 
is determined by its  scattering  rate with atoms  by Coulomb interaction. An 
energy loss is also determined by the 
same cross section. Data on the energy loss   are summarized well  
in particle data summary \cite{particle-data} and are used  for 
the evaluation of the proton's mean free path. 

The proton's energy loss rate at a momentum, $1$\,[GeV/c], for several 
metals such as Pb, Fe, and
others are  
\begin{eqnarray}
-{d E \over d x}=1- 2 ~[\text{MeVg}^{-1}\text{cm}^2],   
\end{eqnarray}
hence we have the  mean free path of the $1\,[\text{GeV}/c]$ proton in
the material of a density $\rho$,
\begin{eqnarray}
\text{L}_\text{proton}={E \over {dE \over d x}\times \rho }= {1 ~[\text{GeV}]
 \over (1 - 2)\times
 10 ~[\text{MeV g}^{-1} \text{cm}^{-1}]} = 50 -  100~[\text{cm}].
\label{mfp-proton-1}
\end{eqnarray}
At a lower energy of the order of $0.2\,[\text{GeV}/c]$, the energy loss rate 
of the proton is about
$10\,[\text{MeVg}^{-1}\text{cm}^2]$ and the mean free path is  
\begin{eqnarray}
\text{L}_\text{proton}=10\,[\text{cm}].
\end{eqnarray}

A wave  which describes a proton  maintains coherence  in matter 
for  a  distance of the  mean free path, hence this wave is
approximately described by a wave packet that has a size of the mean
free path. We use
the  mean free path for  a wave packet size of the proton ${\sqrt \sigma}_\text{proton}$,
\begin{eqnarray}
{\sqrt \sigma}_\text{proton}= \text{L}_\text{proton}.
\end{eqnarray}

When a proton of this size  is emitted into the
vacuum or to a dilute gas from  matter, the wave  keeps the same size. 
The proton that is moving freely has a constant size  in vacuum or
dilute gas. The
size  varies when the proton is accelerated. If the potential energy
$\mathcal{V}$ is added to the proton  of momentum $\vec{p}_\text{before}$,
then the final value of the momentum
becomes $\vec{p}_\text{after}$ and satisfies 
\begin{eqnarray}
\sqrt{\vec{p}_\text{before}^{\,2}+m^2}+\mathcal{V}=\sqrt{\vec{p}_\text{after}^{\,2}+m^2}.
\label{potential-energy}
\end{eqnarray}
From Eq. $(\ref{potential-energy})$  variants of the momentum satisfy
\begin{eqnarray}
& &v_\text{before}\times|\delta \vec{p}_\text{before}|
=v_\text{after}\times|\delta \vec{p}_\text{after}|,\\
& &v_\text{before}=\frac{|\vec{p}_\text{before}|}{\sqrt{\vec{p}_\text{before}^{\,2}+m^2}},~
v_\text{after}=\frac{|\vec{p}_\text{after}|}{\sqrt{\vec{p}_\text{after}^{\,2}+m^2}}. 
\end{eqnarray}
Hence the size of a particle, $\sqrt {\sigma}_\text{before}$,
which is proportional to the inverse of ${|\delta \vec{p}}_\text{before}|$,
 becomes  ${\sqrt \sigma}_\text{after}$ after  the acceleration from a velocity 
$v_\text{before }$
to a velocity $v_\text{after}$. The wave packet size  is determined by the 
velocity ratio,
\begin{eqnarray}
{\sqrt \sigma}_\text{after}={\sqrt \sigma}_\text{before}\times{v_\text{after} \over
 v_\text{before}}.
\end{eqnarray}
The  velocity is bounded
by the light velocity $c$,
and a velocity ratio from $1\,[\text{GeV}/c]$ to $10\,[\text{GeV}/c]$
 is about $1.2$ and that from $0.2\,[\text{GeV}/c]$ to $10\,[\text{GeV}/c]$
 is about $5$. Hence the proton 
 of $10~[\text{GeV}/c]$ regardless of the energy in matter has the mean 
free path
\begin{eqnarray}
{\sqrt \sigma}_\text{proton}\approx 40 - 100~[\text{cm}],
\label{mfp-proton-2}
\end{eqnarray}   
in vacuum or dilute gas.

\subsection{Pion wave packet  }

Wave packet size  of pions which are produced by a proton collision 
with target nucleus is determined by the proton's  initial size
Eq. $(\ref{mfp-proton-2})$ and  a target size $10^{-15}$\,[m], which is
negligibly small. A pion is produced while the proton wave packet passes
through the small target by the strong interaction, hence this pion 
has a size in temporal direction of the proton wave packet. Hence  the 
size of pion wave packet, ${\sqrt \sigma}_\text{pion}$, is given from 
that of the proton,
 ${\sqrt \sigma}_\text{proton}$, in the form
\begin{eqnarray}
\frac{{\sqrt \sigma}_\text{proton}}{v_\text{proton}} = \frac{{\sqrt
 \sigma}_\text{pion}}{v_\text{pion}},~{\sqrt \sigma}_\text{pion}=
 \frac{v_\text{pion}}{v_\text{proton}}{\sqrt
 \sigma}_\text{proton}\approx {\sqrt \sigma}_\text{proton} .
\end{eqnarray}
In relativistic energy regions,
particles have the light velocity.
Consequently from Eq. $(\ref{mfp-proton-2})$,   pion's wave function 
of $1\,[\text{GeV}/c]$ or larger  momentum has the size  
\begin{eqnarray}
{\sqrt \sigma}_\text{pion}\approx 40 - 100~[\text{cm}].
\label{mfp-pion}
\end{eqnarray}   
We use this  value of  
Eq. $(\ref{mfp-pion})$ as the size of the wave packet 
\begin{eqnarray}
\sqrt{\sigma_{\pi}}={\sqrt \sigma}_\text{pion}.
\label{pion-size}
\end{eqnarray}

In vacuum and dilute gas, pions of the size Eq. $(\ref{pion-size})$ propagate freely.   From
Eqs. $(\ref{mfp-proton-1})$, $(\ref{mfp-proton-2})$ and $(\ref{mfp-pion})$, the
proton and pion have the sizes  of the order of $50 - 100$\,[cm].

%%%%%%%%%%%%%%%%%%%%%%%%%%%%%%%%%%%%%%%%%%%%%%%%%%%%%%%%%%%%%%%
\subsection{Neutrino  on target:   neutrino wave packet}
%%%%%%%%%%%%%%%%%%%%%%%%%%%%%%%%%%%%%%%%%%%%%%%%%%%%%%%%%%%%%
A neutrino interacts with a nucleon or an electron in the detector which are
constituent particles in  bound atoms and are expressed with wave 
functions of finite sizes. So the neutrino wave function in the 
amplitude of event that the neutrino is detected has a size of nucleus 
or atom.   
The   size of wave packet for the neutrino, therefore,  is not 
determined with a mean free path but with a size of the target in 
its detection process.   
 They are either   a size of a nucleons in a nucleus or that of an
 electron in an atom.
Nucleus have sizes of the order of $10^{-15}$\,[m] and  electron's wave
functions have 
sizes of the order of $10^{-11}$\,[m].  So neutrino wave packet is either $10^{-11}$\,[m] or $10^{-15}$\,[m].

Interactions of  muon neutrinos in detectors are 
\begin{eqnarray}
& &\nu_{\mu}+e^{-} \rightarrow e^{-}+\nu_{\mu}, 
\label{numu-leptonic}\\
& &\nu_{\mu}+e^{-} \rightarrow \mu^{-}+\nu_{e}, 
\label{numu-lcharged}\\ 
& &\nu_{\mu}+A \rightarrow \mu^{-}+(A+1)+X, 
\label{numu-hcharged}\\ 
& &\nu_{\mu}+A \rightarrow \nu_{\mu} +A  +X,
\label{numu-hneutral}
\end{eqnarray}
hence the size of the neutrino wave packet  $\sqrt{\sigma_{\nu}}$ in 
processes $(\ref{numu-leptonic})$ and
$(\ref{numu-lcharged})$ is of the order of  $10^{-11}$,[m]
\begin{eqnarray}
\sqrt{\sigma_{\nu}} =10^{-11}~[\text{m}],
\label{mfp-neutrino-e}
\end{eqnarray}
 and the neutrino 
wave packet  $\sqrt{\sigma_{\nu}}$ in processes $(\ref{numu-hcharged})$ and
$(\ref{numu-hneutral})$ is of the order of  $10^{-15}$\,[m]
\begin{eqnarray}
\sqrt{\sigma_{\nu}} =10^{-15}~[\text{m}].
\label{mfp-neutrino-N}
\end{eqnarray}

Interactions of  electron neutrinos in detectors are 
\begin{eqnarray}
& &\nu_{e}+e^{-} \rightarrow e^{-}+\nu_{e}, 
\label{nue-leptonic}\\ 
& &\nu_{e}+A \rightarrow e^{-}+(A+1) +X,
\label{nue-hcharged}\\ 
& &\nu_{e}+A \rightarrow e  +A  +X.
\label{nue-hneutral}
\end{eqnarray}
The neutrino 
wave packet  $\sqrt{\sigma_{\nu}}$ in processes $(\ref{nue-leptonic})$
is of the order of  $10^{-11}$\,[m], Eq. $(\ref{mfp-neutrino-e})$, and the neutrino 
wave packet  $\sqrt{\sigma_{\nu}}$ in processes $(\ref{nue-hcharged})$ and
$(\ref{nue-hneutral})$ is of the order of  $10^{-15}$\,[m],
Eq. $(\ref{mfp-neutrino-N})$. They are treated in
the same way as the neutrino from the pion  decay.

From Eqs. $(\ref{mfp-neutrino-e})$ and $(\ref{mfp-neutrino-N})$, 
the neutrino has the wave packet sizes of the order of $10^{-11}$\,[m]
or  $10^{-15}$\,[m].

For various nucleus the sizes are estimated from the nucleus size,
$A^{2/3}/m_{\pi}^2$ as
\begin{eqnarray}
\sigma_{\nu}=\begin{cases}
	      5.2/m_{\pi}^2   ;C,\\
            6.35/m_{\pi}^2;O, \\
     14.3/m_{\pi}^2;Fe,          \\
     18.9/m_{\pi}^2 ;Pb.        
\end{cases}
\end{eqnarray}
In the amplitude of the event that  neutrinos is detected, the neutrino 
wave packet  is determined by the size of nucleus in the detector.  In this 
respect, the neutrino wave packet of the present work is different from 
some  previous works of wave packets that are connected with flavor neutrino
oscillations \cite{Kayser,Giunti,Nussinov,
Kiers,Stodolsky,Lipkin,Akhmedov,Asahara}, where one particle properties of
neutrino at productions are studied and the detection process was not considered.

%%%%%%%%%%%%%%%%%%%%%%%%%%%%%%%%%%%%%%%%%%%%%%%%%%%%%%%%%%%%%%%%%%%%%%%
\subsection{Charged particles on target: wave packets}
%%%%%%%%%%%%%%%%%%%%%%%%%%%%%%%%%%%%%%%%%%%%%%%%%%%%%%%%%%%%%%%%%%%%
Charged particles are detected from signals caused by their
electromagnetic interactions 
with atoms in matter. The electromagnetic interactions are mediated by 
massless photons and the forces are long range and are much stronger
than the weak interaction. Hence  successive
interactions with many atoms, which are correlated quantum mechanically 
each other,
give signals. Thus the wave packet sizes of the charged 
particles would be  much larger than the size of an atom.  It would
be reasonable to assume that the size is  semi-microscopic, some
number of the order of one times $10^{-10}$\,[m]. This size might 
agree to those that
have been considered before in textbooks
\cite{Goldberger,newton,taylor,Sasakawa}. Although these  sizes  are
much larger than those of the neutrinos,
their diffraction components  are extremely small and negligible 
due to their large masses, and vanish at the 
macroscopic distance. 
Consequently the finite-size corrections may be negligible for 
charged particles. If a high-energy electron is detected with
exceptionally large wave packet, an effect may appear.



%% References


%\bibliographystyle{model1a-num-names}

\begin{thebibliography}{00}

\bibitem{LSZ} H.~Lehman, K.~Symanzik, and W.~Zimmermann, Il Nuovo~Cimento~(1955-1965).~
\textbf{1},~205 (1955).  LSZ, text books on quantum fields theory \cite{qft-texts1, qft-texts2}, and
  general arguments  on scatterings 
\cite{Goldberger,newton,taylor,Sasakawa} study the  large wave
packets which 
can be expressed without  positions.
In small wave packets  used here,  the positions are explicitly written.  

\bibitem{Low} F.~Low, Phys. Rev. \textbf{97}, 1392 (1955).

\bibitem{Ishikawa-Tobita-ptep} K.~Ishikawa and Y.~Tobita,
Prog. Theor. Exp. Phys. \textbf{2013}, 073B02 (2013).

\bibitem{goldberger-watson-paper}  M.~L.~Goldberger and K.~M.~Watson, 
   Phys. Rev. \textbf{136},
	1472 (1964).
\bibitem{Ishikawa-Shimomura} K. Ishikawa and T. Shimomura,
Prog. Theor. Phys.  \textbf{ 114}, 1201 (2005) [hep-ph/0508303].


\bibitem{Goldberger}  M.~L.~Goldberger and K.~M.~Watson, 
{\it Collision Theory}~(John Wiley \& Sons, Inc. New York, 1965).

\bibitem{newton}  R.~G.~Newton, 
\textit{Scattering Theory of Waves and Particles}~(Springer-Verlag, New York, 1982).

\bibitem{taylor}  J.~R.~Taylor, 
\textit{Scattering Theory: The Quantum Theory of non-relativistic
	Collisions}~(Dover  Publications, New York, 2006).

\bibitem{Sasakawa} T.~Sasakawa, Prog. Theor. Phys.  Suppl. \textbf{11}, 69 (1959).


\bibitem{qft-texts1}
See for instance,
 M.~E.~Peskin and D. V. Schroeder, 
\textit{An Introduction to Quantum Field Theory}, ~Sect. 7.2, ~p. 222,
~Addison-Wesley Publishing Company, California, 1995;
 S.~Weinberg, \textit{The Quantum Theory of Fields I},~p. 109,
~Press Syndicate of the University of Cambridge,~New York,~1996;
 M.~Srednicki,
\textit{Quantum Field Theory},~p. 37,~Cambridge University press,~Cambridge,~2007.
\bibitem{qft-texts2}
See for instance,
 N.~N.~Bogolubov, et. al.,
 \textit{General Principles of Quantum Field Theory},
~Kluwer Academic Publishers,~Dordrecht (1990); 
R.~Haag, \textit{Local Quantum Physics},~Springer,~Berlin (1992); H.~Araki,
\textit{Mathematical Theory of Quantum Field},~Iwanami,~Tokyo,~2002.


\bibitem{Sakai-1949}
S.~Sasaki,~S.~Oneda,~and~S.~Ozaki,
~The~Science~Reports~of~the~Tohoku~University
~First~series~(Math.,~Phys.,~Chem.,~Astronomy)
~\textbf{XXXIII},~77~(1949).
\bibitem{Jack} J.~Steinberger, Phys. Rev. \textbf{76}, 1180,(1949).

\bibitem{Ruderman} M.~Ruderman and R. Finkelstein,  Phys. Rev. \textbf{76},
	1458 (1949).


\bibitem{Anderson} H.~L.~Anderson, {\it et al.},  Phys. Rev. \textbf{119},
	2050 (1960).
\bibitem{SK-Atom}
J.~Hosaka, {\it et al.},~Phys. Rev.  \textbf{D74},~032002~(2006) [arXiv:hep-ex/0604011].

\bibitem{SK-Solar}
%The Super-Kamiokande Collaboration,
S.~Fukuda, {\it et al.},~Phys.~Lett.~\textbf{B539},~179~(2002) [arXiv:hep-ex/0205075].

\bibitem{SNO-NC}
S.~N. Ahmed, {\it et al.},
Phys. Rev. Lett.~\textbf{92}, 181301~(2004) [arXiv:nucl-ex/0309004].


\bibitem{KamLAND-Reactor}
T.~Araki, {\it et al.},
Phys. Rev. Lett.~\textbf{94},~081801~(2005) [arXiv:hep-ex/0406035].

\bibitem{Borexino}
E.~A. Litvinovich,
Phys. Atom. Nucl.~\textbf{72},~522~(2009); C. Arpesella, \textit{et al.}, Phys. Rev. Lett. \textbf{101}, 091302 (2008) [arXiv:0805.3843].

\bibitem{K2K}
E.~Aliu, {\it et al.},~Phys. Rev. Lett.~\textbf{94},~081802~(2005) [arXiv:hep-ex/0411038].

\bibitem{particle-data}
  J. Beringer, {\it et al.}  [Particle Data Group],
Phys. Rev. \textbf{D86}, 010001 (2012).

\bibitem{Tritium}
V.~N.~Aseev, {\it et al.},~Phys.~Rev.~\textbf{D84},~112003 (2011)
[arXiv:1108.5034[hep-ex]].

\bibitem{WMAP-neutrino}
E. Komatsu, {\it et al.},
Astrophys. J. Suppl.~\textbf{192},~18 (2011) [arXiv:1001.4538[astro-ph.CO]].

\bibitem{dollard} J. D. Dollard, Commun. Math. Phys. \textbf{12}, 193 (1969).

\bibitem{Gell-mann}  M.~Gell-Mann, \textit{The Quark and the Jaguar:Adventures in the Simple and the Complex},
St.Martin's Griffin,London,1995 ,ILL ed. 

\bibitem{tonomura} A.~Tonomura, {\it et al.},~Ameri.~J.~Physics. \textbf{57}
	No2, 117 (1989).


\bibitem{Tomonaga} S.~Tomonaga,~Prog.~Theor.~Phys., \textbf{1}, 27 (1946).

\bibitem{Schwinger} J.~Schwinger,~Phys.~Rev. \textbf{74}, 1439 (1948).

\bibitem{Schiff and Landau}  L. I.~Schiff, \textit{Quantum Mechanics}, p. 197,
	McGRAW-HiLL, New-York, 2003;
~L. D. Landau and E.M. Lifshitz, \textit{Quantum
	Mechanics}  p.157, Butterworth Heine Mann, New York, 2003.

\bibitem{Dirac} P. A. M. Dirac, The Quantum Theory of the Emission and
	Absorption of Radiation. Pro. R. Soc. Lond. A 114, 243 (1927).
\bibitem{Schiff-golden}
L. I. Schiff, 
{\it  Quantum Mechanics}, p. 199, McGRAW-Hill
Book COMPANY,Inc.  New York, 1955.   
\bibitem{peierls} R. Peierls, {\it Surprises in Theoretical
	Physics},p. 121  (Princeton University Press, New Jersey, 1979)


\bibitem{Moller} C.~Moller,~DET DANSKE VID.SEL.MATH-PHY. XXIII, Nr.1,1 

\bibitem{Wilson-OPE} K.~Wilson, in Proceedings of the Fifth International Symposium on Electron and Photon Interactions at High Energies, Ithaca, New York, 1971,~p.115~(1971).
% edited by N. B. Mistry (Cornell Univ. Press, Ithaca, New York, 1971 
See also  N.~N.~Bogoliubov and D.~V.~Shirkov, 
{\it Introduction to the Theory of Quantized  Fields}~(John Wiley \& Sons, Inc. New York, 1976).




\bibitem{Ishikawa-Tobita-ptp} K. Ishikawa and Y. Tobita,
	Prog. Theor. Phys.  \textbf{122}, 1111 (2009) [arXiv:0906.3938[quant-ph]].
\bibitem{Ishikawa-Tobita} K.~Ishikawa and Y.~Tobita, [arXiv:0801.3124];
``Neutrino mass and mixing'' in the 10th Inter. Symp. on ``Origin of Matter
and Evolution of Galaxies'' AIP Conf. proc. 1016, 329 (2008).




\bibitem{Kayser} B.~Kayser,~Phys.~Rev.~\textbf{D24},~110 (1981);~Nucl.~Phys.~\textbf{B19}~(Proc. Suppl),~177 (1991).

\bibitem{Giunti} C.~Giunti, C.~W.~Kim, and U.~W.~Lee, Phys. Rev. \textbf{D44}, 3635 (1991).
\bibitem{Nussinov} S.~Nussinov, Phys. Lett. \textbf{B63}, 201 (1976).
\bibitem{Kiers} K.~Kiers, S.~Nussinov, and N.~Weiss, 
	Phys. Rev. \textbf{D53}, 537 (1996) [hep-ph/9506271].
\bibitem{Stodolsky} L.~Stodolsky,  Phys. Rev. \textbf{D58}, 036006 (1998) [hep-ph/9802387].

\bibitem{Lipkin} H.~J.~Lipkin,~Phys.~Lett.~\textbf{B642},~366 (2006) [hep-ph/0505141].

\bibitem{Akhmedov} E. K.  Akhmedov,~JHEP. \textbf{0709}, 116 (2007) [arXiv:0706.1216 [hep-ph]].

\bibitem{Asahara} A.~Asahara, K.~Ishikawa, T.~Shimomura, and T.~Yabuki,
Prog. Theor. Phys. \textbf{113}, 385 (2005) [hep-ph/0406141]; T.~Yabuki and K.~Ishikawa,
Prog. Theor. Phys. \textbf{108}, 347 (2002).


\bibitem{excess-total-detectorMino}
P. Adamson, \textit{et al.}, 
%Neutrino and Antineutrino Inclusive Charged-current
%	Cross Section Measurements with the MINOS Near
%	Detector.
 Phys. Rev. \textbf{D81}, 072002 (2010) [arXiv:0910.2201[hep-ex]].

\bibitem{excess-total-detectorNOMAD}
Q.~Wu, \textit{et~al.},
% A Precise Measurement of the Muon Neutrino-Nucleon
%	Inclusive Charged Current Cross-Section off an Isoscalar Target
%	in the Energy Range $ 2.5 < E_{\nu} < 40$ GeV by NOMAD. 
 Phys. Lett. \textbf{B660}, 19 (2008) [arXiv:0711.1183 [hep-ex]].

%\bibitem{particle-data}
%  J. Beringer {\it et al.},  [Particle Data Group],
%Phys. Rev. \textbf{D86}, 010001 (2012).

\bibitem{excess-qenear-detectorMini}
A.~A.~Aguilar-Arevalo, \textit{et~al.},~Phys. Rev. \textbf{D81},~092005~(2010) [arXiv:1002.2680[hep-ex]].

\bibitem{T2K-nue}
K. Abe \textit{et al}, Phys. Rev. Lett. \textbf{107}, 041801 (2001) [arXiv:1106:2822 [hep-ex]].

\bibitem{ICARUS-NEAR}
M. Antonello, \textit{et al.}, arXiv:1203:3432 [physics].

\bibitem{excess-LSND}
C.~Athanassopoulos, \textit{et~al.},~Phys. Rev. Lett. \textbf{75},~2650 (1995) [nucl-ex/9504002];
  \textbf{77},~3082 (1996) [nucl-ex/9605003];
	\textbf{81},~1774 (1998) [nucl-ex/9709006].

\bibitem{excess-two-neutrino}
G.~Danby, \textit{et~al.},~Phys. Rev. Lett. \textbf{9},~36~(1962).


\bibitem{excess-near-detectorK2K}
M.~H.~Ahn, \textit{et~al.}, ~Phys. Rev. \textbf{D74},~072003 (2006) [hep-ex/0606032].
\bibitem{excess-near-detectorMini}
A.~A.~Aguilar-Arevalo, \textit{et~al.},~Phys. Rev. \textbf{D79},~072002~(2009) [arXiv:0806.1449 [hep-ex]].
\bibitem{excess-near-detectorMino}
P.~Adamson, \textit{et~al.},~Phys. Rev. \textbf{D77},~072002 (2008) [arXiv:0711.0769[hep-ex]].






 \end{thebibliography}

\end{document}